\documentstyle[psfig]{mn}

\begin{document}

\journal{To Appear in MNRAS}

\title[X-ray emission lines] {X-ray emission lines and multiphase gas
  in elliptical galaxies and galaxy clusters} 
\author[D.~A. Buote et al.]{David A. Buote$^{1,3,4}$,  Claude R.
  Canizares$^2$, and A. C. Fabian$^1$\\  
$^1$ Institute of Astronomy, Madingley Road, Cambridge CB3 0HA\\ 
$^2$ Department of Physics and Center for Space Research 37-241,
  Massachusetts Institute of Technology, 77 Massachusetts Avenue, \\
  Cambridge, MA 02139, U.S.A.\\
$^3$ UCO/Lick Observatory, University of California at Santa Cruz,
  Santa Cruz, CA  95064, U.S.A.\\
$^4$ {\sl Chandra} Fellow}

\maketitle

\begin{abstract}

We examine the K shell emission lines produced by isothermal and
simple multiphase models of the hot gas in elliptical galaxies and
galaxy clusters to determine the most effective means for constraining
the width of the differential emission measure ($\xi(T)$) in these
systems which we characterize by a dimensionless parameter,
$\sigma_\xi$. Comparison of line ratios of two-temperature
$(\sigma_\xi\ll 1)$ and cooling flow $(\sigma_\xi\sim 1)$ models is
presented in detail. We find that a two-temperature model can
approximate very accurately a cooling flow spectrum over 0.5-10 keV.

We have re-analyzed the {\sl ASCA} spectra of three of the brightest
galaxy clusters to assess the evidence for multiphase gas in their
cores: M87 (Virgo), the Centaurus cluster, and the Perseus
cluster. K$\alpha$ emission line blends of Si, S, Ar, Ca, and Fe are
detected in each system as is significant Fe K$\beta$ emission.  The
Fe K$\beta$/K$\alpha$ ratios are consistent with optically thin plasma
models and do not suggest resonance scattering in these systems.
Consideration of both the ratios of H-like to He-like K$\alpha$ lines
and the local continuum temperatures clearly rules out isothermal gas
in each case. To obtain more detailed constraints we fitted plasma
models over 1.6-9 keV where the emission is dominated by these K shell
lines and by continuum. In each case the {\sl ASCA} spectra cannot
determine whether the gas emits at only two temperatures or over a
continuous range of temperatures as expected for a cooling flow. The
metal abundances are near solar for all of the multiphase models.  We
discuss the implications of these results and examine the prospects
for determining the temperature structure in these systems with
upcoming X-ray missions.

\end{abstract}

\begin{keywords}
galaxies: general -- galaxies: evolution -- X-rays: galaxies.
\end{keywords}
 
\section{Introduction}
\label{intro}

There is mounting evidence that the X-ray emission in elliptical
galaxies and galaxy clusters originates from multiphase plasma.  The
first important evidence for multiphase gas in ellipticals and
clusters came from analysis of emission lines in M87 and the Perseus
cluster.  Using ratios of several blends of Fe L shell lines obtained
from the Focal Plane Crystal Spectrometer (FPCS) on the {\it Einstein}
Observatory, Canizares et al.  \shortcite{crc82} found that the hot
gas in M87 emits over a large range of temperatures: from $\sim
2-4\times 10^6$ K to $\sim 3\times 10^7$ K. The strength of the Fe
XVII line measured from FPCS data of the Perseus cluster also suggests
a multiphase plasma (Canizares, Markert, \& Donahue 1988).

Recently, data from the {\sl ROSAT} \cite{rosat} and {\sl ASCA}
\cite{tanaka} satellites have provided new evidence for multiphase hot
gas. Broad-band spectral analysis of {\sl ASCA} data of clusters
(Fukazawa et al. 1994; Fabian et al. 1994a), the brightest ellipticals
(Buote \& Fabian 1998; Buote 1999a), and of the brightest poor galaxy
groups \cite{bgroup} indicates that the hot gas in the central regions
of these systems consists of at least two temperature components. The
equivalent widths and ratios of K$\alpha$ line blends of Si and S
alone strongly favor multitemperature models of the {\sl ASCA} data of
the brightest ellipticals \cite{bmulti}. 

Complementary evidence for multiphase gas in clusters is provided by
the analysis of imaging data from {\sl ROSAT} \cite{rosat} of cooling
flow clusters which infers mass deposition rates which vary with
radius approximately as $\dot{M}\propto r$ (e.g. Peres et al. 1998).
This type of distributed mass deposition is the signature of
multiphase cooling flows (for a review see Fabian 1994). Multiphase
cooling flows are also consistent with the positive temperature
gradients observed in the centers of the brightest elliptical galaxies
with {\sl ROSAT} (e.g. Forman et al. 1993; David et al. 1994;
Trinchieri et al. 1994; Rangarajan et al. 1995; Irwin \& Sarazin 1996;
Jones et al. 1997; Buote 1999a).

It is essential that the temperature structure be accounted for in
order to deduce correctly the properties of the hot gas in these
systems.  The metal abundances are especially sensitive to the assumed
temperature structure.  For example, analyses of the X-ray spectra of
elliptical galaxies where the hot gas is assumed to be isothermal
always obtain very sub-solar metallicities, $Z\la 0.3Z_{\sun}$,
whereas when multi-temperature models are considered the metallicity
of ellipticals becomes consistent with solar (see Buote \& Fabian 1998
and Buote 1999a). A similar situation prevails for poor groups of
galaxies \cite{bgroup}.

Although this ``Fe bias'' is unimportant for clusters of galaxies with
temperatures in excess of $\sim 3$ keV, if an isothermal gas is
assumed for a cooling-flow cluster (or elliptical or group) then the
Si/Fe and S/Fe ratios can be significantly over-estimated (see the
appendix in Buote 1999b). Hence, in order to obtain reliable
measurements of the heavy element abundances crucial to interpreting
models of the formation, evolution, and enrichment of the hot gas in
ellipticals, groups, and clusters (e.g. David, Forman, \& Jones 1991;
Loewenstein \& Mathews 1991; Ciotti et al. 1991; Loewenstein \&
Mushotzky 1996; Renzini 1997; Brighenti \& Mathews 1999), it is
necessary to know the true temperature structure of the hot gas.

Knowledge of the temperature structure is also required to understand
the contribution of discrete sources to the hard X-ray emission in
elliptical galaxies.  For example, if the hot gas is assumed to be
isothermal then the hard spectral component measured by {\sl ASCA} for
the elliptical galaxy NGC 3923 is inconsistent with the expectation
that discrete sources follow the same profile as the optical light
(Buote \& Canizares 1998; see also Pellegrini 1999).  Finally, when
multiphase cooling flows are accounted for in X-ray observations of
clusters the scatter in the X-ray luminosity-temperature relationship
is significantly reduced (e.g. Fabian et al. 1994b).

The Fe L shell emission lines from $\sim 0.7$ - 1.4 keV generally
provide the strongest constraints on the temperature structure of
ellipticals, groups, and clusters. There has been considerable
progress in improving the accuracy of the theoretically computed Fe L
shell transitions (e.g Liedahl et al.  1995), but they are still among
the most uncertain transitions implemented in the currently available
plasma codes (e.g. Savin et al. 1999). There is reason for some
confidence in the codes since recent empirical studies have concluded
that the errors in the Fe L transitions are not overly serious (Hwang
et al. 1997; Buote \& Fabian 1998). (However, see the recent papers by
Savin 1999 and Savin et al. 1999 for discussions of the inaccuracies
in the dielectronic recombination rates used by the popular plasma
codes which affect all types of emission lines.)

Nevertheless, it is important to use lines from the more-accurate K
shell transitions to search for evidence of multiphase gas to
complement the information from Fe L lines.  To obtain the most
accurate constraints on the emission measure of a multiphase plasma it
is necessary to obtain independent measurements of the temperature(s)
over the entire energy range of significant emission. It is a happy
coincidence that the K shell lines of many abundant elements appear
throughout the X-ray wave band (0.1-10 keV).

Line ratios have important advantages over broad-band spectral fitting
for probing the temperature structures of ellipticals and clusters.
Firstly, in an isothermal plasma using ratios of lines of the same
element avoids uncertainties in the relative abundances of the
elements; there is a small dependence on metallicity in a multiphase
cooling flow (section \ref{cfs}).  Secondly, ratios of lines nearby in
energy are essentially independent of Galactic or any intrinsic
absorption. Thirdly, since the strength of a line can be computed
using a local continuum measurement, line ratios are insensitive to
any poorly constrained power-law or bremsstrahlung spectral
components.

For a multiphase {\it coronal} plasma \footnote{The assumptions of the
  standard coronal model are discussed in, e.g., Sarazin
  \shortcite{sarazin} and Mewe \shortcite{mewe91}.}  the luminosity of
an emission line arising from a transition from an upper state $u$ to
a lower state $l$ of an element $X$ in ionization state $r$ may be
expressed as,
\begin{equation}
L_{ul}(X_r) = \int \Lambda_{ul}(X_r,T)\xi(T)\,dT, \label{eqn.line}
\end{equation}
where $\Lambda_{ul}(X_r,T)$ is the plasma emissivity of the line at
temperature $T$ which incorporates the abundance of element $X$, the
relative fraction of ionization state $r$, and other quantities
depending only on the atomic physics. (The integral extends over the
emitting volume $V$.) The quantity,
\begin{equation} 
\xi(T)\equiv n^2_e(T)\frac{dV}{dT}, \label{eqn.dem}
\end{equation}
is called the ``differential emission measure'' and $\xi(T)dT$ the
``emission measure'' at temperature $T$. The emission measure weights
$\Lambda$ as a function of temperature in equation (\ref{eqn.line})
and thus effectively characterizes the temperature structure of the
plasma.  Notice that different phases $T$ in general can have
different electron number densities $n_e$.

It has been known for some time \cite{cb} that precise constraints on
$\xi(T)$ from inversion of equation (\ref{eqn.line}) are difficult to
obtain in practice.  (For a recent discussion of this topic see, e.g.,
Judge, Hubeny, \& Brown 1997.)  That is, for a given line typically
$\Lambda_{ul}(X_r,T)$ is significant over temperature ranges of
factors of $\sim 2$-$5$, which means that the plasma emissivity acts
as a smoothing function which obscures the detailed shape of $\xi(T)$.
Moreover, small uncertainties in the data are amplified by the
inversion. Given these difficulties, it may only be possible to obtain
relatively crude constraints on the shape of $\xi(T)dT$ for
ellipticals and clusters from inversion of equation (\ref{eqn.line});
e.g., width, skewness, etc \footnote{See, e.g., Kaastra et al.
\shortcite{kas} and Schmitt et al.  \shortcite{schmitt} for recent
constraints on $\xi(T)$ of stellar coronae.}.

The primary theoretical goal of this paper is to identify the best
X-ray emission line ratios for determining the gross features of
$\xi(T)$ of ellipticals and clusters with the specific objective of
distinguishing between single-temperature, two-temperature, and
cooling-flow spectra. (Previous discussions of important temperature
sensitive lines for clusters in the context of a different set of
models may be found in Bahcall \& Sarazin \shortcite{bs77}, Sarazin \&
Bahcall \shortcite{sb77}, and Canizares et al. \shortcite{crc82}. Wise
\& Sarazin \shortcite{ws93} discuss the radial variation of the shapes
of emission lines as diagnostics for cooling flow models.) To
facilitate quantitative comparison of the multiphase structure of
different ellipticals and clusters we introduce a quantity,
$\sigma_\xi$, the ``multiphase strength''.  We focus on K shell
emission lines since they are generally the strongest lines of
ellipticals and clusters and the atomic physics is relatively simple
and so the available plasma codes should be most accurate for these
transitions.  Because the Fe L shell lines occurring at energies $\sim
0.7-1.4$ keV are strong for ellipticals and clusters and are powerful
as temperature diagnostics, we also examine them even though at the
present time the accuracy of the plasma codes is less certain for
these transitions (see above). The line ratios allow simple yet
effective discrimination between popular spectral models of
ellipticals and clusters and serve as an important guide to
interpreting results of $\xi(T)$ obtained from inversion of equation
(\ref{eqn.line}).

The paper is organized as follows. In Section \ref{theory} we analyze
theoretical line ratios in single-phase (Section \ref{iso}) and
multiphase (Section \ref{multi}) models. The multiphase strength is
introduced in Section \ref{mps} and it is used to elucidate the
temperature structure of simple multiphase models in Section
\ref{examples} and of cooling flow models in Section \ref{cfs}. We
discuss in detail how to distinguish a two-temperature plasma from a
cooling flow in Section \ref{2tcf}.  In Section \ref{asca} we compute
and analyze line ratios of {\sl ASCA} data of M87, the Centaurus
cluster, and the Perseus cluster. Simulated observations with {\sl
Chandra}, {\sl XMM}, and {\sl ASTRO-E} are considered in Section
\ref{sims}.  Finally, we comment on recent evidence for resonance
scattering in \ref{res} and discuss cosmological application of the
multiphase strengths of clusters in Section \ref{omega}. In Section
\ref{conc} we present our conclusions.

\section{Theoretical line ratios of ellipticals and clusters}
\label{theory}

We examine the theoretical emission lines predicted by coronal plasma
models of ellipticals, groups, and clusters. In order to maintain a
connection to feasible observations within the next 10 years, for the
purposes of identifying lines we consider spectra folded through an
energy resolution of 2 eV which is the targeted performance of the
{\sl Constellation-X} mission
\footnote{http://constellation.gsfc.nasa.gov/}.  To model the emission
of a coronal plasma we use the MEKAL code which is a modification of
the original MEKA code (Mewe, Gronenschild, \& van den Oord 1985;
Kaastra \& Mewe 1993) where the Fe L shell transitions have been
re-calculated \cite{mekal}. We fix the relative abundances of the
elements to the (photospheric) solar values \cite{ag} and, as a
result, we focus on ratios of lines of the same element so that our
results are not sensitive to this assumption.  Finally, we do not
include photo-electric absorption (Galactic or intrinsic) in this
discussion\footnote{We discuss the possible complication of resonance
scattering in Section \ref{res}.} but include its effects when
interpreting line ratios from real data (as in Section \ref{asca}).

\begin{table}
\caption{K Shell Transitions for H-like and He-like Ions}
\label{tab.notation}
\begin{tabular}{cclll}
Isoelectronic &  Transition\\
Sequence & Symbol & \multicolumn{3}{c}{Transition}\\
H  & 1 & $\rm 1s$   $\rm ^2S_{\frac{1}{2}}$ & $-$ & $\rm 2p$  $\rm ^2P_{\frac{1}{2},\frac{3}{2}}$\\ 
   & 2 & $\rm 1s$  $\rm ^2S_{\frac{1}{2}}$ & $-$ & $\rm 3p$  $\rm ^2P_{\frac{1}{2},\frac{3}{2}}$\\ 
   & 3 & $\rm 1s$  $\rm ^2S_{\frac{1}{2}}$ & $-$ & $\rm 4p$  $\rm ^2P_{\frac{1}{2},\frac{3}{2}}$\\ 
   & 4 & $\rm 1s$  $\rm ^2S_{\frac{1}{2}}$ & $-$ & $\rm 5p$  $\rm
^2P_{\frac{1}{2},\frac{3}{2}}$\\
He & 1 & $\rm 1s^2$  $\rm ^1S_0$ & $-$ & $\rm 1s5p$  $\rm ^1P_1$\\ 
   & 2 & $\rm 1s^2$  $\rm ^1S_0$ & $-$ & $\rm 1s4p$  $\rm ^1P_1$\\ 
   & 3 & $\rm 1s^2$  $\rm ^1S_0$ & $-$ & $\rm 1s3p$  $\rm ^1P_1$\\ 
   & 4 & $\rm 1s^2$  $\rm ^1S_0$ & $-$ & $\rm 1s2p$  $\rm ^1P_1$\\ 
   & 5 & $\rm 1s^2$  $\rm ^1S_0$ & $-$ & $\rm 1s2p$  $\rm ^3P_{2,1}$\\ 
   & 6 & $\rm 1s^2$  $\rm ^1S_0$ & $-$ & $\rm 1s2s$  $\rm ^3S_1$\\ 
\end{tabular}

\medskip

Transitions are given in the standard notation of the electron
configuration (orbitals) and the level with the lower state listed
first.  H1 and He4-6 are all K$\alpha$ transitions while H2 and He3
are K$\beta$ transitions.

\end{table}

The emission lines which are perhaps the most useful for our study of
the temperature structure of ellipticals and clusters \footnote{A
summary of standard temperature diagnostics of coronal plasmas may be
found in Mewe \shortcite{mewe91}.}  are the K shell transitions
involving ions of the H-like and He-like isoelectronic sequences
listed in Table \ref{tab.notation} using the standard notation
(e.g. Table 3 of Mewe et al.  1985). Also very useful are the
satellites to the resonance line of the preceding ion in the
ionization sequence. Of particular interest are the satellite lines to
the He4 resonance from dielectronic recombination to autoionizing
states of the Li-like ion. The strongest of these satellite lines
typically come from the multiplet $\rm 1s^22p$ $\rm ^2P$ - $\rm
1s2p^2$ $\rm ^2D$.  Finally, the L shell transitions of Fe between
$\sim 0.7$ - 1.4 keV (especially the 2s-3p lines) are strong and very
temperature sensitive, though most are unresolved at 2 eV resolution
and thus blends of lines are emphasized below.

We identified significant K and Fe L shell lines between 0.4 - 10 keV
from visual inspection of isothermal MEKAL models. These models were
viewed at 2 eV resolution and had temperatures ranging from 0.1 - 10
keV\footnote{Throughout the paper we quote temperatures in units of
  keV $(k_BT)$.}. We restricted our attention to this energy range
because below $\sim 0.4$ keV Galactic photo-electric absorption is
prohibitive, and energies above $\sim 10$ keV will not be effectively
probed by high energy resolution X-ray satellites expected to operate
in the near future.

For lines that are not isolated at the 2 eV resolution, our chosen
energy boundaries in many cases enclose other emission lines from the
same element or others. The result is that the line ratios we discuss
below often behave differently over some temperature ranges than
expected if there were no blending of lines.  However, we enforce that
such peculiarities are not the result of uncertainties in defining the
continuum because the line luminosities are computed from the models
using a finer energy resolution which allows the continuum to be
accurately subtracted.  Within the energy boundaries used to define a
line (or blend) we compute the total emission above the continuum on
an energy grid of 15000 logarithmically spaced energy bins over 0.4 -
10 keV; i.e. $\sim 0.2$ eV bins around 1 keV and $\sim 1.5$ eV bins
around 7 keV.

Our procedure to define the continuum for the lines of interest begins
by identifying from visual examination of the spectrum a nearby bin at
lower energy having no apparent lines. This choice is made after
examining spectra of several temperatures. (Note that we avoid energy
bins having strong recombination edges.) Let us denote the continuum
emission at this energy bin, $j_c(E_i)$, where $E_i$ is the energy of
this bin $i$. Proceeding to the next bin $(i+1)$ with energy
$E_{i+1}$, if there is no line emission in this bin, then the total
emission $j(E_{i+1})$ is all continuum such that $j(E_{i+1})\le
j_c(E_{i})$ since (at a fixed temperature) the (free-free) continuum
is a monotonically decreasing function of energy. In this case,
$j_c(E_{i+1})\equiv j(E_{i+1})$ becomes our new definition of the
continuum. If there is line emission in bin $i+1$ so that $j(E_{i+1})>
j_c(E_{i})$, then we set $j_c(E_{i+1})\equiv j_c(E_{i})$. This
procedure is followed up to and through the lower boundary of the
emission line of interest until the upper boundary of the line region
is reached. The continuum values $(j_c(E_{i}),j_c(E_{i+1}),\ldots)$
are subtracted from the corresponding bins of the line region.

\subsection{Isothermal case}
\label{iso}

\begin{figure*}
\parbox{0.49\textwidth}{
\centerline{\psfig{figure=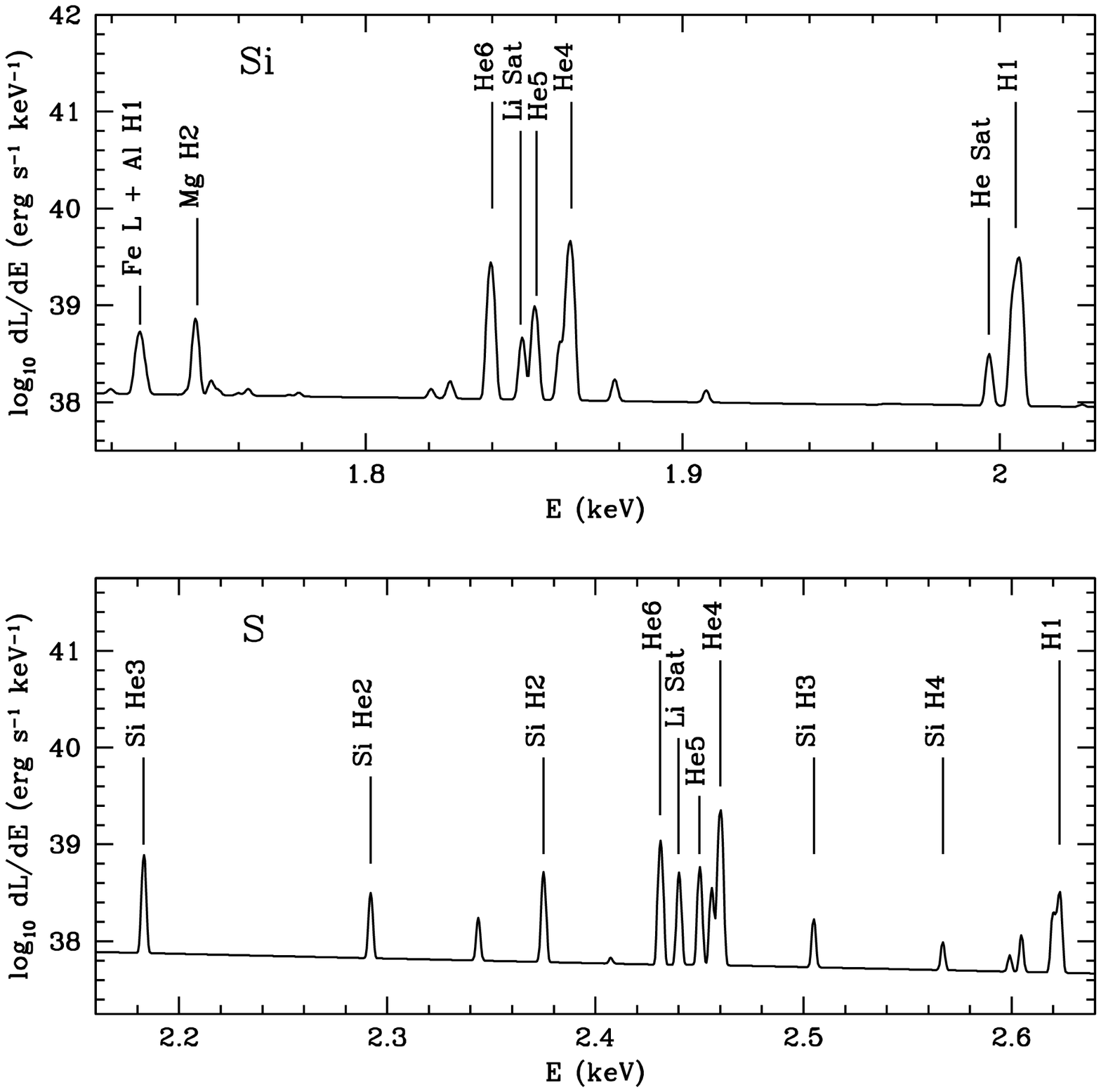,angle=0,height=0.39\textheight}}
}
\parbox{0.49\textwidth}{
\centerline{\psfig{figure=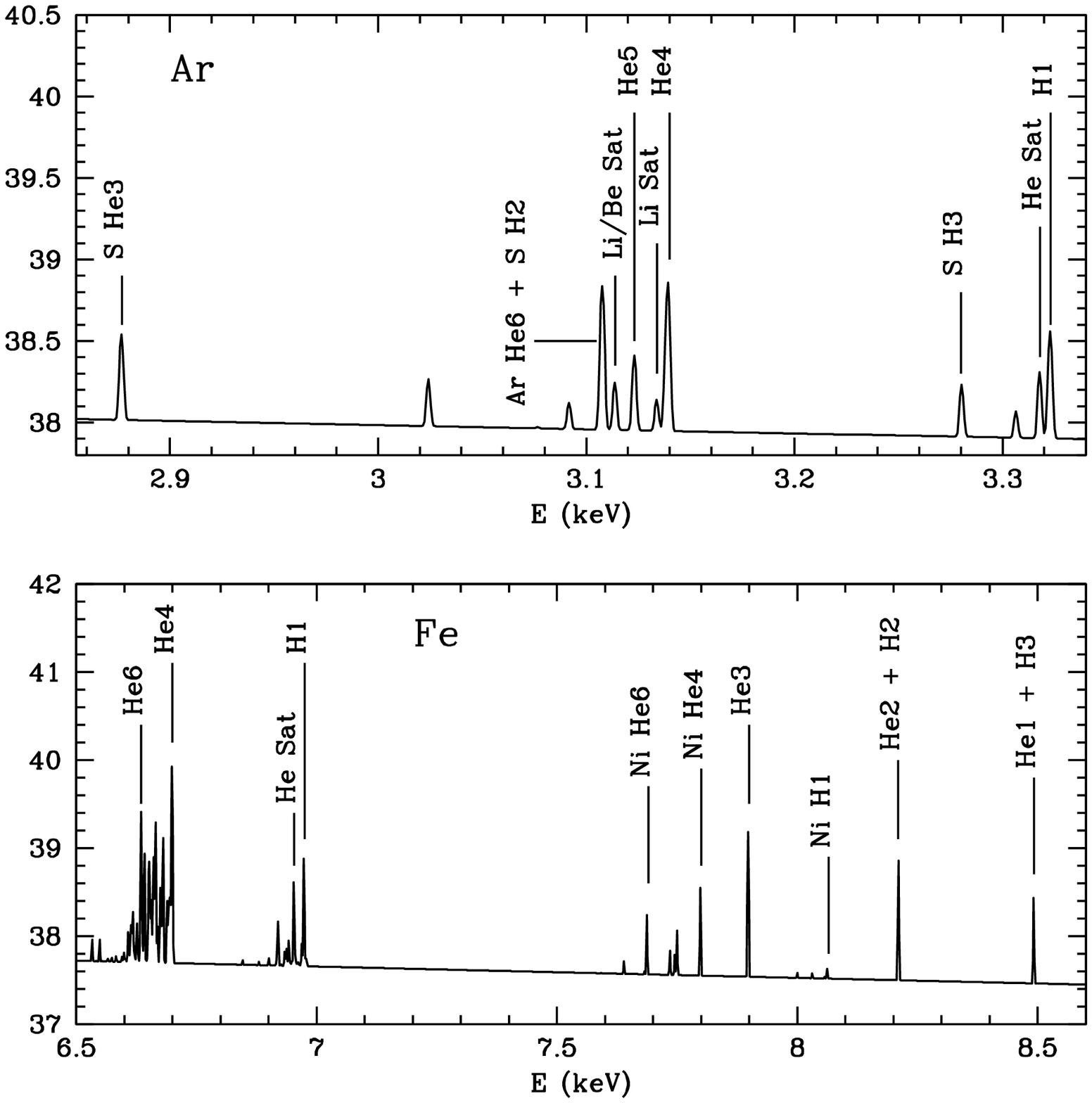,angle=0,height=0.39\textheight}}
}
\caption{\label{fig.specs} Each panel shows a portion of the X-ray
spectrum where the K shell emission lines are prominent for an
important abundant element: Si XII-XIV, S XIV-XVI, Ar XVII-XVIII, and
(mainly) Fe XXV-XXVI. The spectra are computed for isothermal MEKAL
models with $Z=0.5Z_{\sun}$ and are plotted at 2 eV energy resolution;
the temperatures are 1 keV for the Si and S panels, 2 keV for Ar, and
4 keV for Fe. The transitions mentioned in the text are labeled in
each figure according to the notation in Table \ref{tab.notation}; the
labels include the name of the element only if it is not the element
labeled on the panel. The luminosity is calculated for a total
emission measure $n_e^2V=10^{65}$ cm$^{-3}$ appropriate for M87 (e.g
Canizares et al.  1982) which is approximately midway between the
luminosities of ellipticals and clusters.}

\end{figure*}

\begin{figure*}
\parbox{0.49\textwidth}{
\centerline{\psfig{figure=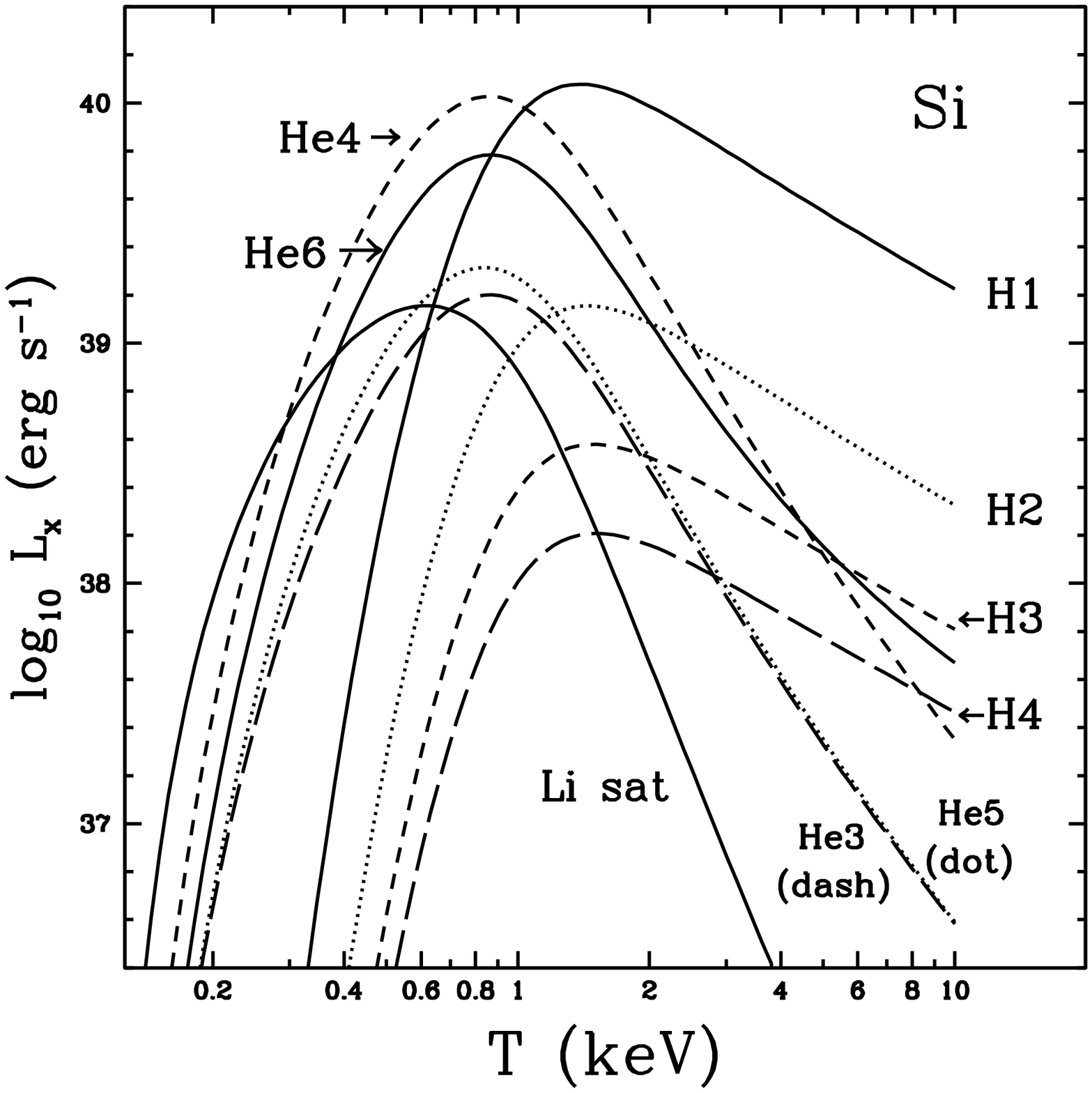,angle=0,height=0.3\textheight}}
}
\parbox{0.49\textwidth}{
\centerline{\psfig{figure=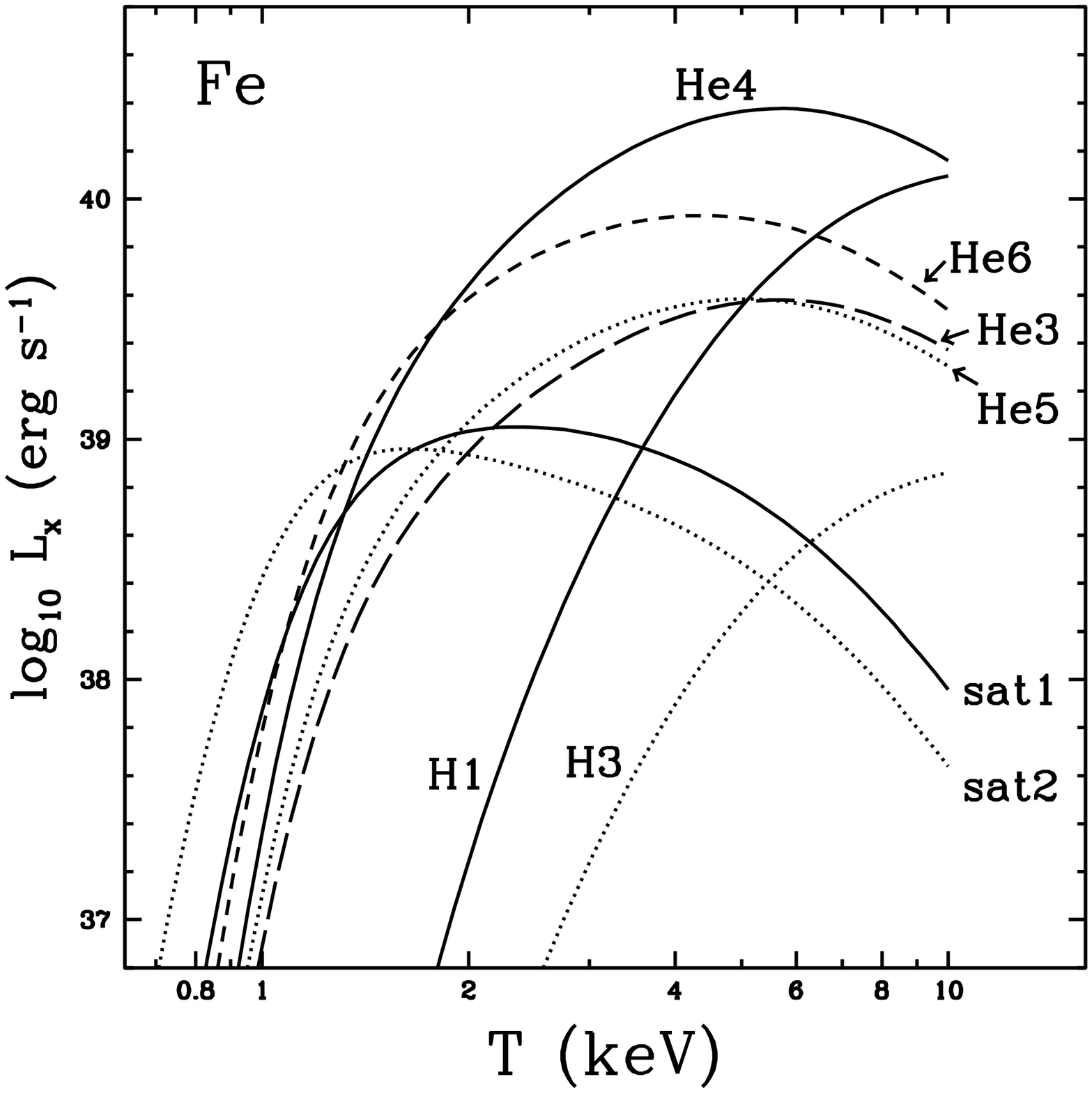,angle=0,height=0.3\textheight}}
}
\caption{\label{fig.sifek} Emission lines of K shell transitions of Si
  XII-XIV (left) and Fe XXIII-XVI (right) computed in isothermal MEKAL
  models with $Z=0.5Z_{\sun}$ and $n_e^2V=10^{65}$ cm$^{-3}$ as in
  Figure \ref{fig.specs}. See text in Sections \ref{theory} and
  \ref{iso} for details regarding the definitions of these lines.}
\end{figure*}

\begin{figure*}
\parbox{0.49\textwidth}{
\centerline{\psfig{figure=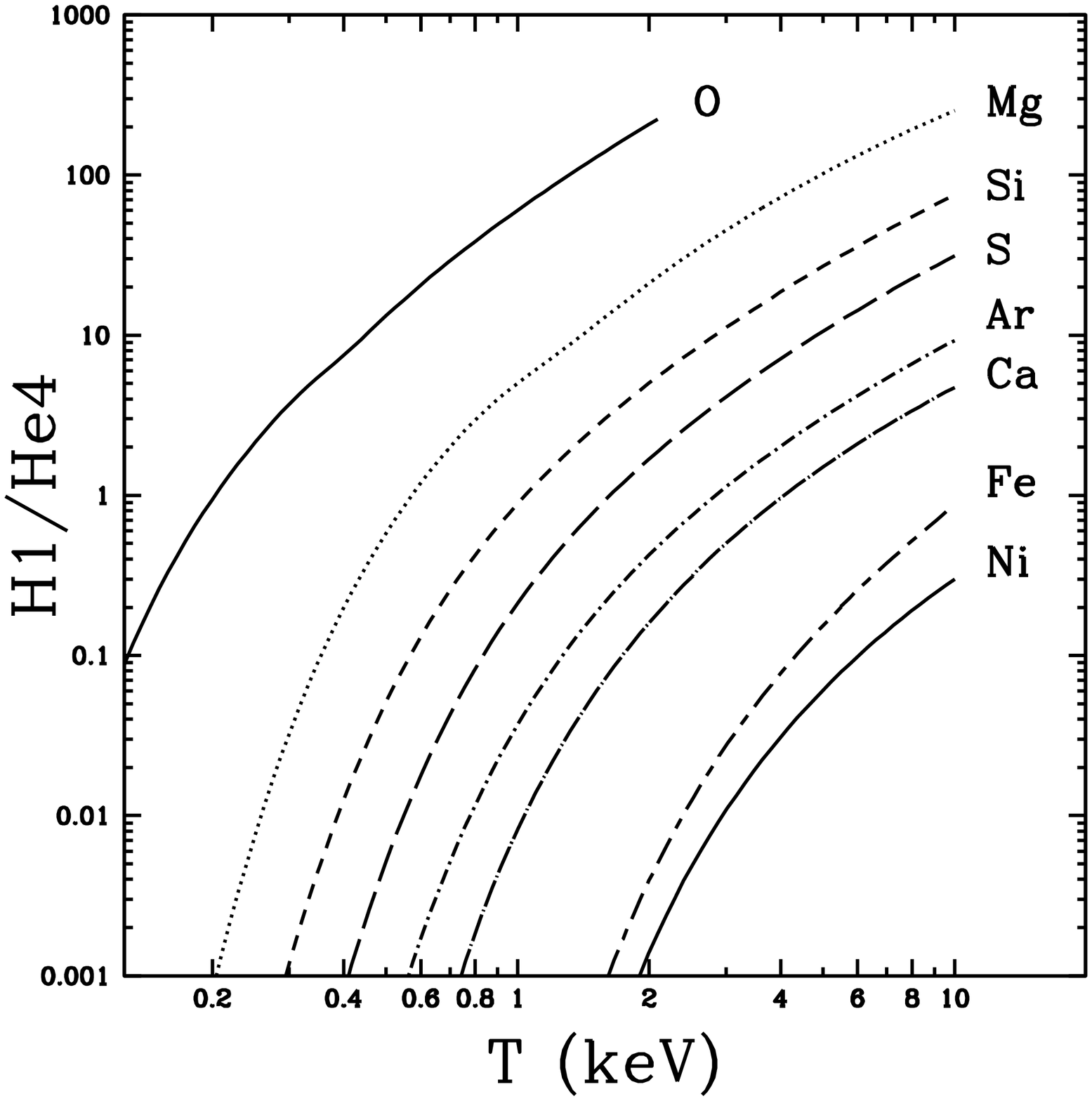,angle=0,height=0.3\textheight}}
}
\parbox{0.49\textwidth}{
\centerline{\psfig{figure=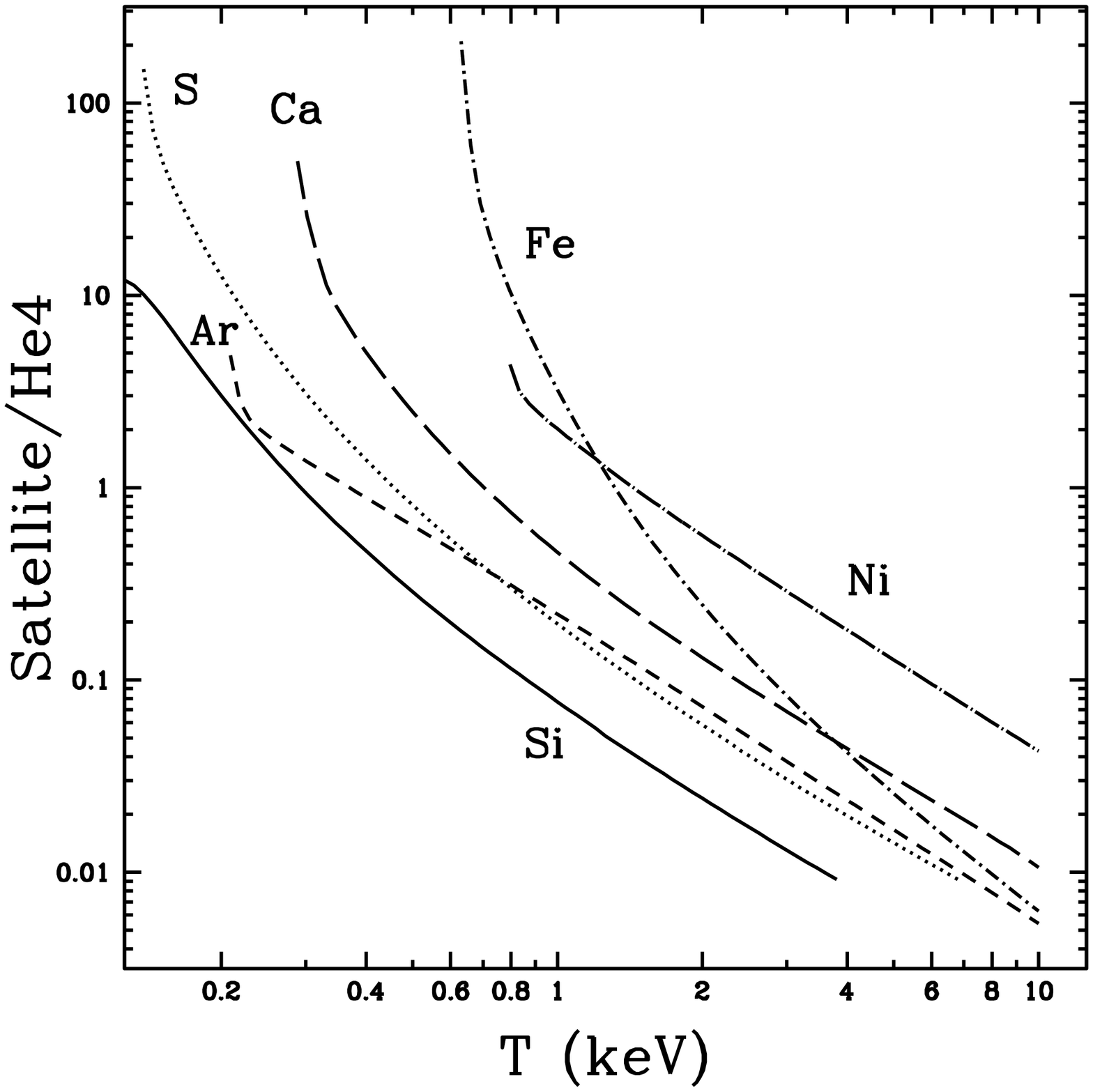,angle=0,height=0.3\textheight}}
}
\vspace{-0.2cm}
\parbox{0.49\textwidth}{
\centerline{\psfig{figure=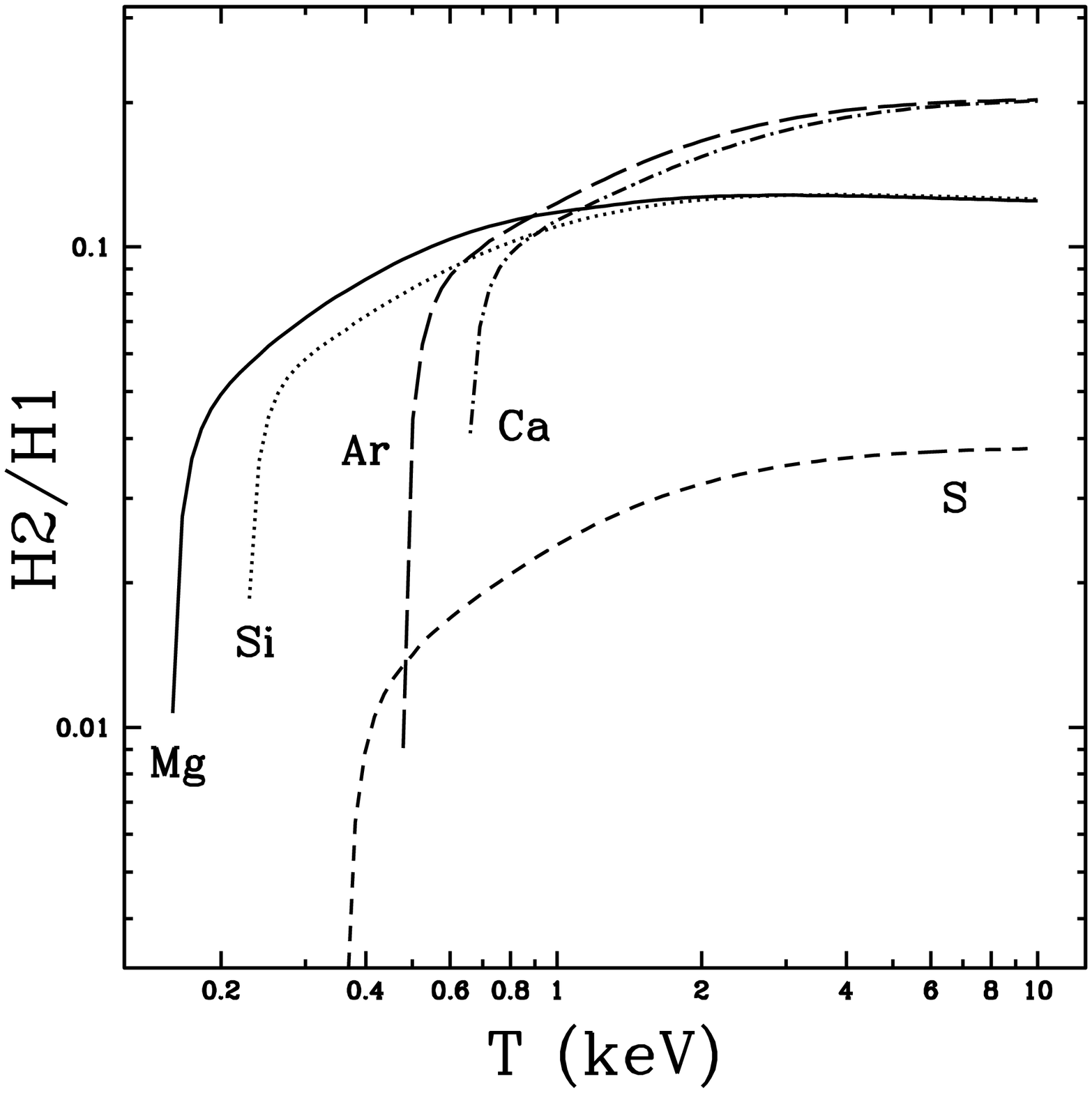,angle=0,height=0.3\textheight}}
}
\parbox{0.49\textwidth}{
\centerline{\psfig{figure=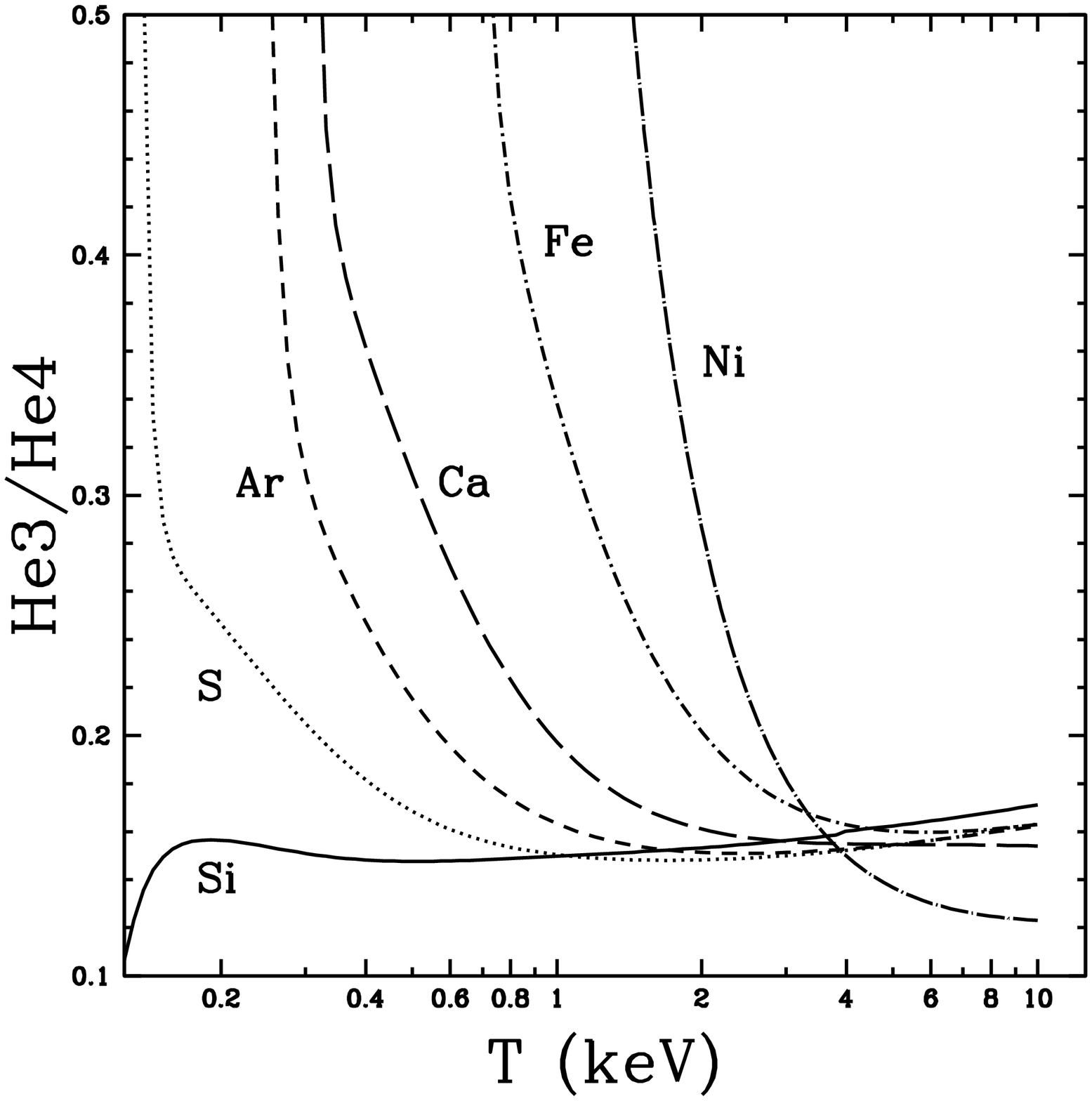,angle=0,height=0.3\textheight}}
}
\caption{\label{fig.ratios} Emission line ratios of K shell
  transitions (Tables \ref{tab.notation} and \ref{tab.sats}) of
  abundant ions in isothermal MEKAL models. Note the linear scale for
He3/He4.}
\end{figure*}

\begin{table}
\caption{Energy Ranges of Satellite Lines}
\label{tab.sats}
\begin{tabular}{lcc}
  & Energy Range & Isoelectronic\\
Ion(s) & (keV) & Sequence\\
Si XII & 1.847-1.852 & Li\\
S XIV & 2.436-2.445 & Li\\
Ar XVI & 3.130-3.137 & Li\\
Ca XVII/XVIII & 3.863-3.875 & Be/Li blend\\
Fe XX/XXI & 6.610-6.630 & N/C blend (sat1)\\
Fe XVIII/XIX & 6.520-6.610 & F/O blend (sat2)\\
Ni XXV & 7.725-7.747 & Li\\

\end{tabular}

\medskip

Energy ranges used to define the dielectronic recombination satellite
lines.

\end{table}

We begin by examining isothermal plasma models in detail since the
temperature dependences of line strengths and line ratios are most
easily understood for these models.  We plot spectral regions in
Figure \ref{fig.specs} corresponding to the strongest K shell
transitions of Si, S, Ar, and Fe. In Figure \ref{fig.sifek} we plot
the luminosities of the K shell emission lines of Si XII-XIV and Fe
XXIII-XXVI obtained from the isothermal models; these ions have among
the strongest K shell lines which are also isolated from strong lines
of other elements.  Since the Li-like satellites of Fe XXIV are
unresolved at 2 eV resolution, the Fe satellites plotted in Figure
\ref{fig.sifek} result from ions in the next few isoelectronic
sequences. In Table \ref{tab.sats} we list the energy ranges used to
define these satellite blends for Fe as well as the Li-like satellites
of the other elements.

In Figure \ref{fig.ratios} we plot as a function of temperature the
ratios of the K shell emission lines of the abundant elements. (Ratios
involving lines that are very contaminated are not plotted.)  The
ratios of lines emitted from subsequent ionization stages of the same
element (H1/He4, Satellite/He4) are overall more useful as temperature
diagnostics than are the ratios involving lines of the same ion
(H2/H1, He3/He4). The former, however, rely on the assumption of
ionization equilibrium which can be checked using the latter ratios.
Ionization equilibrium is expected to be an accurate approximation for
the hot gas in ellipticals, groups, and clusters, even those with
cooling flows \cite{crc88}.

The H1/He4 ratio is probably the most useful because the H1 and He4
transitions give rise to the strongest lines and the ratio $\sim 1$
where the lines have maximum luminosity; see the examples of Si and Fe
in Figure \ref{fig.sifek}. The mostly smooth curves of the H1/He4
ratios as a function of temperature in Figure \ref{fig.ratios}
demonstrate that the H1 and He4 lines of many of the abundant elements
can be measured accurately at 2 eV resolution. The largest distortions
are evident for the Mg ratio which result from contamination in the
measurement of the Mg XI He4 transition by Fe L emission.  Other
significant contamination appears in the computation of the Fe XXV He4
transition which includes unresolved Li satellites at 2 eV resolution.

Since the H1/He4 ratios for the different elements take a value of 1
approximately uniformly distributed over the logarithmic range of
temperatures $\sim 0.5$ - 10 keV, accurate estimation of any
temperature over this range can be obtained by H1/He4 ratios of more
than one element.  For temperatures below $\sim 0.5$ keV the only
useful H1/He4 ratio is that of oxygen.

The next most useful diagnostic is the Satellite/He4 ratio, where
``Satellite'' typically refers to the Li-like satellites; see Table
\ref{tab.sats} for the definitions of the satellite lines used.  The
principal difficulty here is measuring the satellite lines because
they are usually much weaker than the He4 line and are difficult to
resolve from the (forbidden) He5 line, especially for elements with
atomic number $\la 20$. The shapes of the temperature profiles of the
Satellite/He4 ratios of Si, S, Ca, and Ni are similar in Figure
\ref{fig.ratios}.  The ratio of Fe K lines has a steeper profile
because satellites from isoelectronic sequences above Li are used;
i.e. sats1 in Table \ref{tab.sats}. The distorted temperature profile
of the Ar ratio for $T\la 1$ keV arises from unresolved Li satellites
being included in the He4 line.

Over $T\sim 0.5$ - 5 keV the Si Satellite/He4 ratio changes by a
factor $\sim 20$ compared to $\sim 1000$ for the Si H1/He4 ratio. The
steepest Satellite/He4 profile shown is for Fe K.  Even if we do not
consider $T < 2$ keV where the Fe K lines are weak, then the
Satellite/He4 ratio changes by a factor of $\sim 50$ from $T\sim 2$ -
8 keV. This variation is about as large as seen for the Fe K H1/He4
ratio.

The H2/H1 (i.e. Ly $\beta$/Ly $\alpha$) ratios are much less
temperature sensitive than are the ratios described previously. The
temperature dependence of this ratio is simply, ${\rm H2/H1} \propto
\exp[-(E2-E1)/k_BT]$, where $k_B$ is Boltzmann's constant and $E2$ and
$E1$ are the energies of the H2 and H1 lines respectively. For Si,
$E2-E1=0.37$ keV, and as is clear from Figure \ref{fig.ratios}, H2/H1
only varies significantly for temperatures within approximately a
factor of 2 of 0.37 keV. At these temperatures the hydrogenic lines of
Si are also very weak. This behavior holds qualitatively for the other
elements shown. (Note that the plotted S ratio is actually H3/H1
because the H2 transition is blended with the Ar XVIII He6
intercombination line.) Typically, for temperatures where the lines
are at their strongest, the H2/H1 ratio varies by at most a factor of
2; e.g. Si from 0.5-5 keV.

Similar to H2/H1, the ratio He3/He4 (i.e. K$\beta$/K$\alpha$) involves
lines of the same isoelectronic sequence and has a fairly weak
temperature dependence for temperatures where the lines are of
significant strength. Taking S for an example (Figure
\ref{fig.ratios}), for temperatures ranging from 0.5 - 5 keV the
He3/He4 ratio varies by only $\sim 5$ - 10 per cent.  For $T\la 0.5$
keV, the ratio is very sensitive to temperature, but the lines are
also very weak: at $T=0.25$ keV we have $\log_{10} \rm L_x = 36.1$ and
$36.7$ for He3 and He4 of S assuming a total emission measure of
$10^{65}$ cm$^{-3}$ (as in Figure \ref{fig.sifek}).  Probably Fe XXV
is the most useful ion for this ratio because at $\sim 2$ keV where
He3/He4 begins to change noticeably, the emission lines still have
significant emissivity; e.g., at $T=2$ keV we have $\log_{10} \rm L_x
= 39.0$ and $39.6$ for He3 and He4 of Fe (see Figure
\ref{fig.sifek}). Hence, for Fe the He3/He4 ratios varies by about a
factor of 10 over interesting temperatures. (Note that the Si ratio
turns over below 0.2 keV because of weak Al XI contamination in the
He4 line of Si XIII.)

\begin{figure}
\centerline{\psfig{figure=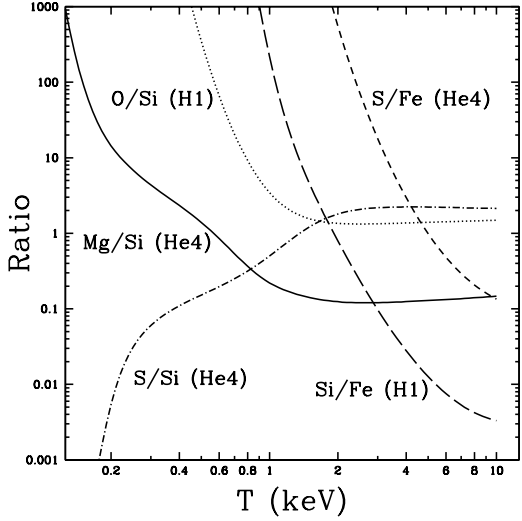,angle=0,height=0.3\textheight}}
\caption{\label{fig.diff} Ratios of K shell emission lines from the
  same isoelectronic sequences of different elements for isothermal
  MEKAL models with $Z=0.5Z_{\sun}$.} 
\end{figure}

For any individual element, the ratios described above involve lines
which differ in energy by $<1$ keV. That is, each ratio gives a
measure of the temperature in a small energy region of the spectrum.
Comparing these ratios for different elements over the entire X-ray
spectrum allows one to examine the consistency of the temperatures
measured by each ratio. Another method to compare lines that are
substantially different in energy is to compare ratios of lines of the
same isoelectronic sequences of different elements (Figure
\ref{fig.diff}).  Although these ratios depend on the relative
abundances of the elements, they can provide an interesting
consistency check of the other ratios. (In fact, these ratios are used
to measure abundances once $\xi(T)$ is known) Typically, these ratios
vary by a factor of 100 over the temperature ranges where the lines
emit significantly (Figure \ref{fig.diff}). (These ratios have a
stronger dependence on $N_{\rm H}$ since the lines are not close
together in energy, but for energies above $\sim 2$ keV the dependence
should be negligible.)

\begin{figure*}
\parbox{0.49\textwidth}{
\centerline{\psfig{figure=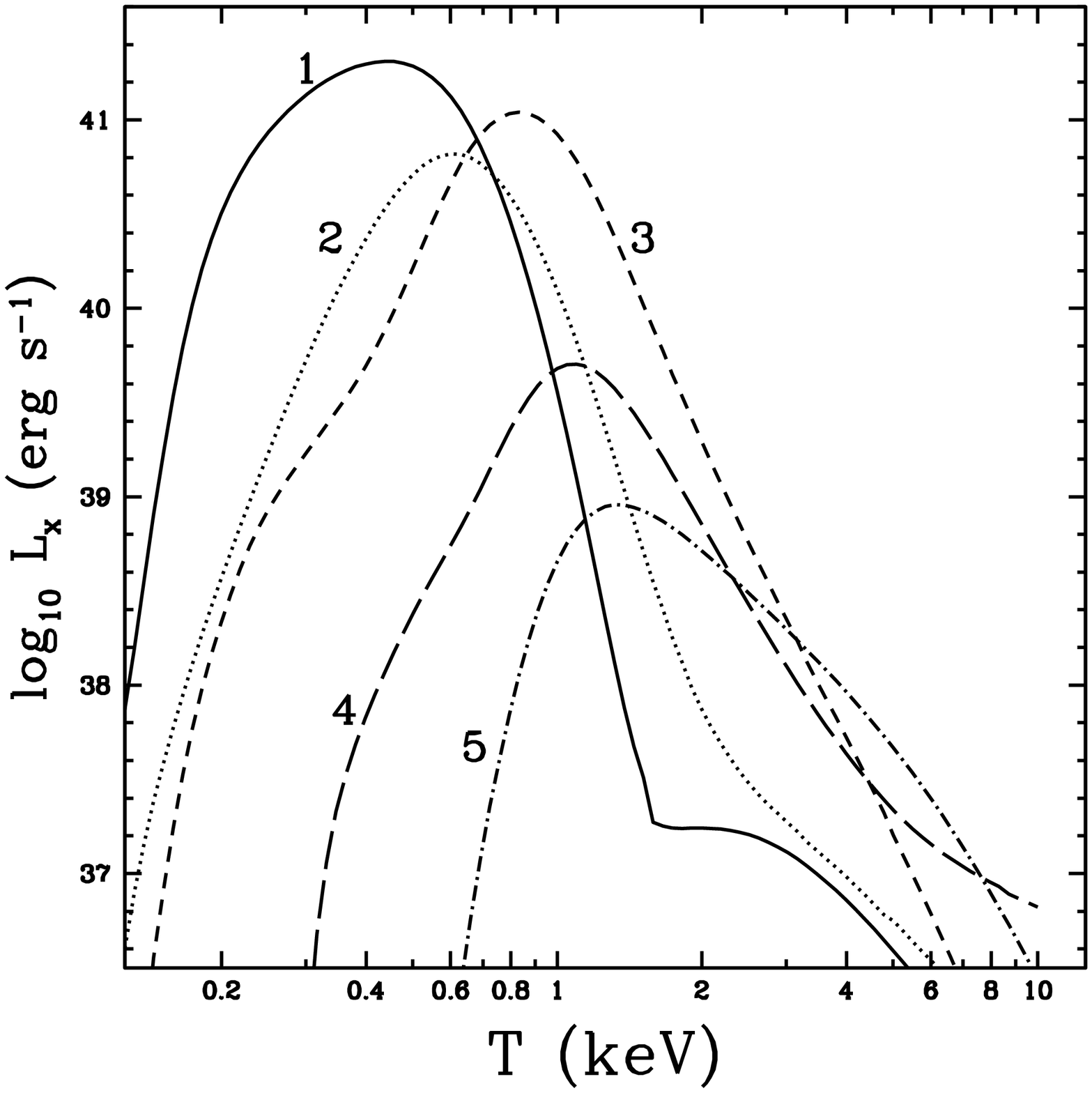,angle=0,height=0.3\textheight}}
}
\parbox{0.49\textwidth}{
\centerline{\psfig{figure=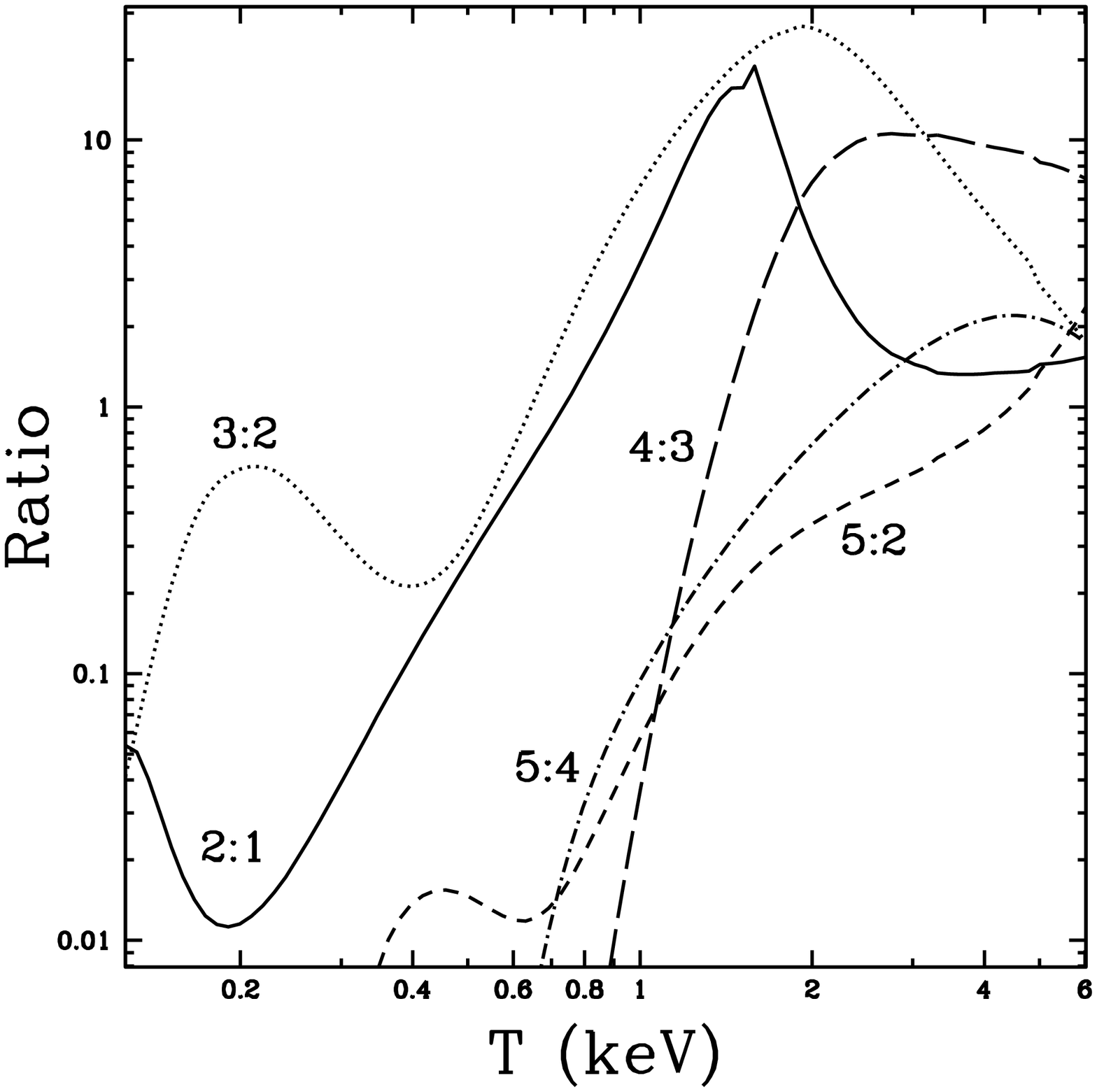,angle=0,height=0.3\textheight}}
}
\caption{\label{fig.fel} Fe L emission line strengths (left)
  and ratios (right) computed in isothermal MEKAL models with
  $Z=0.5Z_{\sun}$. The lines correspond to the definitions in Table
  \ref{tab.fel}. The luminosity is calculated for a total emission
  measure $n_e^2V=10^{65}$ cm$^{-3}$ as in Figure \ref{fig.specs}.}
\end{figure*}

Finally, we consider the Fe L shell emission lines. Even at 2 eV
resolution most of the Fe L lines between 0.7-1.4 keV are not
resolved. From visual inspection of the MEKAL model spectra we
identified blends of nearby lines which varied similarly with
temperature and were least contaminated by lines of other elements.
In Table \ref{tab.fel} we list energy ranges for five blends (or, more
accurately, groups) of lines and plot their strengths and ratios as a
function of temperature in Figure \ref{fig.fel}.  These line groups of
$\sim 10$ - 20 eV width are shown because they have the most regular
behavior over the temperature range investigated.  (The lack of blends
between 1.1-1.3 keV is due to the difficulty of finding groups
uncontaminated by emission lines of other elements.)  It is clear from
comparison to the K shell strengths and ratios discussed previously
that the Fe L lines are very strong and very temperature sensitive
over energies $\sim 0.2$ - 5 keV; e.g.  the 2:1 ratio varies by a
factor of $\sim 1000$ over 0.2 - 1.5 keV.  The superior performance of
the Fe L line ratios highlights the importance of improving the
accuracy of the plasma codes for these transitions. It may be that
future observations will shed light on the accuracy of the codes in
the Fe L region by comparing to the temperatures inferred from K shell
transitions.

\begin{table}
\caption{Energy Ranges of Fe L Blends}
\label{tab.fel}
\begin{tabular}{ccc}
& Energy Range\\
Blend & (keV) & Dominant Ion(s)\\
1 & 0.722-0.741 & Fe XVII\\
2 & 0.867-0.878 & Fe XVII \& XVIII\\
3 & 0.940-0.975 & Fe XVII \& XVIII\\
4 & 1.375-1.387 & Fe XXII\\
5 & 1.435-1.445 & Fe XXIII\\

\end{tabular}

\end{table}

\subsection{Multiphase cases}
\label{multi}

We now examine the ability of the line ratios to quantify the general
multiphase structure of the hot gas in ellipticals, groups, and
clusters.  The temperature structure is fully characterized by the
distribution of emission measure as a function of temperature (see
Introduction).  Of particular relevance to ellipticals, groups, and
clusters are emission measures of a single temperature component, a
cooling flow component (see below), or a mixture of both. The single
temperature component is often modeled as a $\delta$-function
differential emission measure while a multiphase cooling flow emits
over a wide range of temperatures.  Hence, to facilitate our
discussion of the temperature structures of ellipticals and clusters,
we wish to define a dimensionless parameter, $\sigma_\xi$, which
isolates the information contained in $\xi(T)$ indicating the
multiphase ``width'' or ``strength''. (We discuss a possible
cosmological application of $\sigma_\xi$ in Section \ref{omega}.)

\subsubsection{Multiphase strength}
\label{mps}

We first consider the case where $\xi(T)$ does not contain any
$\delta$-function components. This is the physical situation since in
a real elliptical or cluster the hot gas will always have some
temperature fluctuations about the mean as a result of incomplete
relaxation.

Over some temperature range $(T_{\rm min},T_{\rm \max})$ it is
necessary to define the ``maximal multiphase strength'' which we take
to be $\xi(T) =$ constant; i.e. for $T$ within this range the emission
measure, $\xi(T)dT$, is constant for fixed intervals $dT$.  For
constant $\xi$ the total emission measure equals $\xi\Delta T$, where
$\Delta T= T_{\rm max} - T_{\rm min}$. If $\xi(T)$ is centrally peaked
(i.e. approximately a single phase) then $\xi_{\rm max}\Delta T$ will
greatly exceed the total integrated emission measure.  Thus, a
sensible definition for the ``multiphase strength'' of an
emission measure distribution is,
\begin{eqnarray}
\sigma_\xi & \equiv & \left( {\int_{T_{\rm min}}^{T_{\rm max}}
    \xi(T)\, dT \over \xi_{\rm max}\Delta T}\right) {\Delta T
    \over 2 \langle T\rangle} \label{eqn.mps.def}\\
& = & { 1 \over 2 \langle T\rangle \xi_{\rm max} } {\int_{T_{\rm
    min}}^{T_{\rm max}} \xi(T)\, dT}, \label{eqn.mps}
\end{eqnarray}
where $\xi_{\rm max}$ is the maximum value of $\xi$, $\langle
T\rangle$ is the emission measure weighted temperature,
\begin{equation}
\langle T\rangle = {\int_{T_{\rm min}}^{T_{\rm max}}
    T\xi(T)\, dT \over \int_{T_{\rm min}}^{T_{\rm max}} \xi(T)\, dT}, 
\end{equation}
and $(T_{\rm min},T_{\rm \max})$ are any two temperatures which
bracket the region in temperature space where $\xi\ne 0$.  The factor
$\Delta T/2\langle T\rangle$ in equation (\ref{eqn.mps.def}) takes
into account the size of the temperature interval; i.e. for uniform
$\xi$ we have $\sigma_\xi=\Delta T/2\langle T\rangle \ll 1$ for a
narrow interval $T_{\rm \max} \approx T_{\rm min}$ (i.e.
approximately single phase) but $\sigma_\xi=\Delta T/2\langle T\rangle
\sim 1$ for $T_{\rm \max} \gg T_{\rm min}$. Thus, $\sigma_\xi$ is a
measure of the fractional width of $\xi$.

To illustrate the meaning of $\sigma_\xi$ let us for the moment
consider $\xi$ to be a Gaussian with standard deviation $\sigma$ and
mean temperature $\langle T\rangle$. In this case there is an analytic
expression for the multiphase strength,
\begin{equation}
\sigma_\xi = \sqrt{\frac{\pi}{2}} {\sigma \over \langle T\rangle }
\hskip 0.5cm (\rm Gaussian),  
\end{equation}
which implicitly assumes $(T_{\rm min},T_{\rm max})=(-\infty,+\infty)$
but is accurate provided $\Delta T\gg\sigma$.  As would be expected,
$\sigma_\xi \propto \sigma$, indicating that $\sigma_\xi$ measures the
width of the emission measure distribution, but $\sigma_\xi$ applies
equally well to ellipticals and clusters because the mean temperature,
$\langle T\rangle$, is divided out.

The definition of multiphase strength just described is applicable for
all ``well behaved'' $\xi(T)$. However, it is often convenient to fit
models consisting of individual $\delta$-function temperature
components to X-ray data of ellipticals and clusters. This widespread
practice is well justified for nearly isothermal systems because the
true finite width of $\xi(T)$ cannot be distinguished from a
$\delta$-function due to the limitations in the energy resolution of
current X-ray data. Hence, since $\delta$-function temperature
components (even though they are not quite physical) are often fitted
to data of ellipticals, groups, and clusters, we consider how to
modify the definition of $\sigma_\xi$ so that it applies for emission
measure distributions having only $\delta$-function temperature
components or having both a $\delta$-function and ``well behaved''
temperature components.

For $\xi(T)$ consisting of a single $\delta$-function temperature
component we have $\sigma_\xi=0$ because $\xi_{\rm max}=\infty$. This
indicates correctly that the width of the emission measure
distribution is zero. However, if we consider $\xi(T)$ to be composed
of a single $\delta$-function component plus some ``well behaved''
component such as a Gaussian, then $\sigma_\xi$ is still zero
regardless of the shape of the ``well behaved'' component.  When
considering emission measure distributions composed of both a
$\delta$-function temperature component and other non-singular
components, it is reasonable to modify $\sigma_\xi$ according to the
fractional contribution of the ``well behaved'' component to the total
integrated emission measure. That is, if $\rm EM_{ns}$ and $\rm
EM_\delta$ are the total emission measures of the non-singular and
$\delta$-function components respectively, then the total multiphase
strength is constructed by replacing $\sigma_\xi$ of the non-singular
component by $f\sigma_\xi$, where $f= {\rm EM_{ns}}/({\rm EM_{ns} +
EM_\delta})$.  This form of the multiphase strength is especially
useful when considering models consisting of a mixture of a cooling
flow and ambient gas (as in Sections \ref{cfs}, \ref{2tcf}, and
\ref{omega}).

The definition of $\sigma_\xi$ cannot be so easily modified if
$\xi(T)$ consists of multiple $\delta$-functions.  A differential
emission measure consisting of a finite sum of $N$ $\delta$-function
temperatures, $\xi(T) = \sum_{i=1}^N \xi_i\delta(T-T_i)$, always has
$\sigma_\xi = 0$ independent of the $\xi_i$ and $N$.  Clearly
$\sigma_\xi$ should depend on $N$ and on the temperature difference
between each $\delta$-function. To simplify let us restrict ourselves
to the case where the $\delta$-functions are evenly spaced in
temperature. In the limit of large $N$ we can generalize $\sigma_\xi$
to the discrete case $\sigma_\xi^N$ by simply replacing $\xi_{\rm
max}\Delta T$ in equation (\ref{eqn.mps.def}) with $N\xi_{\rm
max}$. This is appropriate for $N\gg N_c$, where $N_c$ is some
critical number of $\delta$-functions above which a discrete number of
phases is a good approximation to a continuous $\xi(T)$. In the other
limit we require that $\sigma_\xi^N\rightarrow 0$ as $N\rightarrow
1$. Considering these limits we define $\sigma_\xi^N$ as follows,
\begin{equation}
\sigma_\xi^N  \equiv  \left( {\sum_{i=1}^N \xi_i 
    \over N\xi_{\rm max}}\right) {\Delta T
    \over 2 \langle T\rangle} \left(1-e^{-(N-1)/N_c}\right).
    \label{eqn.mps.disc} 
\end{equation}
The first two terms are just the straightforward generalization of
$\sigma_\xi$ for large $N$ while the last term is an arbitrary means
to smoothly connect the limiting cases. A reasonable choice for $N_c$
is $\sim 20$.

\subsubsection{Simple examples}
\label{examples}

\begin{figure*}
\parbox{0.49\textwidth}{
\centerline{\psfig{figure=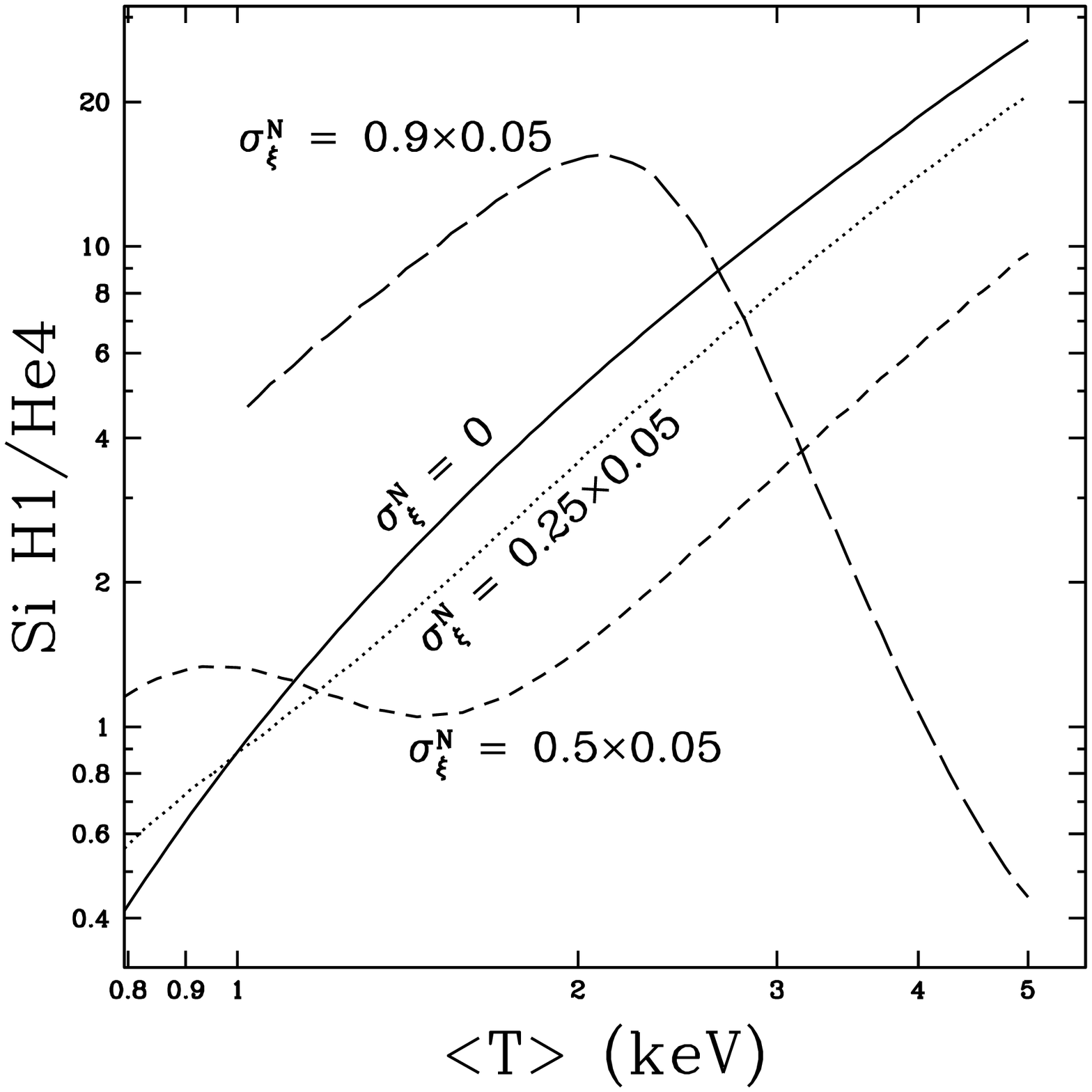,angle=0,height=0.3\textheight}}
}
\parbox{0.49\textwidth}{
\centerline{\psfig{figure=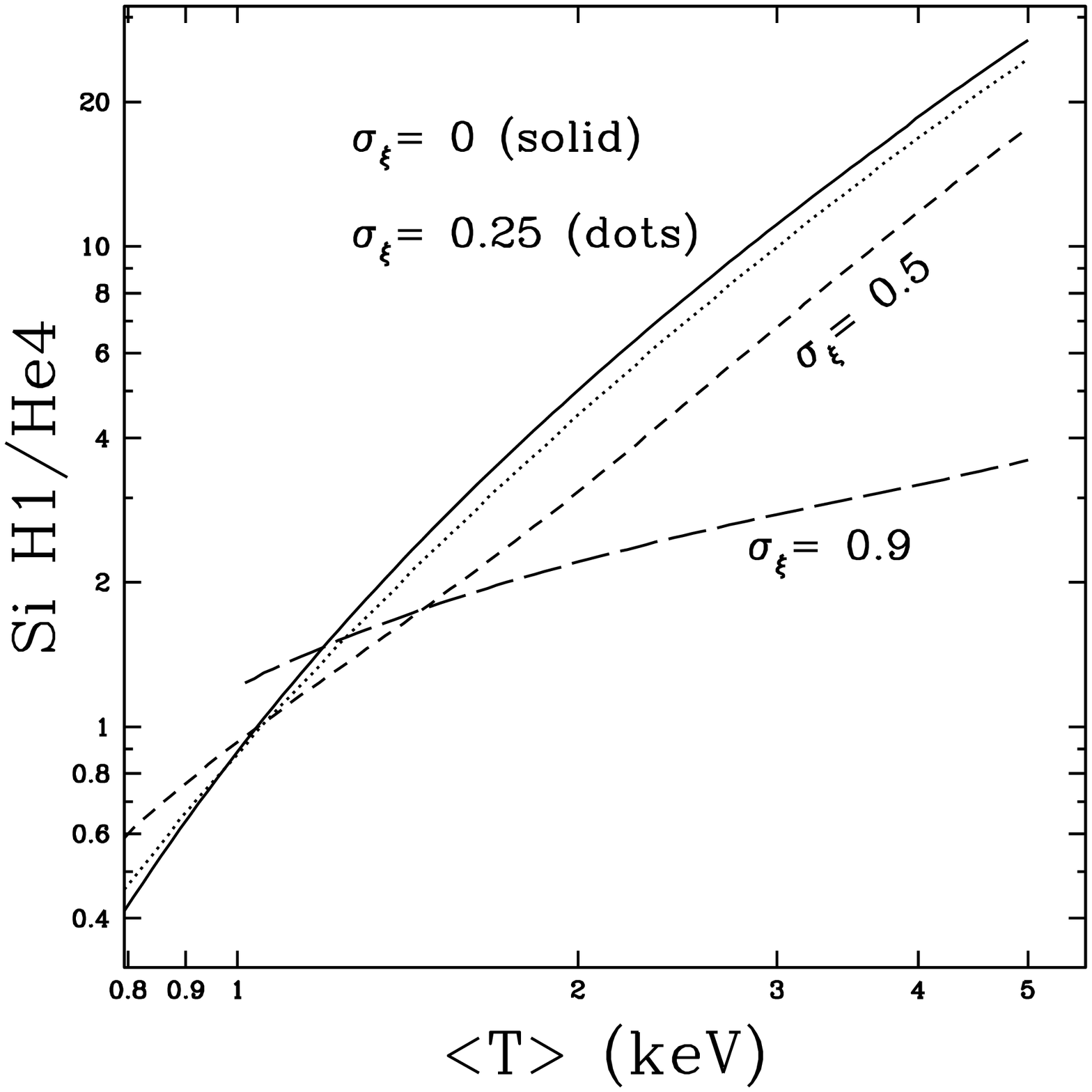,angle=0,height=0.3\textheight}}
}
\caption{\label{fig.multi} H1/He4 ratio for Si versus emission measure
weighted temperature for two-temperature models, $\xi(T)
\propto\delta(T-T_1) + \delta(T-T_2)$ (left), and models with a
constant differential emission measure, $\xi =$ constant, over some
temperature range (right). Each curve corresponds to a different
multiphase strength. For the constant $\xi$ case (right) the
multiphase strength is given by $\sigma_\xi$ (equation \ref{eqn.mps})
and for (left) the $\delta$-function case $\sigma_\xi^N$ (equation
\ref{eqn.mps.disc}) for $N=2$ and $N_c=20$.}
\end{figure*}

We plot the H1/He4 ratio of Si as a function of temperature in Figure
\ref{fig.multi} for MEKAL models consisting of two temperatures with
equal emission measures. For clarity we express $\sigma_\xi^N$ as the
product of two factors: (1) the terms which are the generalization of
$\sigma_\xi$ for large $N$, and (2) the factor involving the number of
temperature components which is 0.05 for $N=2$ and $N_c=20$.  Even for
relatively small $\sigma_\xi^N=0.25\times 0.05$ there are considerable
departures from the isothermal case. Of particular interest is that
there are values of $\langle T\rangle$ where very different
$\sigma_\xi^N$ have very similar line ratios. Just above 1 keV the
ratios for models with $\sigma_\xi^N=0$ and $0.5\times 0.05$ differ by
only $\sim 15$ per cent. Just below $\langle T\rangle = 3$ keV the
curves of $\sigma_\xi^N=0$ and $0.9\times 0.05$ intersect!

The key implication of these curves is that one line ratio does not
uniquely determine $\sigma_\xi^N$. The situation is even more
complicated than shown in Figure \ref{fig.multi} if the two
temperature components are allowed to have different emission
measures. In that case models having the same $\langle T\rangle$ and
$\sigma_\xi^N$ can also have different line ratios. These degeneracies
are related to the ambiguity in determining $\xi(T)$ from inversion of
equation (\ref{eqn.dem}) (see Introduction).  To narrow the region of
$\sigma_\xi^N$ space, one must add constraints from similar ratios of
other elements and use as many temperature sensitive ratios as
possible (e.g. satellite/He4).

The situation is more tractable for a simple non-singular differential
emission measure such as a constant or a Gaussian. In Figure
\ref{fig.multi} we show the Si H1/He4 ratio versus temperature for the
case $\xi=$ constant. The ratio profile changes slowly away from the
isothermal curve and flattens for larger $\sigma_\xi$ which is just
$\Delta T/2\langle T\rangle$ for a constant $\xi$. Again, for the Si
ratio there is a $\langle T\rangle$ near 1 keV where a large range of
$\sigma_\xi$ have nearly the same H1/He4 ratio; i.e. even for this
rigid model, more than one line ratio is required to constrain
$\sigma_\xi$. Notice, however, that there are large regions of
ratio-$\langle T\rangle$ space which are uninhabited. In particular,
the accessible region of the $\xi=$ constant model ends not much below
the line $\sigma_\xi=0.9$. This also applies for any continuous
$\xi(T)$ since $\xi=$ constant is defined to be the limiting case for
maximal multiphase strength. (Note that a Gaussian model has behavior
similar to that of the constant $\xi$ in Figure \ref{fig.multi}.)

From examination of plots like Figure \ref{fig.multi} for different
elements we can locate the critical value of $\sigma_\xi$ that
represents the boundary between approximately single-phase and
strongly multiphase $\xi$.  If we define $\xi$ to be significantly
multiphase when the H1/He4 line ratios of Si (and others) typically
differ by more than 10 per cent from the isothermal case, then
$\sigma_\xi\sim 0.2$ represents the boundary between approximately
single-phase and significantly multiphase $\xi$. This criteria applies
for a constant and Gaussian $\xi$. When a $\delta$-function is added
to these models we find that $\sigma_\xi\sim 0.1$ is a better
representation of the transition between a single-phase and multiphase
gas.

\subsubsection{Cooling flows}
\label{cfs}

\begin{figure}
\centerline{\psfig{figure=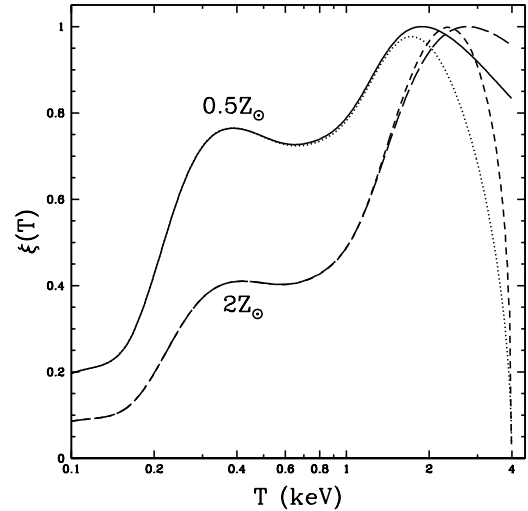,angle=0,height=0.3\textheight}}
\caption{\label{fig.dem_cf} The differential emission measures for a
  constant pressure cooling flow ($Z=0.5Z_{\sun}$ -- solid,
  $Z=2Z_{\sun}$ -- long dash) and for a cooling flow with
  $\dot{M}\propto r$ in the gravitational potential of a singular
  isothermal sphere ($Z=0.5Z_{\sun}$ -- dotted, $Z=2Z_{\sun}$ -- short
  dash). All models have $T_{\rm max}=4$ keV.  The units of $\xi$ (for
  each metallicity) are normalized to the maximum value of $\xi$ for
  the constant pressure cooling flow. }
\end{figure}

\begin{figure}
\centerline{\psfig{figure=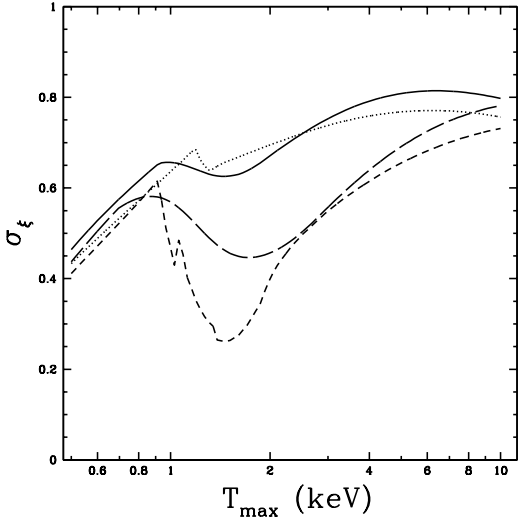,angle=0,height=0.3\textheight}}
\caption{\label{fig.sigxi_cf} Multiphase strengths for the cooling
  flow models in Figure \ref{fig.dem_cf}. See text for discussion of the
  behavior near 1 keV.}
\end{figure}

A physical mechanism for producing multiphase gas in ellipticals,
groups, and clusters is provided by cooling flows (Fabian, Nulsen, \&
Canizares 1984; Nulsen 1986, 1997; Thomas, Fabian, \& Nulsen 1987;
White \& Sarazin 1987; for a recent review see Fabian 1994). In these
models gas cools out of the flow in regions where the cooling time is
sufficiently short (usually taken to be less than the assumed age of
the system). The simplest example is for gas cooling at constant
pressure from some upper temperature $T_{\rm max}$ in which case the
conservation of enthalpy implies an energy release $dL$ within a
temperature interval $dT$ of, $dL=5\dot{M}k_BdT/2\mu m_p$, where $\mu$
is the mean atomic weight, $m_p$ the proton mass, and $\dot{M}$ is the
rate of mass deposition. Comparing this expression to equation
(\ref{eqn.line}) one obtains the differential emission measure,
\begin{equation}
\xi(T) = {5\over 2}{\dot{M}k_B\over \mu m_p} {1 \over \Lambda(T)}
\hskip 0.5cm (\rm Constant \hskip 0.1cm Pressure \hskip 0.1cm CF),
\label{eqn.cfdem} 
\end{equation}
for $T\le T_{\rm max}$ which we plot in Figure \ref{fig.dem_cf} for
$T_{\rm max}=4$ keV with abundances $Z=0.5,2Z_{\sun}$. The shape of
$\xi(T)$ is sensitive to $Z$, particularly for $T\la 1$ keV where the
Fe L shell lines dominate $\Lambda(T)$. However, since most of the
emission measure, $\int\xi(T)dT$, in the models in Figure
\ref{fig.dem_cf} result from temperatures between $\sim 2$ - 4 keV,
these metallicity differences do not translate to large differences in
the K shell line ratios as we show below.

The emission measure weighted temperatures for the models with
$Z=0.5Z_{\sun}$ range from $\langle T\rangle \approx 0.6T_{\rm max}$
for $T_{\rm max}=0.5$ keV to $\langle T\rangle \approx 0.5T_{\rm max}$
for $T_{\rm max}= 10$ keV.  The luminosity weighted temperatures are
$\approx 15$ per cent larger than $\langle T\rangle$. For
$Z=2Z_{\sun}$, $\langle T\rangle$ ranges from $\langle T\rangle
\approx 0.4T_{\rm max}$ for $T_{\rm max}=0.5$ keV to $\langle T\rangle
\approx 0.8T_{\rm max}$ for $T_{\rm max}= 10$ keV.

The multiphase strength as a function of $T_{\rm max}$ is shown in
Figure \ref{fig.sigxi_cf}. Over $T_{\rm max} = 0.5$ - 10 keV we have
$\sigma_\xi\sim 0.45$ - 0.8 for both $Z=0.5Z_{\sun}$ and $Z=2Z_{\sun}$
and thus the constant pressure cooling flow has nearly maximal
multiphase strength for the temperatures of ellipticals, groups, and
clusters.  The $Z=2Z_{\sun}$ case lies below the $Z=0.5Z_{\sun}$ model
but they approach each other at the largest values of $T_{\rm max}$.
Note that the wiggles in $\sigma_\xi$ are simply due to the same
wiggles in $\Lambda$.

\begin{figure*}
\parbox{0.49\textwidth}{
\centerline{\psfig{figure=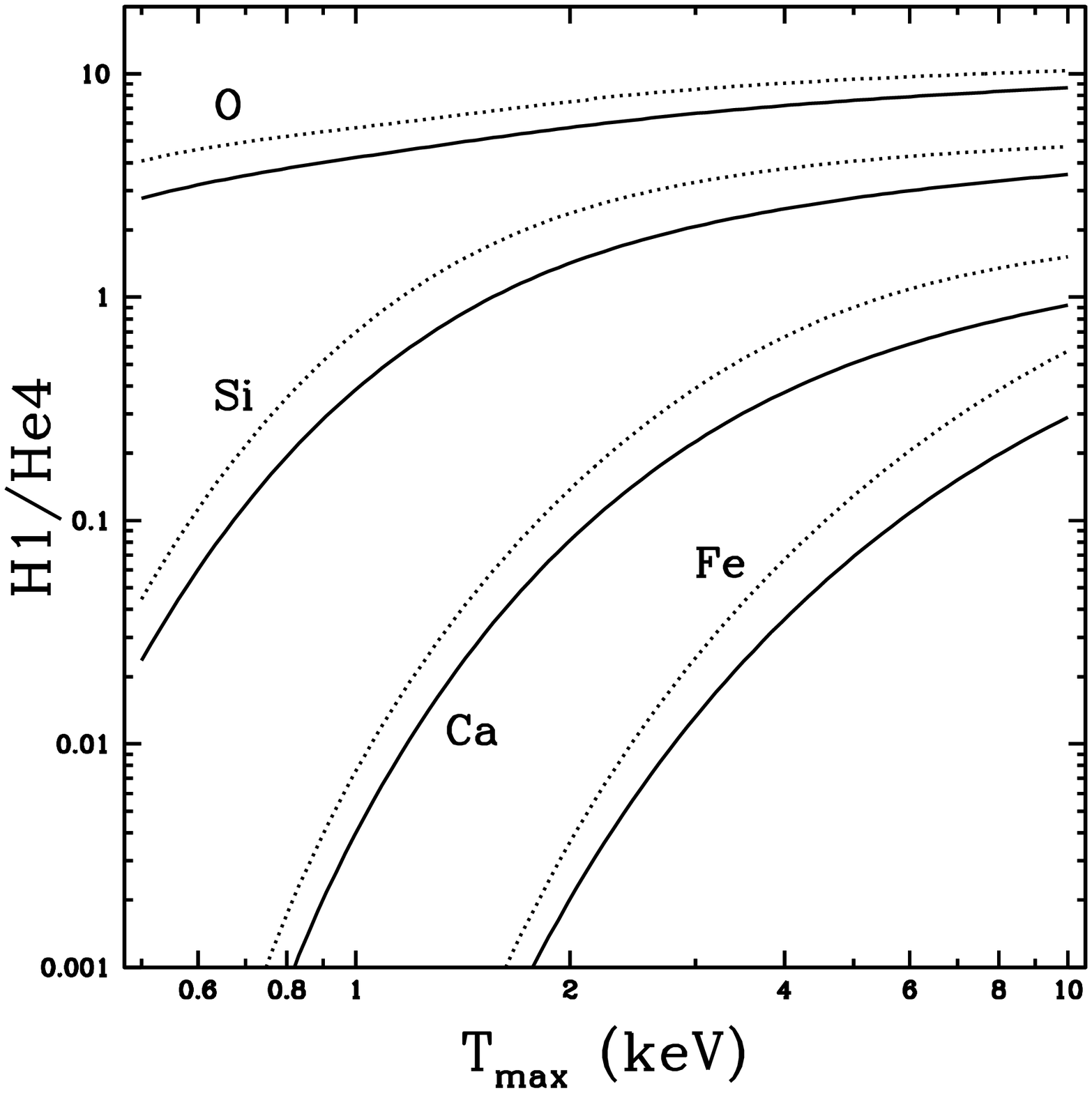,angle=0,height=0.3\textheight}}
}
\parbox{0.49\textwidth}{
\centerline{\psfig{figure=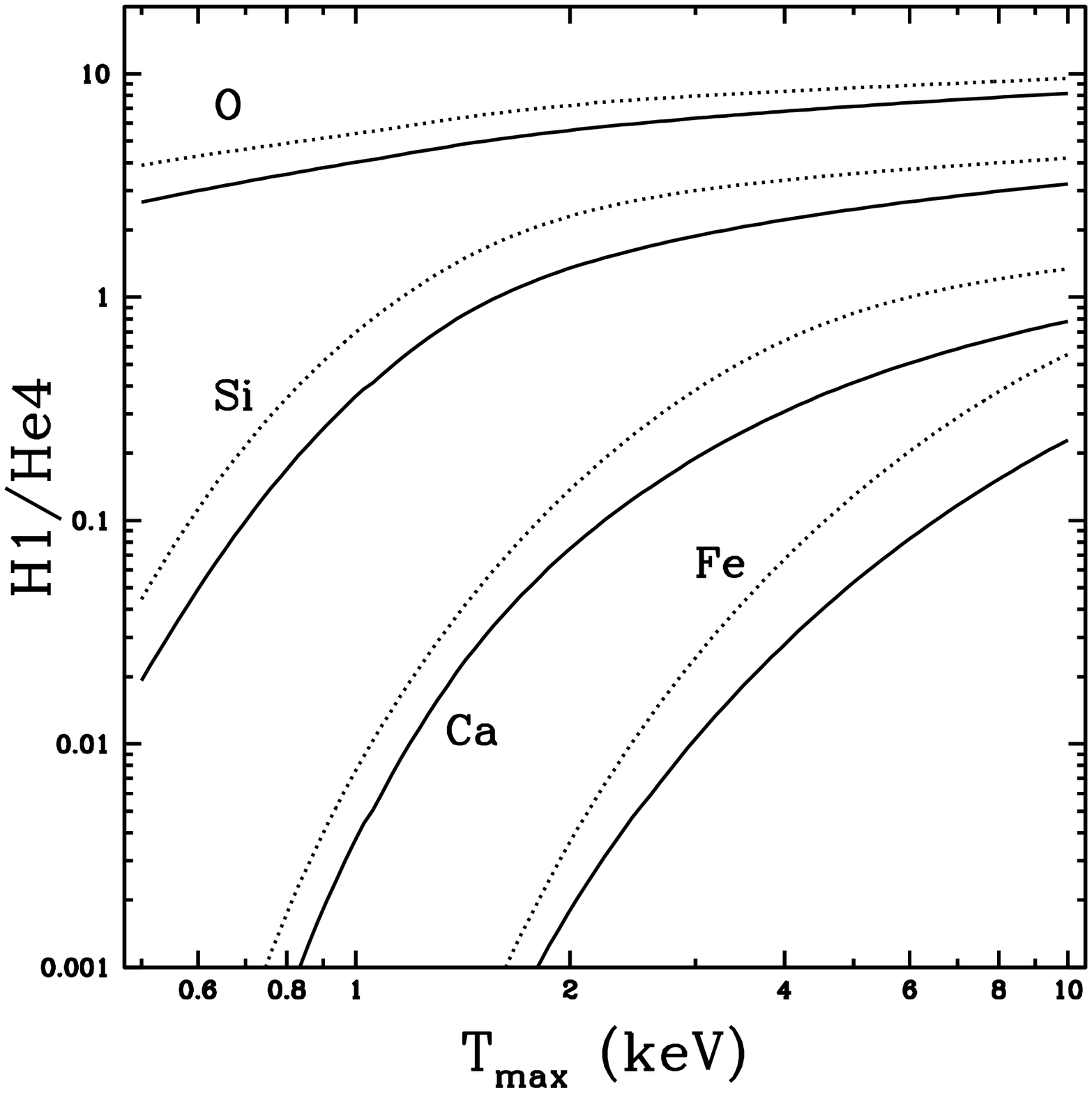,angle=0,height=0.3\textheight}}
}
\caption{\label{fig.cf} H1/He4 ratios for selected elements in models
with a constant pressure cooling flow (left) and isothermal sphere
cooling flow with $\dot{M}\propto r$ (right). In each case the line
ratios for the cooling flows are given by solid lines while composite
models consisting of a cooling flow and a single temperature component
(each with equal total emission measure) are given by the dotted
lines. All models have $Z=0.5Z_{\sun}$ and relative abundances are
fixed at the (photospheric) solar values.}
\end{figure*}

In Figure \ref{fig.cf} we show the H1/He4 line ratios of the constant
pressure cooling flow (with $Z=0.5Z_{\sun}$) as a function of $T_{\rm
max}$. The H1/He4 profiles are markedly different from the single
temperature case shown in Figure \ref{fig.ratios}. The differences are
most striking for those lines with $E\la 2$ keV. In particular, the O
H1/He4 profile changes by only a factor of $\sim 3$ while the single
temperature profile changes by several orders of magnitude over the
temperature range shown. (The O lines are much stronger in the cooling
flow models as well.) In addition to the K shell lines, the Fe L line
ratios (not shown) also have very different profiles from the
isothermal case making them very powerful for constraining the
properties of cooling flows (e.g. Canizares et al. 1982, 1988).

The line ratios are very similar for the $Z=2Z_{\sun}$ case where the
Ca and Fe H1/He4 profiles deviate from those in Figure \ref{fig.cf} by
less than $\sim 5$ per cent over the whole temperature range plotted.
The O and Si ratios show larger deviations which increase with
increasing $T_{\rm max}$; i.e. above 2 keV the O and Si ratios are
larger by $\sim 20$ - 40 per cent and $\sim 10$ - 20 per cent
respectively for the $Z=2Z_{\sun}$ case.

Since only regions where the cooling time is less than the age of the
system participate in the cooling flow, the observed X-ray flux will
generally include emission from the non-cooling ``ambient'' phase. In
Figure \ref{fig.cf} over the constant pressure model we also plot the
H1/He4 ratios of a model consisting of a constant pressure cooling
flow of temperature, $T_{\rm max}$ and a single temperature model with
temperature, $T=T_{\rm max}$ where the cooling-flow and isothermal
components have equal total emission measure.  The line ratio profiles
of the composite models have reasonably similar shapes to the case of
the cooling flow alone, but they are shifted up by $\sim 10$ - 40 per
cent.  The largest differences occur at the largest temperatures for
the lines with the highest energies.

The reason why the composite model still has line ratios markedly
different from the isothermal case is that each component has the same
total emission measure which implies $f=0.5$ and
$\sigma_\xi\rightarrow 0.5\sigma_\xi$ for the composite cooling flow
and $\delta$-function model (see section \ref{mps}); i.e. the values
of $\sigma_\xi$ range from $\sim 0.3$ - 0.4. As noted at the end of
the previous section, the boundary between approximately single-phase
and significantly multiphase emission measure distributions occurs for
$\sigma_\xi\sim 0.1$ when $\xi$ includes a $\delta$-function
component. Thus, a model with a cooling flow component and a
single-phase component has line ratios similar to an isothermal gas
for $f\la 1/7$ and $\rm EM_\delta/EM_{cf} \ga 6$.  (Note this applies
for $Z=0.5Z_{\sun}$ but is similar for $Z=2Z_{\sun}$ where we deduce
$f\la 1/5$ and $\rm EM_\delta/EM_{cf} \ga 4$.)

It is worth considering how much these properties change for a more
sophisticated multiphase cooling flow model that allows for
gravitational work to be done on the flow as it cools. Nulsen
\shortcite{pejn98} has derived $\xi(T)$ for a cooling flow in the
gravitational potential of a singular isothermal sphere. For $T\ll
T_{\rm max}$ the differential emission measure of this isothermal
sphere cooling flow model is equivalent to the constant pressure case
(equation \ref{eqn.cfdem}).  In contrast, $\xi(T_{\rm max}) = 0$ for
the isothermal sphere cooling flow.  The detailed shape for
intermediate temperatures depends on the slope $\eta$ of the mass
deposition profile, $\dot{M}\propto r^\eta$.

We plot $\xi(T)$ and $\sigma_\xi$ for $\eta=1$ in Figures
\ref{fig.dem_cf} and \ref{fig.sigxi_cf} for comparison to the constant
pressure case. Except for $T_{\rm max}\sim 1$ - 2 keV, the multiphase
strength for the $Z=0.5Z_{\sun}$ case is slightly smaller for the
isothermal sphere cooling flow; the differences are more drastic for
the $Z=2Z_{\sun}$ case. The emission measure weighted temperatures for
these models are $\sim 5$ - 10 per cent smaller than those of the
constant pressure cooling flow.

There is an abrupt change in $\sigma_\xi$ near 1 keV for the
isothermal sphere cooling flow models. Let us focus on the
$Z=0.5Z_{\sun}$ case first. Near 1.2 keV $\xi_{\rm max}$ jumps from
its location near the local maximum at 0.36 keV to a temperature just
below $T_{\rm max}$ (see Figure \ref{fig.dem_cf}). As $T_{\rm max}$
increases over the next $\sim 0.3$ keV we have $\xi_{\rm max}$
increasing faster than the total emission measure translating to a
decrease in $\sigma_\xi$.  This change is much more abrupt for the
isothermal sphere cooling flow because of the enforced condition that
$\xi(T_{\rm max}) = 0$. This overall behavior is more pronounced for
the $Z=2Z_{\sun}$ case because of the details of the strong Fe L shell
lines are proportionally more important.

The H1/He4 line ratios for both the isothermal sphere cooling flow
component and a composite model of an isothermal sphere cooling flow
component with a single temperature component (all with
$Z=0.5Z_{\sun}$) are plotted along side the corresponding constant
pressure models in Figure \ref{fig.cf}. For the case of only a cooling
flow the line ratios are very similar, especially for $T_{\rm max}\la
2$ keV. The greatest discrepancies between the isothermal and
constant-pressure cases are seen at the largest temperatures for the
lines with the highest energies; e.g. the Fe H1/He4 ratios differ by
$\sim 40$ per cent for $T_{\rm max}=10$ keV.  Adding the single
temperature to the isothermal cooling flow has a slightly larger
effect than found for the constant pressure cooling flow.  However,
the composite models of the isothermal and constant pressure cases
actually differ from each other by $\la 10$ per cent over the entire
temperature range shown.  (Note that the line ratios for models with
$Z=2Z_{\sun}$ differ by essentially the same small amounts described
above for the constant pressure models.)

\subsection{Distinguishing a two-temperature plasma from a cooling flow}
\label{2tcf}

\begin{table*}
\caption{Spectral Models for NGC 4472 and the Centaurus Cluster}
\label{tab.models}
\begin{tabular}{lccccccccccc}

& $N_{\rm H}^{\rm c}$ & $N_{\rm H}^{\rm h}$ & $T_{\rm c}$ & $T_{\rm
h}$ & $Z$ & $EM_{\rm c}$ & $EM_{\rm h}$ & $\dot{M}$ & $T_{\rm B}$ &
$EM_{\rm B}$ & $\sigma_\xi$\\ 
Model & \multicolumn{2}{c}{($10^{21}$ cm$^{-2}$)} & (keV) & (keV) &
$(Z_{\sun})$ & \multicolumn{2}{c}{($10^{65}$ cm$^{-3}$)} & ($M_{\sun}$
yr$^{-1}$) & (keV) & ($10^{65}$ cm$^{-3}$)\\ \\[-2 pt]
\multicolumn{10}{l}{NGC 4472:}\\
2T+BREM     &  2.9 & 0.04 & 0.72 & 1.39 & 2.00 & 0.065 & 0.050 & $\cdots$ &
3.23 & 0.0069 & 0.44\\
CF+1T+BREM & 2.9 & 0.00 & 1.30 & 1.30 & 1.41 & 0.185 & 0.032 & 2.54 &
3.71 & 0.0045 & 0.03\\ \\[-2 pt]
\multicolumn{10}{l}{Centaurus:}\\
2T & 2.3 & 1.4 & 1.04 & 3.07 & 1.10 & 1.41 & 17.0 & $\cdots$ &
$\cdots$ & $\cdots$ & 0.02\\
CF+1T & 3.0 & 1.3 & 3.30 & 3.30 & 1.19 & 2.45 & 11.8 & 45.4 & $\cdots$ &
$\cdots$ & 0.11\\ 
\end{tabular}

\medskip

\raggedright

Best-fitting spectral models to {\sl ASCA} data for the elliptical
galaxy NGC 4472 and for the Centaurus cluster. The CF components are
constant pressure cooling flow models, the 1T components are
single-temperature MEKAL models, and B stands for a bremsstrahlung
component. See text for further explanation of these models. The
emission measures are computed using the distances described in the
notes to Table \ref{tab.obs}. (Note that $\sigma_\xi$ for the 2T and
2T+BREM models is actually $\sigma_\xi^N$ evaluated for $N_c=20$.)

\end{table*}

\begin{figure*}
\parbox{0.49\textwidth}{
\centerline{\psfig{figure=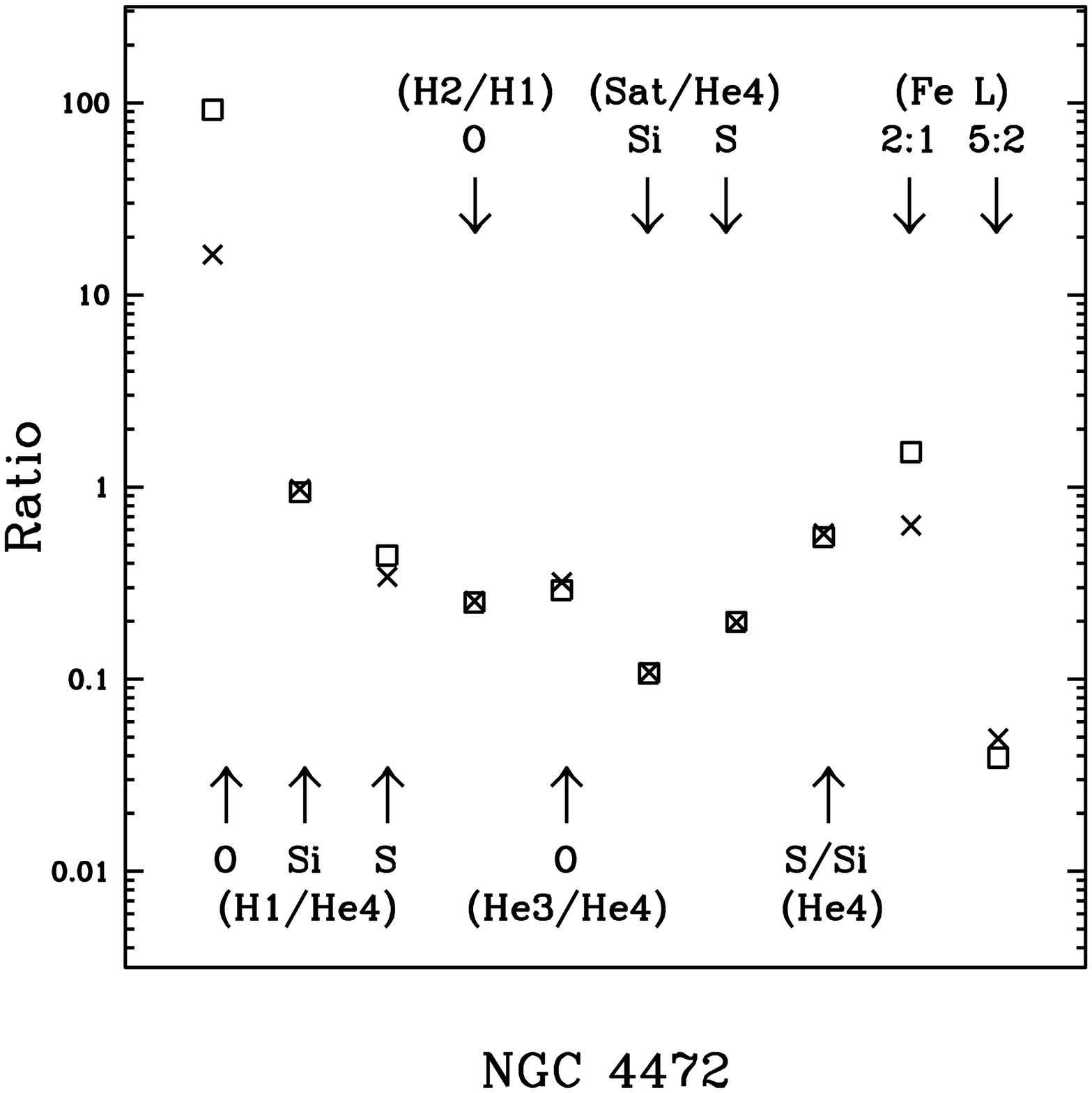,angle=0,height=0.3\textheight}}
}
\parbox{0.49\textwidth}{
\centerline{\psfig{figure=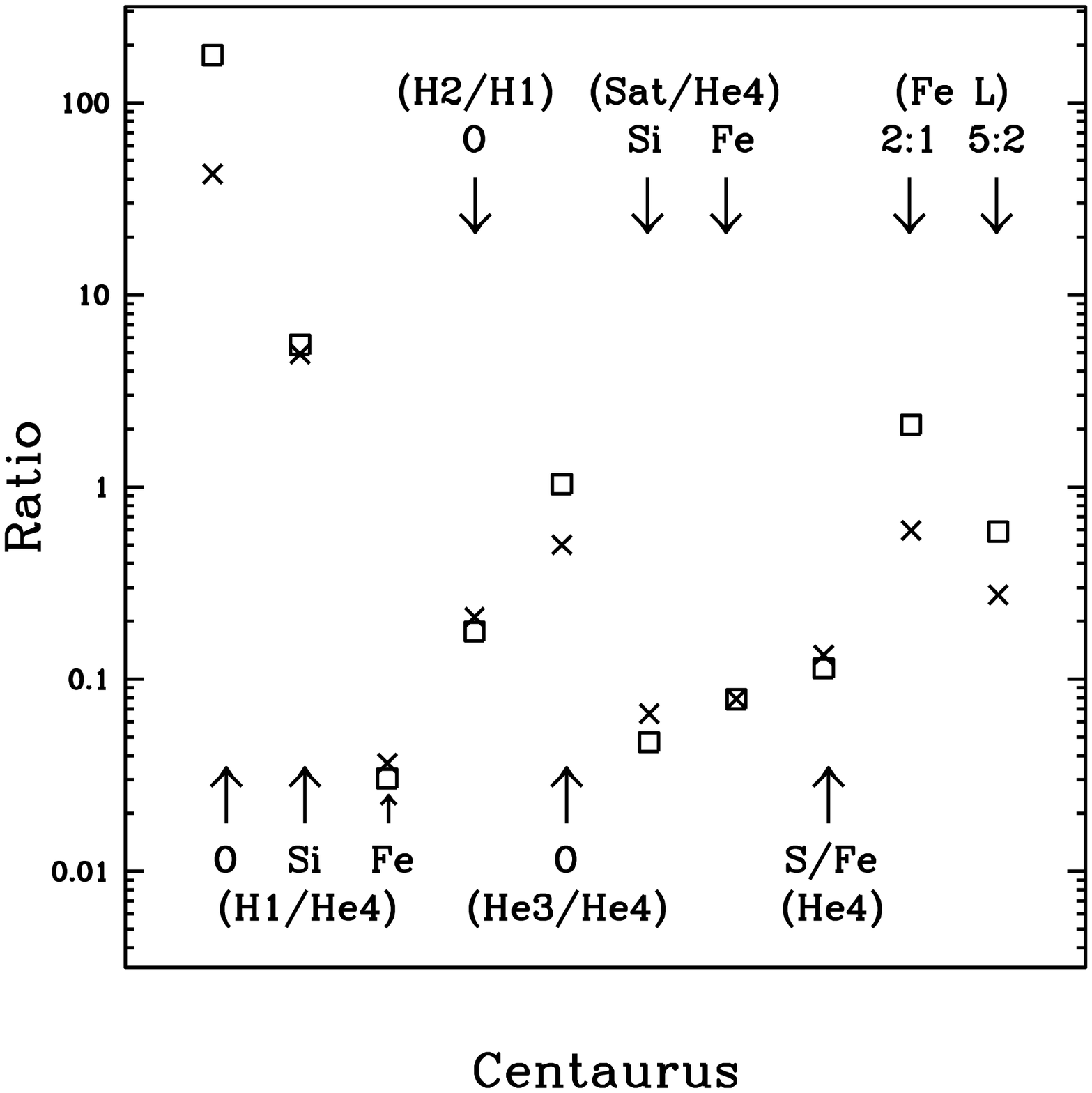,angle=0,height=0.3\textheight}}
}
\caption{\label{fig.2tvscf} Selected line ratios of two-temperature
models (boxes) and models consisting of a cooling flow and a single
temperature (crosses) for NGC 4472 and Centaurus (see Table
\ref{tab.models}). (Note that the line ratios are computed as
described in Section \ref{theory} except that the absorption is
included.)}
\end{figure*}

We have seen in the previous section that for a particular value of a
line ratio (e.g. Si H1/He4) the two-temperature models appear to have
a larger variety of possible $\sigma_\xi$ which can produce a line
ratio in comparison to the simple continuous emission measure models
examined. As a result, for moderate resolution spectra like {\sl ASCA}
\cite{tanaka} where at most only a limited number of line blends are
resolved, it should be expected that a two-temperature model can
provide an acceptable fit even if the true spectrum is that of a
cooling flow. In fact, analyses of {\sl ASCA} spectra of the brightest
ellipticals (Buote \& Fabian 1998; Buote 1999a), poor groups
\cite{bgroup}, and clusters of galaxies (e.g. Fabian et al. 1994a)
have found that two-temperature models and models composed of a
cooling flow with an ambient temperature component essentially
describe the data equally well.

To examine how to distinguish between the two-temperature and cooling
flow models we first examine the case of the bright elliptical galaxy
NGC 4472. In Table \ref{tab.models} we list the results of the
best-fitting two-temperature and cooling flow models to the{\sl ASCA}
spectra accumulated within a radius of $\sim 30$ kpc from the center
of NGC 4472 by Buote \shortcite{bmulti}.  The ``2T+BREM'' model
consists of two MEKAL plasma components and one component of thermal
bremsstrahlung to account for possible emission from discrete sources;
$T_{\rm c}$ and $T_{\rm h}$ represent the temperatures of the
``cooler'' and ``hotter'' MEKAL components respectively.  The
``CF+1T+BREM'' model replaces the cooler MEKAL component with a
constant pressure cooling flow component (equation \ref{eqn.cfdem}).
Each model component is modified by photo-electric absorption
\cite{phabs} although the absorption on the bremsstrahlung components
is fixed to the Galactic value. We refer the reader to Buote
\shortcite{bmulti} for full details of these models.

In Figure \ref{fig.2tvscf} we plot a selection of line ratios computed
from the models in Table \ref{tab.models}. (We show a related plot of
the ratios of the spectra of two temperature and cooling flow models
in Figure \ref{fig.astroe.ratio} in Section \ref{astroe}.) Overall, it
is remarkable how similar the line ratios are for the two models.  The
O H1/He4 and Fe L ratios show the most differences between the
two-temperature and cooling flow models; i.e. the O H1/He4 ratios
differ by a factor of $\sim 4$ and the 2:1 Fe L ratios differ by a
factor of $\sim 2$. The ratios shown involving other elements differ
by $\la 10$ per cent. Note that we expect the H1/He4 ratio of Si to be
similar for the models since the blends involving the He4 and H1
transitions are reasonably resolved by {\sl ASCA} (see next section)
and thus the best-fitting models in Table \ref{tab.models} attempt to
match the observed Si ratio as accurately as possible.

For comparison to NGC 4472 where the cooling flow component dominates
the emission measure and $T_{\rm max}< 2$ keV we consider the
Centaurus cluster of galaxies. In Table \ref{tab.models} we list the
best-fitting 2T and CF+1T models to the {\sl ASCA} data accumulated
within a radius of $\sim 100$ kpc. The details of the data reduction
and spectral fitting follow those in section \ref{asca} with the
following important exceptions. Firstly, we fitted the {\sl ASCA} SIS
spectra over the energy range 0.5-9. keV. Secondly, we used the
photospheric solar abundances according to Anders \& Grevesse
\shortcite{ag} because of our current implementation of the cooling
flow model for use in {\sc xspec} \cite{xspec} is defined for those
solar abundances. Although this restriction limits the quality of the
fits, the 2T and CF+1T models give very similar quality fits which is
all that we require for this discussion. We refer the reader to
section \ref{asca} for a thorough discussion of successful spectral
models of Centaurus.

In contrast to NGC 4472 the isothermal component dominates the
emission measure of the cooling flow model for Centaurus, $\rm EM_{\rm
h}/EM_{\rm c} \approx 5$. Nevertheless, the CF+1T model has
$\sigma_\xi=0.11$ indicating that it is significantly multiphase (see
previous section). We plot selected line ratios of the 2T and CF+1T
models in Figure \ref{fig.2tvscf}. Since Centaurus has a higher
temperature than NGC 4472, in the figure we replace the S H1/He4 and
Sat/He4 ratios with the corresponding Fe ratios; S/Si (He4) is
replaced with S/Fe (He4). (Again, we show a related plot in Figure
\ref{fig.astroe.ratio} in Section \ref{astroe}.)  Similar to NGC 4472
the O H1/He4 and Fe L ratios show the most substantial differences
(factors of $\sim 5$). However, unlike NGC 4472 the O He3/He4 and Si
Sat/He4 show noticeable differences. (The Sat/He4 ratios of S, Ar, and
Ca have quite similar differences as well.) Again the Si H1/He4 ratios
are very similar for the three models because the Si H1/He4 ratio is a
key constraint of the {\sl ASCA} spectral fits.

Hence, the most substantial differences between two-temperature and
cooling flow models occur for O and Fe L line ratios. Analyses that
exclude energies below 1.4 keV (e.g. to avoid small inaccuracies in
the plasma codes) will have a very difficult time differentiating
two-temperature and cooling flow models, particularly for moderate
spectral resolution data provided by {\sl ASCA, Chandra, and XMM}.
Moreover, because of their overall flexibility, an observed
consistency of a two-temperature model with data of low-moderate
spectral resolution of ellipticals, groups, and clusters should not be
regarded as decisive evidence that the system has only two phases.

 \section{Application to {\sl ASCA} data}
\label{asca}

\begin{figure*}
\parbox{0.49\textwidth}{
\centerline{\psfig{figure=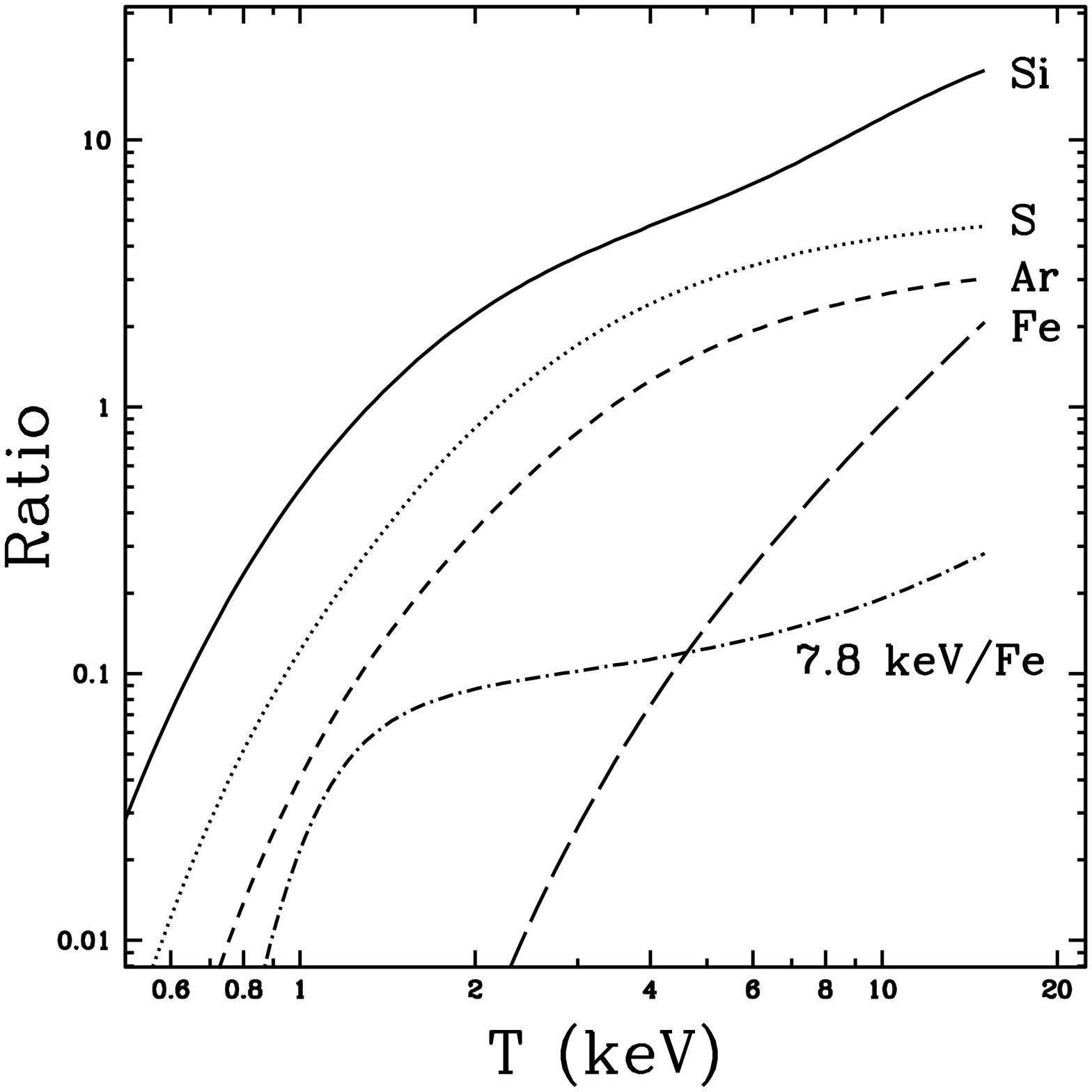,angle=0,height=0.3\textheight}}
}
\parbox{0.49\textwidth}{
\centerline{\psfig{figure=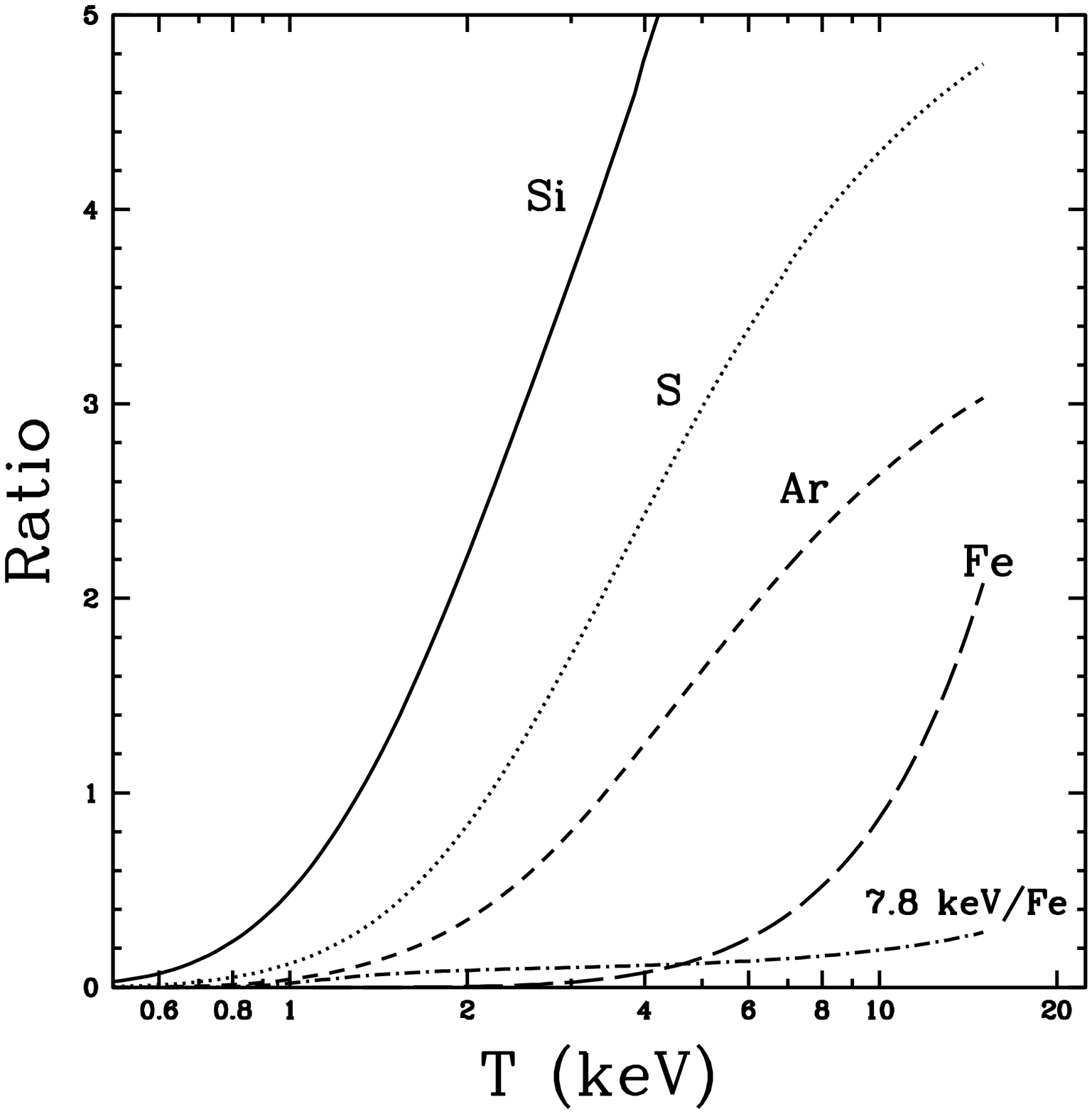,angle=0,height=0.3\textheight}}
}
\caption{\label{fig.ratios.asca} (Left) Line ratios as a function of
temperature in isothermal MEKAL models (with $Z=0.5Z_{\sun}$) where
the emission lines have been computed over large energy ranges
appropriate for the energy resolution of {\sl ASCA} (see Table
\ref{tab.asca.def}). The ratios for Si, S, Ar, and Fe refer to ratios
of H-like to He-like K$\alpha$ transitions as in Figure
\ref{fig.ratios} except that now additional transitions are included
in the blends.  The 7.8 keV blend consists mostly of K$\beta$ Fe XXV
(He3) although K$\alpha$ emission from Ni XXVII and Ni XXVIII
contribute as well. (Right) The same thing but plotted on a linear
scale.}
\end{figure*}

\begin{table}
\caption{Energy Ranges of {\sl ASCA} Line Blends}
\label{tab.asca.def}
\begin{tabular}{lc}
  & Energy Range\\
Dominant Ion & (keV)\\
Si XIII & 1.80 -  1.90\\
Si XIV & 1.95 -  2.05\\
S XV & 2.35 -  2.53\\
S XVI & 2.55 -  2.70\\
Ar XVII & 3.00 - 3.25\\
Ar XVIII & 3.25 - 3.40\\
Fe XXV & 6.40 -  6.80\\
Fe XXVI & 6.86 -  7.14\\
7.8 keV Blend & 7.60 - 8.10\\

\end{tabular}

\medskip

Energy ranges used to define the lines in Figure \ref{fig.ratios.asca}
appropriate for the energy resolution of the {\sl ASCA} SIS during
approximately 1994.
\end{table}

Recently the {\sl ASCA} satellite is able to resolve emission line
blends of K shell transitions of Mg, Si, S, Ar, Ca, and Fe, and also
has the effective area and imaging capabilities lacking in previous
X-ray satellites having similar or better spectral resolution.  The
energy resolution of {\sl ASCA} (e.g. $\Delta E \sim 100$ eV FWHM at 2
keV) allows only blends of lines to be measured, and we show in Figure
\ref{fig.ratios.asca} the temperature profiles of ratios of the most
important line blends relevant to {\sl ASCA} observations of
ellipticals, groups, and clusters; the definitions of the blends are
given in Table \ref{tab.asca.def}. These profiles should be compared
to Figure \ref{fig.ratios}.

Even though they are the most powerful diagnostics of multiphase
cooling flows (see Section \ref{2tcf}), we do not show oxygen or Fe L
blends in Figure \ref{fig.ratios.asca} for the following reasons.  The
strongest transitions of the He-like O VII ion appear at energies
$\sim 0.57$ keV which lie in a region where the {\sl ASCA} SIS have
small effective area and problems with the calibration. The energy
resolution of {\sl ASCA} does not allow useful separation of blends of
Fe L lines between $\sim 0.7$ - 1.4 keV for individual analysis. This
is not crucial to our present analysis since it is our intention to
investigate the K shell emission lines which do not suffer from the
possible inaccuracies in the plasma codes associated with the Fe L
shell transitions.

As indicated by the discussion in Section \ref{2tcf} line ratios other
than those of oxygen or Fe L in rival multiphase models generally
differ by only $\sim 10$ - 20 per cent. Only a handful of {\sl ASCA}
observations of the brightest ellipticals, groups, and clusters have
sufficient S/N to even approach this level of precision. The K$\alpha$
line blends of both H-like and He-like Si and S have been analyzed
with {\sl ASCA} by Buote \shortcite{bmulti} for three of the brightest
elliptical galaxies (NGC 1399, 4472, and 4636) and one of the
brightest galaxy groups (NGC 5044). Although the properties of these
line blends favor multitemperature models obtained from broad-band
(0.5-9 keV) fits to the {\sl ASCA} spectra, the line blends are not
measured precisely enough to place interesting constraints on
$\xi(T)$ by themselves.

Clusters of galaxies are hotter and more luminous than ellipticals and
groups, and thus the K shell emission lines above $\sim 2$ keV in
principle can be measured more precisely for the brightest
clusters. Let us consider the bright nearby clusters M87 (Virgo),
Centaurus (A3526), and Perseus (A426) which have among the highest S/N
data available for clusters observed with {\sl ASCA}. Broad-band
spectral fitting of the {\sl ASCA} data (which includes the Fe L
lines) has shown that at least two temperature components are required
in the cores of each of these clusters (Fabian et al. 1994a; Fukazawa
et al. 1994; Matsumoto et al. 1996). It is our intention to examine
what can be learned from analysis of only the K shell emission line
blends and, in particular, to examine whether they too require
multiple temperature components in M87, Centaurus, and Perseus.
 
\subsection{{\sl ASCA} observations and data reduction}
\label{obs}

\begin{table*}
\caption{Cluster Properties and {\sl ASCA} Observations}
\label{tab.obs}
\begin{tabular}{lcccccccccccrr}
Cluster & $z$ & $N_{\rm H}$ & Sequence & \multicolumn{1}{c}{Date} &
SIS & SIS & RDD & \multicolumn{2}{c}{Radius} &
\multicolumn{2}{c}{Radius}\\  
& & ($10^{21}$ cm$^{-2}$) & & \multicolumn{1}{c}{Mo/yr} & CCD & Data &
Correction & \multicolumn{2}{c}{(arcmin)} & \multicolumn{2}{c}{(kpc)}\\  
& & & & & Mode & Mode & &  SIS & GIS & SIS & GIS\\
M87       & 0.0043 & 0.25 & 60033000 & 6/93 & 4 & BRIGHT  & N & 6.2 &
7.5 & 32.5 & 39.3\\
Centaurus & 0.0110 & 0.81 & 83026000 & 7/95 & 1 & BRIGHT2 & Y & 5.6 &
6.5 & 77.2 & 89.6 \\
Perseus   & 0.0183 & 1.41 & 80007000 & 8/93 & 4 & BRIGHT  & N & 5.6 &
6.3 & 128.7 & 144.8\\
\end{tabular}

\medskip
\raggedright

Selected properties of the {\sl ASCA} SIS and GIS observations.
Galactic Hydrogen column densities $(N_{\rm H})$ are from Dickey \&
Lockman \shortcite{dl} using the HEASARC w3nh tool. Extraction radii
in kpc are computed using luminosity distances of 18 Mpc for M87, 47
Mpc for Centaurus, and 79 Mpc for Perseus assuming $H_0=70$ km
s$^{-1}$ Mpc$^{-1}$ and $\Omega_0=0.3$.

\raggedright

\end{table*}

\begin{table*}
\caption{{\em ASCA} Exposures and Count Rates}
\label{tab.exp}
\begin{tabular}{lrrrrrrrr}
Cluster  & \multicolumn{2}{c}{Exposure} &
      \multicolumn{2}{c}{Count Rate} & \multicolumn{2}{c}{Exposure} &
      \multicolumn{2}{c}{Count Rate}\\
      & \multicolumn{2}{c}{($10^{3}$s)}  &
      \multicolumn{2}{c}{(ct s$^{-1}$)} &
      \multicolumn{2}{c}{($10^{3}$s)}  & \multicolumn{2}{c}{(ct s$^{-1}$)} \\
      & SIS0 & SIS1 & SIS0 & SIS1 & GIS2 & GIS3 & GIS2 & GIS3\\
M87       & 12.8 & 9.79 & 2.59 & 1.87 & 16.7 & 16.7 & 2.33 & 2.39\\
Centaurus & 64.6 & 64.4 & 1.22 & 0.94 & 70.8 & 70.8 & 0.86 & 1.06\\
Perseus   & 16.1 & 16.1 & 7.15 & 6.16 & 15.0 & 14.9 & 5.53 & 6.62\\
\end{tabular}

\medskip
\raggedright

Exposures include any time filtering. The 1.6-9 keV count rates are
background-subtracted within the particular aperture (see Table
\ref{tab.obs}). 

\end{table*}

The {\sl ASCA} satellite consists of two X-ray CCD cameras (Solid
State Imaging Spectrometers -- SIS0 and SIS1) and two proportional
counters (Gas Imaging Spectrometers -- GIS2 and GIS3) each of which is
illuminated by its own X-ray telescope (XRT). The SIS has superior
energy resolution and effective area below $\sim 7$ keV while the GIS
has a larger field of view. Although the point spread function (PSF)
of each XRT has a relatively sharp core, the wings of the PSF are
quite broad (half power diameter $\sim 3\arcmin$) and increase
markedly for energies above a few keV (e.g. Kunieda et al. 1995). As
in our previous studies we do not attempt to analyze the spatial
distribution of the {\sl ASCA} data of these sources because of the
large, asymmetric, energy-dependent PSF. Rather, we analyze the {\sl
ASCA} X-ray emission within a single large aperture for each cluster
which encloses the region of the brightest emission (see below). For
clusters of galaxies the spectrum within this central region is not
very much affected by scattered emission from large radii
(e.g. Boehringer et al. 1998).

We obtained {\sl ASCA} data for the clusters from the public data
archive maintained by the High Energy Astrophysics Science Archive
Research Center (HEASARC). The properties of the observations are
listed in Tables \ref{tab.obs} and \ref{tab.exp}. We reduced these
data with the standard {\sc FTOOLS} (v4.1) software according to the
procedures described in The {\sl ASCA} Data Reduction Guide and the
WWW pages of the {\sl ASCA} Guest Observer Facility (GOF)\footnote{See
http://heasarc.gsfc.nasa.gov/docs/asca/abc/abc.html and
http://heasarc.gsfc.nasa.gov/docs/asca/.}.

For the observations of M87 and Perseus (taken during the
Performance-Verification phase) we used the events files generated by
the default screening criteria processed under the Revision 2 Data
Processing (REV2).  As is standard procedure for analysis of SIS data,
only data in BRIGHT mode taken in medium or high bit rate were used.
Since the Centaurus observation was performed in 1995 its SIS data are
degraded to some extent because of radiation damage to the CCDs (see
Dotani et al. 1997). We corrected the Centaurus data for this ``RDD
effect'' following the procedure outlined on the {\sl ASCA} GOF WWW
pages which produces corrected SIS event files in BRIGHT2 mode
constructed using the default REV2 screening criteria. (Again, only
data with medium or high bit rate were used.)  Since we focus on
energies above 1.6 keV this correction is actually not very important
for our study.

Our final screening of the data involved visual examination of the
light curves for each observation and removing intervals of
anomalously high (or low) count rate. We note that only data with the
standard events grades (0234) were used, and we made the required dead
time corrections for the GIS data.

The final processed events were then extracted from a region centered
on the emission peak for each detector of each sequence.  We selected
a particular extraction region using the following general
guidelines. Our primary concern is to select a region that encloses
most of the X-ray emission yet is symmetrically distributed about the
origin of the region.  We limited the size of the aperture to ensure
that the entire aperture fit on the detector, an issue more important
for the SIS because of its smaller field-of-view. Moreover, for the
SIS0 and SIS1 we tried to limit the apertures to as small a number of
chips as possible to reduce the effects of residual errors in
chip-to-chip calibration. The GIS regions were chosen to be of similar
size to the corresponding SIS regions of a given sequence for
consistency. Since, however, the GIS+XRT PSF is somewhat larger than
that of the SIS+XRT the extraction regions for the GIS are usually
$\sim 20\%$ larger.

Since the SIS was operated in 1-ccd mode for the Centaurus
observation we extracted all of the emission from that ccd which
amounts to a square region with sides of $\sim 11\arcmin$ width. The
observation of Perseus was positioned very near the center of the
default chips of the SIS and so we also extracted its SIS data from
the whole chip. The observation of M87 was centered between two chips,
and thus we used a circular region. In Table \ref{tab.obs} we list the
region sizes for the SIS and GIS data of each cluster. For the SIS
data of Centaurus and Perseus the radii actually are half-widths of
the default ccds. (Note that the data between the ccds are excluded.)
All of the GIS radii refer to circles. We extracted the events using
regions defined in detector coordinates as is recommended in the {\it
ASCA} ABC GUIDE because the spectral response depends on the location
on the detector not the position on the sky.

We computed a background spectrum for each detector using the standard
deep observations of blank fields. It is necessary to use these
background templates because the cluster emission fills the entire
field of view in each case The background templates for the SIS are
most appropriate for SIS data taken in 4-ccd mode early in the mission
which applies to the M87 and Perseus observations. For more recent
data taken in 1-cdd or 2-ccd mode, using these background templates
can lead to spurious effects particularly for energies above $\sim 7$
keV where instrumental background dominates the cosmic X-ray
background. The Centaurus observation was performed in 1995 and thus
these effects are not expected to be as serious as for more recent
data. At any rate, the Centaurus data are not dominated by background
so small systematic effects in the background should not be apparent
in our analysis. The final screened background-subtracted exposures
and count rates are listed in Table \ref{tab.exp}.

The instrument response matrix required for spectral analysis of {\sl
ASCA} data is the product of a spectral Redistribution Matrix File
(RMF) and an Auxiliary Response File (ARF). The RMF specifies the
channel probability distribution for a photon (i.e. energy resolution
information) while the ARF contains the information on the effective
area.  An RMF needs to be generated specifically for each SIS of each
observation because, among other reasons, each chip of each SIS
requires its own RMF and the spectral resolution of the SIS is
degrading with time. We generated the responses for each SIS using the
{\sc FTOOL sisrmg} selecting for the standard event grades
(0234). Using the response matrix and spectral PI (Pulse Invariant)
file we constructed an ARF file with the {\sc FTOOL ascaarf}.

The source aperture used for the SIS overlapped more than one chip for
M87. In order to analyze the spectra of such regions we followed the
standard procedure of creating a new response matrix that is the
average of the individual response matrices of each chip weighted by
the number of source counts of each chip (i.e. within the source
aperture). (Actually, the current software only allows the RMFs to be
averaged and then an ARF is generated using the averaged RMF. This is
considered to be a good approximation for most sources.)
Unfortunately, some energy resolution is lost as a result of this
averaging. 

Since the K shell emission lines for energies above $\sim 2$ keV are
relatively faint in these clusters we maximize the S/N of the data by
combining the spectra from each SIS and similarly for each GIS;
i.e. SIS=SIS0+SIS1 and GIS=GIS2+GIS3. For the GIS data this does not
result in significant loss of information because the RMFs of the GIS2
and GIS3 are the same and time-independent; only the ARFs need to be
averaged for the GIS2 and GIS3. However, the responses of the SIS0 and
SIS1 detectors are different and thus the required averaging of the
RMFs and ARFs leads again to loss of information. Because the gain in
S/N is important for our study, we examine the combined SIS and
combined GIS data of each cluster where the source spectra, RMF, ARF,
and background for each detector are summed and scaled appropriately
using the {\sc FTOOL addascaspec} (v1.27).

Overall we found that the results obtained from analysis of these
summed SIS spectra agreed very well with results obtained without
summing the data. Henceforth we shall focus on results obtained from
the summed SIS data though we will (for completeness) mention some of
the results obtained from separate analysis of the SIS0 and SIS1 data.

\subsection{Analysis of individual emission line blends}
\label{id}

\begin{table*}
\caption{Emission Lines Measured from {\sl ASCA} Data}
\label{tab.lines}
\begin{tabular}{lccccc}
& $E$ & $T$ & EW & \multicolumn{2}{c}{Flux}\\ 
Ion & (keV) & (keV) & (eV) & ($10^{-5}$ ph cm$^{-2}$ s$^{-1}$) & ($10^{-13}$ erg
cm$^{-2}$ s$^{-1}$)\\
\\M87:\\[+3pt]
Mg XII   (H)     & $1.48_{-0.01}^{+0.01}$ & $1.06_{-0.14}^{+0.18}$ & $    28_{-     5}^{+     6}$ & $ 62.86_{- 10.99}^{+ 11.07}$ & $ 14.86_{-  2.60}^{+  2.62}$\\
1.7 keV Feature    & $1.74_{-0.02}^{+0.02}$ & $1.23_{-0.21}^{+0.29}$ & $    13_{-     6}^{+     6}$ & $ 19.91_{-  9.17}^{+  9.16}$ & $  5.54_{-  2.55}^{+  2.55}$\\
Si XIII  (He)    & $1.84_{-0.01}^{+0.01}$ & $1.23_{-0.21}^{+0.29}$ & $    54_{-     8}^{+     9}$ & $ 68.68_{-  9.34}^{+  9.49}$ & $ 20.29_{-  2.76}^{+  2.80}$\\
Si XIV   (H)     & $1.99_{-0.00}^{+0.00}$ & $1.23_{-0.21}^{+0.29}$ & $   110_{-    16}^{+    17}$ & $110.40_{- 11.00}^{+ 11.22}$ & $ 35.26_{-  3.51}^{+  3.58}$\\
2.2 keV Feature  & $2.18_{-0.01}^{+0.02}$ & $1.23_{-0.21}^{+0.29}$ & $    28_{-    14}^{+    16}$ & $ 21.94_{-  9.97}^{+ 10.07}$ & $  7.66_{-  3.48}^{+  3.52}$\\
S XV     (He)    & $2.44_{-0.01}^{+0.01}$ & $2.19_{-0.50}^{+0.89}$ & $    68_{-    13}^{+    14}$ & $ 47.25_{-  7.94}^{+  7.84}$ & $ 18.49_{-  3.11}^{+  3.07}$\\
S XVI    (H)     & $2.60_{-0.01}^{+0.01}$ & $2.19_{-0.50}^{+0.89}$ & $    68_{-    11}^{+    12}$ & $ 40.52_{-  6.17}^{+  6.17}$ & $ 16.90_{-  2.57}^{+  2.57}$\\
Ar XVII  (He)    & $3.09_{-0.02}^{+0.02}$ & $2.43_{-0.19}^{+0.22}$ & $    44_{-    12}^{+    12}$ & $ 16.86_{-  4.38}^{+  4.41}$ & $  8.35_{-  2.17}^{+  2.18}$\\
$^{\ast}$Ar XVIII (H)     & $3.28_{-0.05}^{+0.05}$ & $2.43_{-0.19}^{+0.22}$ & $11_{-7}^{+7}$ & $3.54_{-2.27}^{+2.18}$ & $1.86_{-1.19}^{+1.15}$\\
Ca XIX   (He)    & $3.89_{-0.03}^{+0.03}$ & $2.43_{-0.19}^{+0.22}$ & $    43_{-    15}^{+    15}$ & $  8.70_{-  3.04}^{+  2.98}$ & $  5.42_{-  1.90}^{+  1.86}$\\
Fe XXV   (He)    & $6.63_{-0.02}^{+0.01}$ & $2.61_{-0.45}^{+0.62}$ & $  1010_{-   147}^{+   160}$ & $ 32.57_{-  4.11}^{+  4.09}$ & $ 34.61_{-  4.37}^{+  4.35}$\\
7.8 keV Feature  & $7.82_{-0.23}^{+0.15}$ & $2.61_{-0.45}^{+0.62}$ & $   308_{-   204}^{+   228}$ & $  5.02_{-  3.22}^{+  3.19}$ & $  6.29_{-  4.03}^{+  4.00}$\\
\\Centaurus:\\[+3pt]
Mg XII   (H)     & $1.47_{-0.01}^{+0.01}$ & $3.06_{-0.64}^{+1.06}$  & $    27_{-     4}^{+     4}$ & $ 26.22_{-  3.34}^{+  3.35}$ & $  6.18_{-  0.79}^{+  0.79}$\\
Si XIII  (He)    & $1.81_{-0.01}^{+0.01}$ & $1.61_{-0.19}^{+0.24}$  & $    25_{-     4}^{+     4}$ & $ 16.52_{-  2.29}^{+  2.30}$ & $  4.78_{-  0.66}^{+  0.66}$\\
Si XIV   (H)     & $1.98_{-0.00}^{+0.00}$ & $1.61_{-0.19}^{+0.24}$  & $    83_{-     8}^{+     8}$ & $ 42.43_{-  3.22}^{+  3.22}$ & $ 13.47_{-  1.02}^{+  1.02}$\\
2.2 keV Feature  & $2.22_{-0.02}^{+0.02}$ & $1.61_{-0.19}^{+0.24}$  & $    34_{-    12}^{+    13}$ & $ 13.17_{-  4.36}^{+  4.26}$ & $  4.68_{-  1.55}^{+  1.51}$\\
S XV     (He)    & $2.44_{-0.01}^{+0.01}$ & $2.97_{-0.61}^{+1.08}$  & $    44_{-     9}^{+    10}$ & $ 16.52_{-  3.00}^{+  3.01}$ & $  6.45_{-  1.17}^{+  1.18}$\\
S XVI    (H)     & $2.60_{-0.01}^{+0.01}$ & $2.97_{-0.61}^{+1.08}$  & $    50_{-     7}^{+     8}$ & $ 16.09_{-  2.15}^{+  2.14}$ & $  6.71_{-  0.90}^{+  0.89}$\\
Ar XVII  (He)    & $3.09_{-0.02}^{+0.02}$ & $4.12_{-0.35}^{+0.41}$  & $    40_{-     9}^{+     9}$ & $  8.58_{-  1.80}^{+  1.78}$ & $  4.25_{-  0.89}^{+  0.88}$\\
Ar XVIII (H)     & $3.31_{-0.52}^{+0.58}$ & $4.12_{-0.35}^{+0.41}$ & $    22_{-     8}^{+     8}$ & $  4.15_{-  1.78}^{+  0.53}$ & $  2.20_{-  0.79}^{+  0.79}$\\
Ca XIX   (He)    & $3.86_{-0.58}^{+0.51}$ & $4.12_{-0.35}^{+0.41}$ & $    26_{-     9}^{+     9}$ & $  3.54_{-  0.87}^{+  2.10}$ & $  2.19_{-  0.72}^{+  0.71}$\\
Ca XX    (H)     & $4.04_{-0.07}^{+0.07}$ & $4.12_{-0.35}^{+0.41}$  & $    12_{-     8}^{+     9}$ & $  1.53_{-  1.08}^{+  1.09}$ & $  0.99_{-  0.70}^{+  0.70}$\\
Fe XXV   (He)    & $6.63_{-0.01}^{+0.01}$ & $4.37_{-0.64}^{+0.85}$ & $   972_{-    62}^{+    68}$ & $ 33.81_{-  1.81}^{+  1.83}$ & $ 35.89_{-  1.92}^{+  1.94}$\\
Fe XXVI  (H)     & $6.97_{-0.07}^{+0.07}$ & $4.37_{-0.64}^{+0.85}$ & $    41_{-    21}^{+    21}$ & $  2.40_{-  1.22}^{+  1.22}$ & $  2.68_{-  1.37}^{+  1.36}$\\
7.8 keV Feature  & $7.81_{-0.10}^{+0.07}$ & $4.37_{-0.64}^{+0.85}$ & $   211_{-    84}^{+    84}$ & $  4.21_{-  1.58}^{+  1.47}$ & $  5.26_{-  1.97}^{+  1.83}$\\
8.4 keV Feature  & $8.35_{-0.06}^{+0.06}$ & $4.37_{-0.64}^{+0.85}$ & $   241_{-   120}^{+   126}$ & $  3.86_{-  1.83}^{+  1.78}$ & $  5.17_{-  2.45}^{+  2.38}$\\
\\Perseus:\\[+3pt]
Mg XII   (H)     & $1.48_{-0.01}^{+0.01}$ & $10.96_{-4.27}^{+15.60}$ & $    13_{-     3}^{+     3}$ & $ 74.19_{- 15.07}^{+ 16.28}$ & $ 17.62_{-  3.58}^{+  3.87}$\\
1.7 keV Feature  & $1.72_{-0.02}^{+0.03}$ & $5.01_{-1.15}^{+2.16}$   & $     9_{-     4}^{+     4}$ & $ 42.13_{- 18.20}^{+ 18.22}$ & $ 11.60_{-  5.01}^{+  5.02}$\\
Si XIII  (He)    & $1.84_{-0.13}^{+0.14}$ & $5.01_{-1.15}^{+2.16}$   & $    13_{-     3}^{+     3}$ & $ 54.43_{-33.32}^{+ 61.35}$  & $ 16.02_{-  3.38}^{+  3.38}$\\
Si XIV   (H)     & $1.97_{-0.15}^{+0.13}$ & $5.01_{-1.15}^{+2.16}$   & $    27_{-     4}^{+     4}$ & $101.73_{- 58.79}^{+35.82}$  & $ 32.18_{-  4.42}^{+  4.44}$\\
2.2 keV Feature  & $2.23_{-0.01}^{+0.02}$ & $5.01_{-1.15}^{+2.16}$   & $    25_{-     7}^{+     7}$ & $ 75.19_{- 18.88}^{+ 18.67}$ & $ 26.82_{-  6.73}^{+  6.66}$\\
S XVI    (H)     & $2.60_{-0.01}^{+0.02}$ & $2.92_{-0.61}^{+0.96}$   & $    20_{-     5}^{+     5}$ & $ 47.15_{- 11.92}^{+ 12.08}$ & $ 19.68_{-  4.97}^{+  5.04}$\\
$^{\ast}$2.8 keV Feature  & $2.83_{-0.03}^{+0.03}$ & $2.92_{-0.61}^{+0.96}$   & $     7_{-     5}^{+     5}$ & $ 13.23_{-  9.30}^{+  9.53}$ & $  6.00_{-  4.21}^{+  4.32}$\\
Ar XVII  (He)    & $3.10_{-0.03}^{+0.03}$ & $5.90_{-0.51}^{+0.62}$   & $    12_{-     5}^{+     6}$ & $ 21.08_{-  8.92}^{+  8.86}$ & $ 10.45_{-  4.42}^{+  4.39}$\\
Ar XVIII (H)     & $3.26_{-0.04}^{+0.03}$ & $5.90_{-0.51}^{+0.62}$   & $    12_{-     5}^{+     5}$ & $ 18.61_{-  7.94}^{+  7.91}$ & $  9.72_{-  4.15}^{+  4.13}$\\
Ca XIX   (He)    & $3.82_{-0.07}^{+0.06}$ & $5.90_{-0.51}^{+0.62}$   & $    10_{-     5}^{+     6}$ & $ 11.40_{-  5.97}^{+  6.05}$ & $  6.97_{-  3.65}^{+  3.70}$\\
Ca XX    (H)     & $4.04_{-0.04}^{+0.04}$ & $5.90_{-0.51}^{+0.62}$   & $    10_{-     6}^{+     6}$ & $  9.32_{-  5.88}^{+  5.46}$ & $  6.03_{-  3.80}^{+  3.54}$\\
Fe XXV   (He)    & $6.59_{-0.01}^{+0.01}$ & $5.71_{-0.53}^{+0.61}$ & $   402_{-    27}^{+    27}$ & $157.71_{-  9.22}^{+  9.22}$ & $166.55_{-  9.73}^{+  9.73}$\\
Fe XXVI  (H)     & $6.81_{-0.03}^{+0.03}$ & $5.71_{-0.53}^{+0.61}$ & $    69_{-    14}^{+    14}$ & $ 37.14_{-  7.46}^{+  7.49}$ & $ 40.54_{-  8.15}^{+  8.17}$\\
7.8 keV Feature  & $7.74_{-0.06}^{+0.07}$ & $5.71_{-0.53}^{+0.61}$ & $    61_{-    35}^{+    37}$ & $ 13.55_{-  7.68}^{+  8.01}$ & $ 16.80_{-  9.52}^{+  9.93}$\\
\end{tabular}
\medskip

\raggedright

Results of fitting thermal bremsstrahlung and Gaussian models to {\sl
ASCA} SIS data of ellipticals and clusters having the largest
available S/N. Except for Fe XXV (He) the intrinsic widths of the
Gaussians are set to zero. Lines that have the same continuum
temperature, $T$, are cases where multiple gaussians are joined by one
bremsstrahlung model. See text for additional details regarding the
model fitting. Quoted errors are 90 per cent confidence on one
interesting parameter except for those lines with an asterisk for
which 68 per cent limits are quoted.

\end{table*}

\begin{figure*}
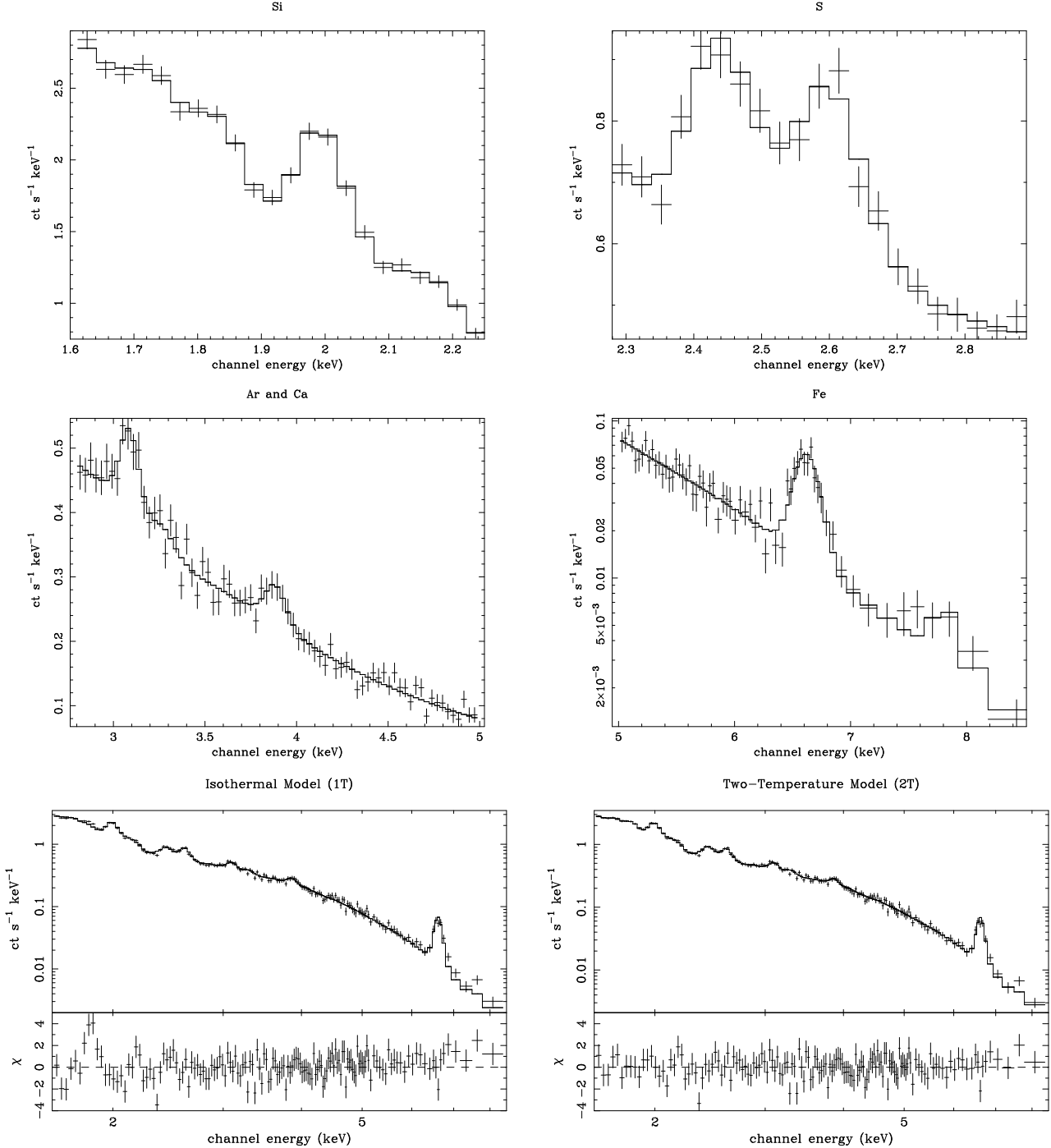

\parbox{0.49\textwidth}{
\centerline{\psfig{figure=fig12a.ps,angle=-90,height=0.25\textheight}}
}
\parbox{0.49\textwidth}{
\centerline{\psfig{figure=fig12b.ps,angle=-90,height=0.25\textheight}}
}
\vskip 0.25cm
\parbox{0.49\textwidth}{
\centerline{\psfig{figure=fig12c.ps,angle=-90,height=0.25\textheight}}
}
\parbox{0.49\textwidth}{
\centerline{\psfig{figure=fig12d.ps,angle=-90,height=0.25\textheight}}
}
\vskip 0.25cm
\parbox{0.49\textwidth}{
\centerline{\psfig{figure=fig12e.ps,angle=-90,height=0.25\textheight}}
}
\parbox{0.49\textwidth}{
\centerline{\psfig{figure=fig12f.ps,angle=-90,height=0.25\textheight}}
}
\caption{\label{fig.m87} {\sl ASCA} SIS spectral data for M87. The top
four panels show the energy ranges and best-fitting Gaussian+continuum
models used to obtain the line properties listed in Table
\ref{tab.lines}. See text in section \ref{id} for further explanation
of these models. The bottom two panels show the best-fitting
isothermal (left) and two-temperature (right) models fitted to the SIS
data over 1.6-9 keV as listed in Table \ref{tab.fits}. These broad-band
models are discussed in section \ref{broad}.}
\end{figure*}

\begin{figure*}
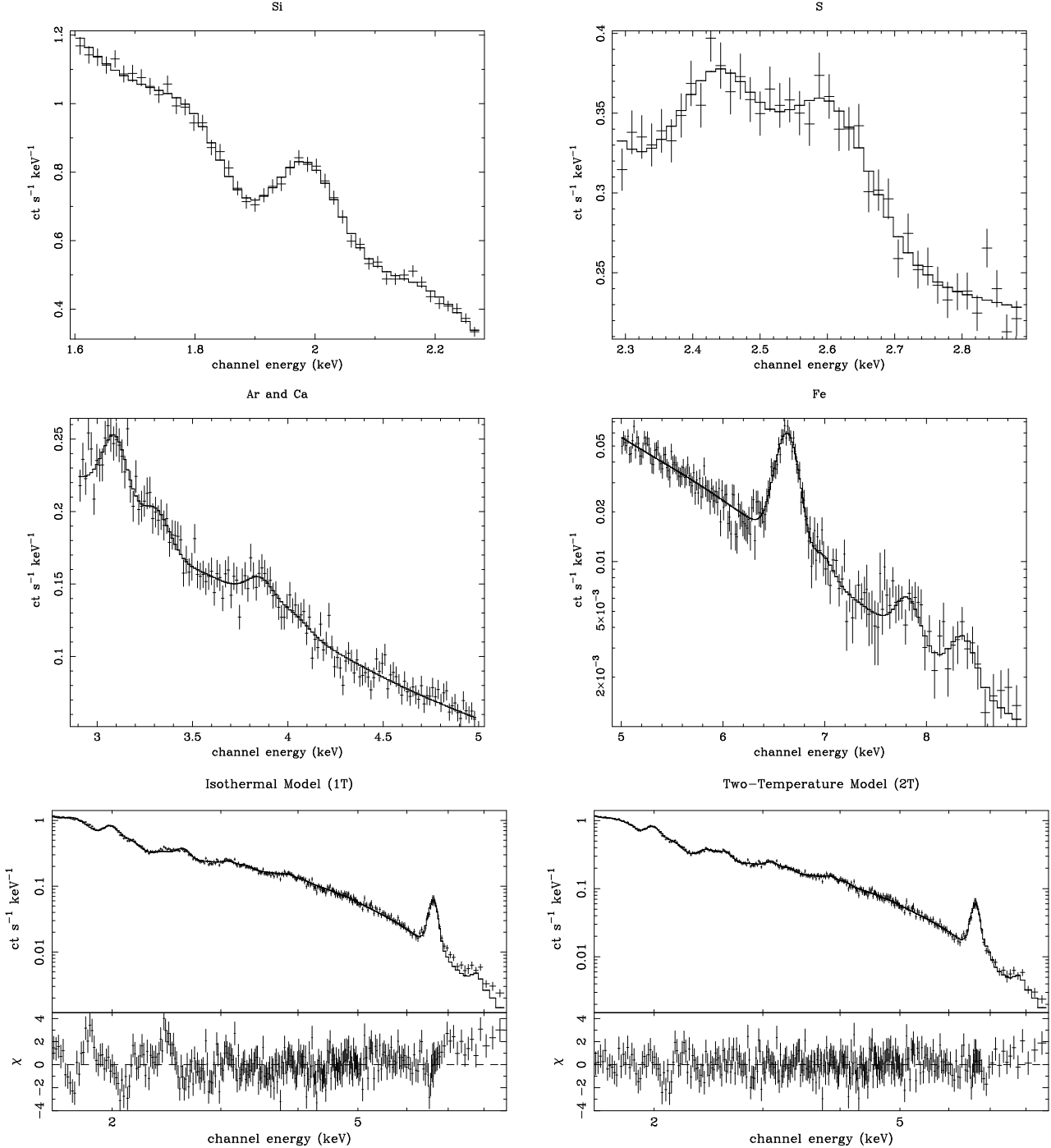

\parbox{0.49\textwidth}{
\centerline{\psfig{figure=fig13a.ps,angle=-90,height=0.25\textheight}}
}
\parbox{0.49\textwidth}{
\centerline{\psfig{figure=fig13b.ps,angle=-90,height=0.25\textheight}}
}
\vskip 0.25cm
\parbox{0.49\textwidth}{
\centerline{\psfig{figure=fig13c.ps,angle=-90,height=0.25\textheight}}
}
\parbox{0.49\textwidth}{
\centerline{\psfig{figure=fig13d.ps,angle=-90,height=0.25\textheight}}
}
\vskip 0.25cm
\parbox{0.49\textwidth}{
\centerline{\psfig{figure=fig13e.ps,angle=-90,height=0.25\textheight}}
}
\parbox{0.49\textwidth}{
\centerline{\psfig{figure=fig13f.ps,angle=-90,height=0.25\textheight}}
}
\caption{\label{fig.cen} As Figure \ref{fig.m87} but for Centaurus}
\end{figure*}

\begin{figure*}
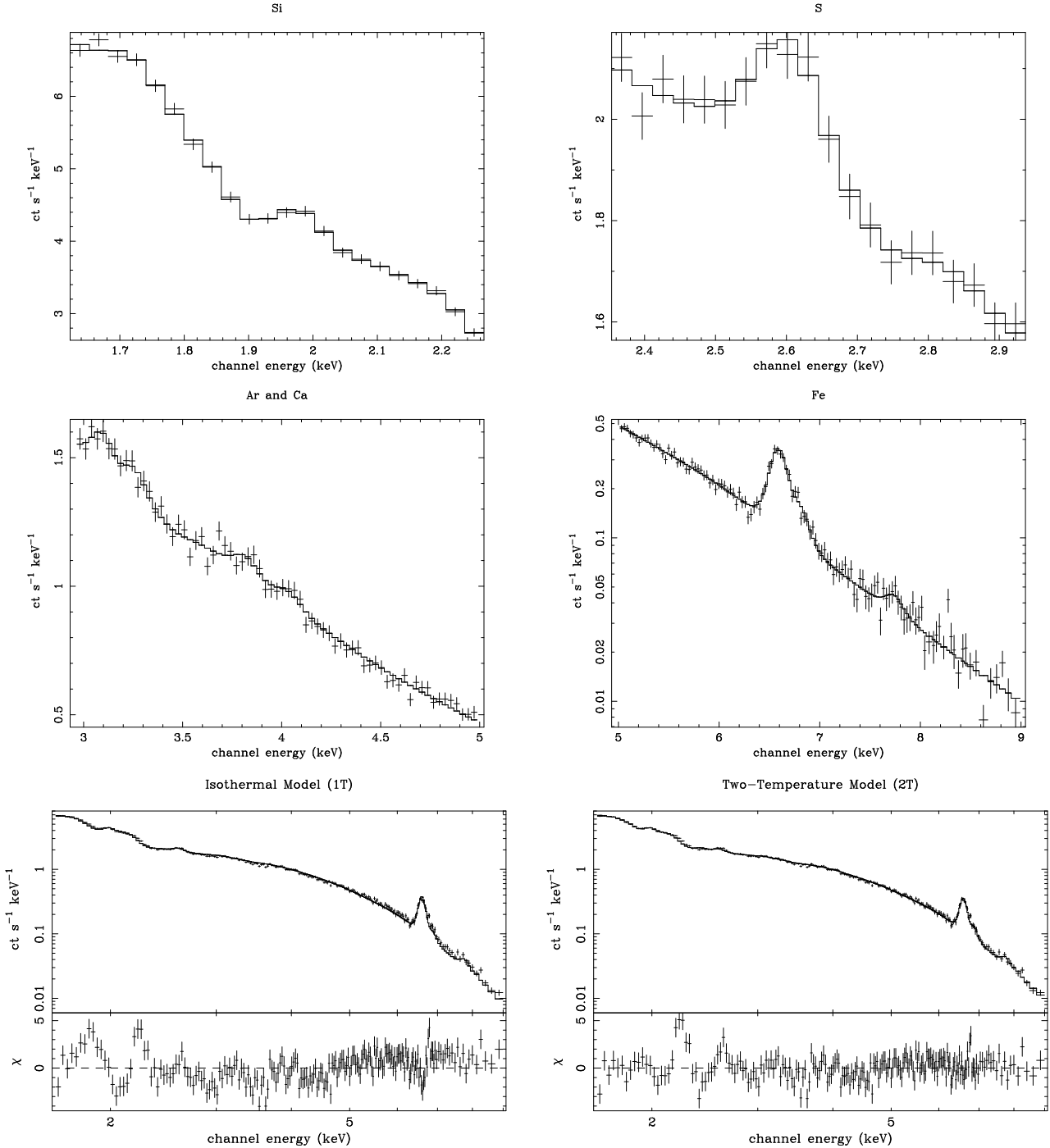

\parbox{0.49\textwidth}{
\centerline{\psfig{figure=fig14a.ps,angle=-90,height=0.25\textheight}}
}
\parbox{0.49\textwidth}{
\centerline{\psfig{figure=fig14b.ps,angle=-90,height=0.25\textheight}}
}
\vskip 0.25cm
\parbox{0.49\textwidth}{
\centerline{\psfig{figure=fig14c.ps,angle=-90,height=0.25\textheight}}
}
\parbox{0.49\textwidth}{
\centerline{\psfig{figure=fig14d.ps,angle=-90,height=0.25\textheight}}
}
\vskip 0.25cm
\parbox{0.49\textwidth}{
\centerline{\psfig{figure=fig14e.ps,angle=-90,height=0.25\textheight}}
}
\parbox{0.49\textwidth}{
\centerline{\psfig{figure=fig14f.ps,angle=-90,height=0.25\textheight}}
}
\caption{\label{fig.per} As Figure \ref{fig.m87} but for Perseus.}
\end{figure*}

We focus our analysis of individual line blends on the SIS data
because, unlike the GIS, the SIS is able to resolve both the He-like
and H-like blends of the K shell lines of the abundant elements
(e.g. see Figure 1 of Buote 1999a). The GIS data can provide useful
supplementary constraints on the continuum for energies above $\sim 5$
keV which we mention below.

Our procedure for obtaining the properties of the line blends
commences by fitting a thermal bremsstrahlung model to represent the
continuum emission. We desire a measurement of the local continuum
near the line blends of interest so that the assumption of a single
temperature is reasonable even if the spectrum actually consists of
multiple temperature components. However, because of S/N
considerations and the finite energy resolution of the SIS we allowed
the continuum component to extend over multiple line blends that are
nearby in energy. The line blends of interest were then modeled as
Gaussians of zero intrinsic width on top of the continuum component.

We fitted only line blends that are either obvious from visual
inspection of a spectrum or significantly improved the fit as judged
in terms of $\chi^2$. All spectral fitting was performed with the
software package {\sc XSPEC} \cite{xspec}.  In order for the weights
to be valid for the ${\chi^2}$ method we followed the standard
procedure and regrouped the PI bins for each source spectrum so that
each group had at least 20 counts. The background templates also
generally have at least 20 counts when their energy bins are grouped
similarly to the source spectra.  The background-subtracted count rate
in a particular group can be small, but the uncertainties in the
source and background are correctly propagated by {\sc XSPEC} to
guarantee approximately gaussian statistics for the statistical
weights (K. Arnaud 1998, private communication).

The energies of the lines were free parameters in the fits. When
determining the error bars on the equivalent widths and fluxes for a
particular line the energies of the other lines were fixed at their
best-fitting values. This was done to insure that the other lines did
not mix during the error search.

\subsubsection{Properties of K$\alpha$ and K$\beta$ lines}
\label{lineprop}

In Table \ref{tab.lines} we summarize the properties of the line
blends measured from the summed SIS data. The spectral regions and
best-fitting models used to measure the line properties are displayed
for each cluster in Figures \ref{fig.m87}-\ref{fig.per}. 

We detect most of the strong K$\alpha$ lines of the abundant elements
lying between the energies 1.6 and 9 keV in each cluster . In
Centaurus both the H- and He-like K$\alpha$ line blends of Si, S, Ar,
Ca, and Fe are detected.  These lines are also detected in the other
clusters except for S XV (He) in Perseus and for Ca XX (H) and Fe XXVI
(H) in M87. Mg XII (H) is also detected in each case but we advise
caution in interpreting the derived EWs and fluxes of this line
because the nearby unresolved Fe L lines undoubtedly affect our
estimates of the local continuum and the flux arising from only Mg.

In addition to the expected strong K$\alpha$ lines we detect other
significant lines in each cluster. We label these as ``Features'' in
Table \ref{tab.lines} because they tend to be weaker than the
K$\alpha$ lines and, more importantly, some of these line features may
result in part or entirely from systematic errors in the SIS
response. In particular, for each cluster we measure a significant
line feature near 2.2 keV which in principle could be attributed to
the 2.18 keV K$\beta$ line Si XIII (He3). However, this is precisely
the location of a known calibration problem owing to the optical
constants in the XRT \cite{gendreau} making it difficult to determine
reliably the origin of the 2.2 keV features.

A significant line feature is also detected near 1.7 keV in M87 and
Perseus. Assuming the feature is not instrumental then it is a blend
of K$\beta$ Mg (H), K$\alpha$ Al (H), and some Fe L emission (see
Figure \ref{fig.specs}). Although we are not aware of any problems
currently in the SIS response near 1.7 keV, we stress that these
features (and those at 2.2 keV) are not especially prominent (see
Figures \ref{fig.m87} and \ref{fig.per}) and their existence is thus
very dependent on the calibration.

In each cluster we also detect line features near 7.8 keV.  If these
features originate from hot plasma, they are primarily the result of
K$\beta$ emission from Fe XXV (He3) although some Ni K$\alpha$ should
contribute as well. This feature is most prominent in Centaurus though
it is clearly apparent to the eye in the SIS spectra of M87 (though it
is noisy) and is least noticeable for Perseus (see below). Visual
examination of the M87 spectrum (Figure \ref{fig.m87}) near 8 keV
suggests that the line feature is broader than the instrumental
resolution. A slightly better fit can be achieved either be adding
another narrow Gaussian or by allowing the width of the single
Gaussian to take a non-zero value ($\sigma\sim 200$ eV). Since the
parameters of these models are not well-constrained we prefer to
present the results for the single Gaussian of zero width. (Also the
other models imply continuum temperatures of $\sim 2.2$ keV which do
not follow the trend of increasing continuum temperature with energy.)

Although no specific calibration features at 7.8 keV have been
reported we do mention that there exist some calibration errors at
high energies (i.e. above $\sim 5$ keV) due to the uncertainty in the
optical axis, satellite wandering, and finite size of the Crab Nebula
which is the key calibration source (see Gendreau \& Yaqoob
1997). These warnings also apply to the line feature detected at 8.4
keV for Centaurus which, if of plasma origin, implies higher order K
shell transitions from Fe (see Figure \ref{fig.specs}). 

Since the effects of calibration impact the various detectors
differently, we have also measured the line properties for the SIS0
and SIS1 detectors separately. Generally we find excellent agreement
for the line properties between the SIS0 and SIS1. The line fluxes are
usually slightly larger for the SIS0, but the relative fluxes between
the lines are generally consistent within the estimated 90 per cent
errors for each detector. Although the SIS0 and SIS1 results are
generally consistent, some features are easier to see in the SIS0
data; e.g. 8.4 keV feature of Centaurus.

We have also verified the SIS results by simultaneously analyzing the
SIS and GIS data. Only for the Fe K emission is some tangible
improvement in the constraints offered by adding the GIS data.  The
key improvement is in the continuum temperature which is consistent
with the values determined from the SIS data but typically has about
half the statistical error. However, because the SIS and GIS responses
differ we must let the line energies (as well as normalizations) of
both the SIS and GIS be free parameters in the fits (i.e. only the
continuum temperature is tied between the detectors). The result is
that adding the GIS does not appreciably improve the SIS constraints
on the line fluxes.

Although the GIS also does not detect the Fe XXVI (H) line near 6.9
keV in Centaurus and Perseus because it is blended with the Fe XXV
complex, we do confirm the features near 8 keV in M87 and Centaurus
with the GIS. The GIS does not resolve the 7.8 and 8.4 keV features in
Centaurus, but the parameters of two gaussian features obtained at
those energies with the GIS agrees with the SIS results.  Similar to
the SIS the 7.8 keV ``feature'' in Perseus is not obvious from visual
examination of the GIS spectrum, though similar constraints on the
feature are achieved with the GIS. The principal reasons why we
present results for a 7.8 keV line in Perseus are (1) because such a
feature is required in M87 and Centaurus, and (2) for comparison to
previous work (see below).

The results for the F XXV (He) line listed in Table \ref{tab.lines}
reflect a non-zero width on the Gaussian component. We obtained 90 per
cent confidence $\sigma$ values of $48.20_{-10.86}^{+33.16}$ eV for
M87, $17.64_{-17.64}^{+21.52}$ eV for Centaurus, and
$26.27_{-26.27}^{+18.33}$ eV for Perseus; i.e. $\sigma$ is consistent
with zero for Centaurus and Perseus. When analyzed separately we found
that $\sigma\sim 60(30)$ eV for the SIS0 (SIS1) data for the Fe line
of M87. Similarly, $\sigma\sim 24(7)$ eV for the SIS0 (SIS1) data of
Perseus. For Centaurus the best-fitting $\sigma\sim 0$ eV for each
SIS. The significant $\sigma$ obtained for M87 may be due to finite
intrinsic width of the Fe K complex, errors in the response matrix, or
(most likely) a contribution to the emission from the (undetected) Fe
XXVI (H) line.

Our results represent the first detections of many of the line blends
in these clusters. The most complete catalogue of X-ray detections was
previously reported by Matsumoto et al. \shortcite{mat_m87} who used
{\sl ASCA} SIS data to constrain the K$\alpha$ lines of H- and He-like
Si and S as well as Fe XXV (He) for M87. Fukazawa et
al. \shortcite{fukazawa} presented equivalent widths for the Si XIV
(H), S XV (He), S XVI (H), and Fe XXV (He) lines which are consistent
with our results (Table \ref{tab.lines}).  Another notable paper is
that of Molendi et al. \shortcite{molendi} who presented results for
both the K$\alpha$ and K$\beta$ Fe XXV emission lines in the center of
Perseus using data from the {\sl SAX} satellite \cite{sax}. Therefore
we have presented the first constraints on (1) individual Ar and Ca
lines in each of these systems, (2) the H-like Fe XXVI lines in
Centaurus and Perseus, and (3) Fe K$\beta$ emission in M87 and
Centaurus. Our results for Mg XII (H) are also novel though our
reported Mg flux is uncertain due to contamination from Fe L emission
as noted above. However, it is reassuring that the energies of the
fitted Mg lines lie exactly where Mg XII should be.

\subsubsection{Temperatures from line ratios and local continuum}
\label{ratio}

\begin{table*}
\caption{Temperatures Implied by {\sl ASCA} Line Ratios}
\label{tab.ratios}
\begin{tabular}{lcccccc}
Cluster & Si & S & Ar & Ca & Fe & Fe K$\beta$/K$\alpha$\\
M87 & $1.7_{-0.3}^{+0.3}$ & $2.1_{-0.2}^{+0.2}$ &
$\cdots$ & $\cdots$ & $\cdots$ & $9.3_{-7.9}^{+5.8}$\\
Centaurus & $2.4_{-0.3}^{+0.3}$ & $2.2_{-0.2}^{+0.3}$ &
$2.4_{-1.}^{+0.6}$ &  $3.1_{-1.1}^{+0.7}$ & $4.0_{-0.8}^{+0.4}$ & $6.9_{-4.7}^{+3.4}$\\
Perseus & $1.9_{-0.2}^{+0.4}$ & $\cdots$ & $3.3_{-0.9}^{+0.9}$ &
$4.2_{-1.5}^{+1.2}$ & $5.9_{-0.4}^{+0.5}$ & $2.8_{-1.6}^{+5.2}$\\
\end{tabular}
\medskip
\raggedright

Isothermal temperatures in units of keV corresponding to H/He flux
ratios of the K$\alpha$ lines listed in Table \ref{tab.lines}; also
displayed are the Fe K$\beta$/K$\alpha$ ratios where Fe K$\beta$
refers to the 7.8 keV feature and Fe K$\alpha$ refers to the strong Fe
XXV (He) line. For each line ratio we computed the (MEKAL) temperature
appropriate for the resolution of the {\sl ASCA} SIS as described in
Figure \ref{fig.ratios.asca}. The error bars reflect 90 per cent
confidence limits.
\end{table*}

In Table \ref{tab.ratios} we list the temperatures implied by the H/He
ratios of K$\alpha$ lines of the same element for each element where
90 per cent confidence limits on the fluxes are reported in Table
\ref{tab.lines}. Recall that for isothermal gas these ratios are very
temperature sensitive (Figure \ref{fig.ratios}) and are independent of
the elemental abundances. At the end of this section we shall also
consider the temperatures suggested by the Fe K$\beta$ lines.  Since
the temperatures listed in Table \ref{tab.ratios} are not obtained by
fitting plasma models directly to the data they should be regarded as
estimates which are sufficient for the qualitative discussion we
present in this section. A quantitative assessment of the
goodness-of-fit of different spectral models is given below in section
\ref{broad}.

For M87 the plasma temperature indicated by the Si ratio is $\sim 1.7$
keV while the S ratio gives $T\sim 2.1$ keV. If the gas is isothermal
then the temperature should be the same for the Si and S
ratios. Instead the increase in temperature from the Si lines near
$\sim 2.0$ keV to the S lines near $\sim 2.5$ keV indicates that the
emission measure consists of at least two components where the cooler
component dominates the energy spectrum for $E\la 2$ keV while the
hotter component dominates for higher energies. These results are also
consistent with emission from a continuous range of temperatures where
again the cooler components contribute most significantly to the Si
ratios and the hotter components dominate the S ratio; note that for
the integrated {\sl ASCA} spectrum within the circular aperture a
continuous range of temperatures would be consistent with, e.g., the
sum of different temperature components at different radii for a
single-phase gas, or the sum of different phases in a multiphase
cooling flow.

The variation in local continuum temperature as a function of energy
(Table \ref{tab.lines}) gives further support for multiple temperature
components in the {\sl ASCA} spectrum of M87; i.e. the continuum
temperature varies from $\sim 1.2$ keV near Si to $\sim 2.6$ keV near
Fe XXV.  This variation in continuum temperature does not arise from
an increasing proportion of bound-free emission since free-free
emission dominates except near the strong Fe L lines near 1 keV (and
for lower energies). Although the local continuum temperature is
consistent with the S ratio, it is somewhat lower than the Si
ratio. This difference for Si could be due to an inaccurate continuum
estimate arising from unresolved lines or from other systematics in
the method we use to infer continuum temperature and line fluxes
(section \ref{id}). However, for a multiphase plasma it is not
necessary that the local continuum temperature agree with the
temperature inferred from a line ratio assuming an isothermal gas.  We
also remark that the continuum temperature can be affected by
non-thermal components whereas the line ratios are determined only by
the hot plasma.

Our results are not inconsistent with those obtained by Matsumoto et
al. \shortcite{mat_m87} but a direct comparison of our results is not
possible because they did not use a local continuum when computing
their line properties. These authors also examined the spectra within
annuli of $2\arcmin$ width which may have been too small since the
{\sl ASCA} PSF is $3\arcmin$ in half-power diameter. Nevertheless,
within our $\sim 6\arcmin$ aperture Matsumoto et al. also found
evidence for non-isothermal gas. They estimated that the continuum
temperature increased from 2.1 to 2.3 keV within $\sim 6\arcmin$ which
is similar to the temperature difference we infer from the Si and S
line ratios.

The evidence for multiple temperature components is even stronger for
Centaurus. The K$\alpha$ line ratios imply temperatures of $\sim 2.4$
keV over energies $\sim 2$-3 keV, $T\sim 3$ keV near 4 keV, and $T\sim
4$ keV near 7 keV. The continuum temperature increases from $T\sim
1.6$ keV near 2 keV to $T\sim 4$ keV near 7 keV. The good agreement
between the temperatures implied by the Fe line ratio and the
continuum near 7 keV indicates that a high-energy non-thermal
component (e.g. AGN, discrete sources) does not dominate the continuum
of Centaurus for energies $\sim 5$-9 keV.

Hence, the variation of continuum temperature with energy should
reflect a related variation in the plasma temperature, and when
considering these temperatures and those inferred from the line ratios
we conclude that the SIS spectrum of Centaurus clearly requires
multiple temperature components for the hot gas within the {\sl ASCA}
aperture. (As with M87 the slight differences between continuum and
line ratio temperatures are expected for a general spectrum with
multiple temperature component.) A variation in temperature between
$T\sim 2$-4 keV is consistent with the two-temperature models fitted
to the broad-band {\sl ASCA} spectra by Fukazawa et
al. \shortcite{fukazawa}.

The K$\alpha$ line ratios for Perseus give perhaps the strongest
evidence for multitemperature hot gas of the three clusters in our
sample. The $\sim 2$ keV temperature inferred from the Si ratio is
significantly less than $T\sim 6$ keV implied by the Fe ratio. Similar
to Centaurus the excellent agreement between the temperatures of the
Fe ratio and the continuum near 7 keV argues against a strong
contribution from a high-energy non-thermal component typical of,
e.g. an AGN. In contrast to Centaurus the local continuum temperatures
of Perseus are nearly constant with $T\sim 5$-6 keV over approximately
2-8 keV; the small glitch near S likely reflects a miscalculation of
the continuum due to our inability to compensate for the undetected
He-like S XV line and the calibration error near 2.2 keV due to the
optical constants in the XRT \cite{gendreau}.

This near-isothermality of the continuum is not inconsistent with the
temperature variations implied by the K$\alpha$ line ratios in
Perseus. If the hottest temperature components near 6 keV dominate the
continuum it is still possible for weaker cooler components to
contribute to the line emission at lower energies. For example, in
Figure \ref{fig.sifek} we show that the H1 Si XIV line is $\sim 5$
times stronger for $T=2$ keV than for $T=6$ keV for an isothermal gas;
i.e. a cooler component can give a significant contribution to the
line emission even if the continuum is dominated by hotter
components. Moreover, the cooler component contributes negligibly to
the the Fe K$\alpha$ ratio since the H1 Fe XXVI line is over 100 times
stronger for $T=6$ keV than for $T=2$ keV (Figure
\ref{fig.sifek}). Thus, the K$\alpha$ line ratios provide the key
evidence for multitemperature gas in Perseus within the {\sl ASCA}
aperture.

Finally, the ratios of K$\beta$/K$\alpha$ lines also probe the
temperature of the gas (Figure \ref{fig.ratios}) but since our
estimates of the fluxes of these lines may be dominated by systematic
errors (see section \ref{lineprop}) we advise extra caution when
interpreting K$\beta$/K$\alpha$ ratios. In particular, since the 2.2
keV features are probably dominated by systematic errors
\cite{gendreau}, and the ratios of the fluxes of these features to
those of the Si XIII (He) lines generally imply much larger
temperatures than the other lines, we do not discuss these lines
further. We also do not discuss the 1.7 keV features since (if of
plasma origin) they are blends of lines of three different elements.

The remaining candidate K$\beta$ lines are the 7.8 keV features. As
indicated in Table \ref{tab.ratios} the Fe K$\beta$/K$\alpha$ ratios
are not very well constrained and in each case give temperatures
consistent with the (H/He) Fe K$\alpha$ ratio and/or the continuum
temperature near 7 keV; note that the ratio ``Fe K$\beta$/K$\alpha$''
means the ratio of the flux of the 7.8 keV feature to that of the
He-like Fe XXV line. For Centaurus the ratio of the flux of the 8.4
keV feature to that of the He-like Fe XXV line implies $T\sim 9$ keV
for the best-fitting line ratio, but the 90 per cent confidence lower
limit is $T\sim 1.3$ keV.

Hence, these Fe K$\beta$/K$\alpha$ ratios do not require anomalously
high temperatures at the 90 per cent confidence level and thus do not
implicate resonance scattering as suggested by Molendi et
al. \shortcite{molendi} for Perseus from analysis of {\sl SAX}
data. Molendi et al. obtained an Fe K$\beta$/K$\alpha$ ratio of $\sim
0.20$ within $r=6.4\arcmin$ which is about twice our value of $\sim
0.10$ within $r=5.6\arcmin$. However, the energy resolution of {\sl
SAX} is poorer than the {\sl ASCA} SIS and thus to make a fair
comparison we need to add the H-like Fe XXVI flux to the Fe K$\alpha$
emission; this reduces our Fe K$\beta$/K$\alpha$ ratio to $\sim 0.08$
corresponding to $T\sim 1.7$ keV. Although there may be systematic
errors in the {\sl ASCA} data at these high energies as we discussed
in section \ref{lineprop} (see Gendreau \& Yaqoob 1997), the agreement
we have have obtained between the continuum temperatures and the Fe
K$\alpha$ ratio for Perseus argues against a large systematic
error. Therefore, we conclude that the Fe K$\beta$/K$\alpha$ ratio in
Perseus deduced from {\sl ASCA} is consistent with optically thin
plasma and that the large ratio found by Molendi et
al. \shortcite{molendi} with {\sl SAX} must be due to an unrecognized
systematic error.

We mention that the Fe K$\beta$/K$\alpha$ results we have quoted
assume a photospheric solar Ni/Fe ratio. If instead we use the
meteoritic solar ratios (see section \ref{broad}) we find that the
temperatures inferred for the Fe K$\beta$/K$\alpha$ ratios in Table
\ref{tab.ratios} are reduced by a modest amount ($\approx 25$ per
cent). To raise the inferred temperatures it is necessary to reduce
the Ni/Fe ratio. Because the uncertainties are large we find that the
temperatures implied by the 7.8 keV/Fe K$\alpha$ ratios are consistent
with the local continuum temperatures even for $Z_{\rm Ni}/Z_{\rm Fe}$
ratios as small as $\approx 1/5$ (meteoritic solar).

\subsection{Broad-band spectral fitting}
\label{broad}

\begin{table*}
\caption{Broad-Band {\sl ASCA} Spectral Fits}
\label{tab.fits}
\begin{tabular}{ccccccccccc}

$T_{\rm c}$ & $T_{\rm h}$ & $EM_{\rm c}$ & $EM_{\rm h}$ & PL & Fe &
Mg & Si & S & Ar & Ca\\ 
(keV) & (keV) & \multicolumn{2}{c}{(see notes)} & (see notes)\\ \\[-2 pt]
\multicolumn{10}{l}{M87:}\\ \\[-2 pt]
\multicolumn{10}{l}{1T: (193.5/142/2.7e-3)}\\ 
$2.16_{-0.04}^{+0.04}$ & $\cdots$ & $1.56_{-0.07}^{+0.07}$  & $\cdots$ & $\cdots$ &
$0.85_{-0.13}^{+0.14}$ & $\cdots$ & $1.12_{-0.10}^{+0.11}$ &
$0.87_{-0.10}^{+0.10}$ & $0.51_{-0.17}^{+0.17}$ &
$0.81_{-0.30}^{+0.31}$\\ \\[-5 pt]
\multicolumn{10}{l}{1T+PL: (156.6/141/0.18)}\\
$1.92_{-0.07}^{+0.07}$ & $\cdots$ & $1.37_{-0.10}^{+0.10}$ & $\cdots$
& $0.39_{-0.12}^{+0.11}$ &
$1.01_{-0.18}^{+0.21}$ & $\cdots$ & $1.14_{-0.13}^{+0.14}$ &
$0.94_{-0.11}^{+0.13}$ & $0.68_{-0.19}^{+0.21}$ &
$1.12_{-0.37}^{+0.39}$\\ \\[-5 pt]
\multicolumn{10}{l}{2T: (130.3/140/0.71)}\\
$1.06_{-0.42}^{+0.23}$ & $2.48_{-0.14}^{+0.23}$ &
$0.36_{-0.12}^{+0.22}$ & $1.25_{-0.21}^{+0.16}$ & $\cdots$ &
$0.78_{-0.13}^{+0.13}$ & $\cdots$ & $1.09_{-0.12}^{+0.15}$ &
$1.01_{-0.11}^{+0.12}$ & $0.68_{-0.20}^{+0.21}$ &
$0.90_{-0.33}^{+0.34}$\\ \\[-5 pt]
\multicolumn{10}{l}{2T+PL: (125.8/139/0.78)}\\
$0.85_{-0.35}^{+0.31}$ & $2.21_{-0.14}^{+0.28}$ &
$0.22_{-0.06}^{+0.20}$ & $1.29_{-0.18}^{+0.11}$ &
$0.20_{-0.19}^{+0.15}$ &
$0.87_{-0.15}^{+0.18}$ & $\cdots$ & $1.16_{-0.14}^{+0.16}$ &
$1.04_{-0.12}^{+0.13}$ & $0.72_{-0.20}^{+0.21}$ &
$0.99_{-0.35}^{+0.37}$\\ \\[-5 pt]
\multicolumn{10}{l}{Constant $\xi$: (131.9/141/0.70)}\\
$2.19_{-0.05}^{+0.05}$ & $2.54_{-0.26}^{+0.24}$ &
$1.56_{-0.07}^{+0.07}$ & $\cdots$ & 
$\cdots$ & $0.75_{-0.12}^{+0.13}$ & $\cdots$ & $1.04_{-0.10}^{+0.11}$
& $0.96_{-0.11}^{+0.11}$ & $0.72_{-0.20}^{+0.21}$ & $0.97_{-0.35}^{+0.36}$\\[+20pt]
\multicolumn{10}{l}{Centaurus:}\\ \\[-2 pt]
\multicolumn{10}{l}{1T: (489.1/301/3.9e-11)}\\ 
$3.23_{-0.05}^{+0.05}$ & $\cdots$ & $0.60_{-0.02}^{+0.02}$ & $\cdots$ & $\cdots$ &
$1.28_{-0.10}^{+0.09}$ & $6.25_{-1.70}^{+1.71}$ &
$1.76_{-0.14}^{+0.13}$ & $1.28_{-0.13}^{+0.13}$ &
$0.68_{-0.21}^{+0.21}$ & $0.74_{-0.32}^{+0.30}$\\ \\[-5 pt]
\multicolumn{10}{l}{1T+PL: (373.2/300/2.6e-3)}\\
$2.49_{-0.10}^{+0.10}$ & $\cdots$ & $0.39_{-0.04}^{+0.04}$ & $\cdots$
& $0.49_{-0.07}^{+0.06}$ & $2.56_{-0.36}^{+0.44}$ & $1.75_{-1.75}^{+2.09}$ &
$2.16_{-0.22}^{+0.27}$ & $1.73_{-0.21}^{+0.24}$ &
$1.36_{-0.27}^{+0.33}$ & $1.50_{-0.44}^{+0.48}$\\ \\[-5 pt] 
\multicolumn{10}{l}{2T: (295.0/300/0.57)}\\
$1.59_{-0.12}^{+0.12}$ & $4.80_{-0.35}^{+0.39}$ & $0.26_{-0.04}^{+0.04}$ & $0.39_{-0.04}^{+0.04}$ & $\cdots$ &
$1.24_{-0.08}^{+0.09}$ & $\cdots$ & $1.31_{-0.09}^{+0.09}$ & $1.23_{-0.10}^{+0.11}$ & $1.24_{-0.23}^{+0.24}$ & $1.47_{-0.39}^{+0.39}$\\ \\[-5 pt]
\multicolumn{10}{l}{2T+PL: (286.8/299/0.68)}\\
$1.56_{-0.17}^{+0.17}$ & $4.09_{-0.56}^{+0.64}$ & $0.21_{-0.07}^{+0.06}$ & $0.33_{-0.06}^{+0.06}$ & $0.22_{-0.12}^{+0.10}$ &
$1.55_{-0.21}^{+0.26}$ & $\cdots$ & $1.54_{-0.09}^{+0.20}$ & $1.46_{-0.19}^{+0.22}$ & $1.45_{-0.28}^{+0.30}$ & $1.62_{-0.42}^{+0.45}$\\ \\[-5 pt]
\multicolumn{10}{l}{Constant $\xi$: (303.2/301/0.45)}\\
$3.46_{-0.06}^{+0.06}$ & $5.02_{-0.24}^{+0.24}$ & $0.62_{-0.01}^{+0.01}$ & $\cdots$ & $\cdots$ &
$1.26_{-0.08}^{+0.08}$ & $\cdots$ & $1.39_{-0.09}^{+0.10}$ & $1.34_{-0.11}^{+0.11}$ & $1.14_{-0.22}^{+0.22}$ & $1.27_{-0.34}^{+0.34}$\\ \\[-5 pt]
\multicolumn{10}{l}{Perseus:}\\ \\[-2 pt]
\multicolumn{10}{l}{1T: (367.8/167/3.7e-17)}\\ 
$4.62_{-0.06}^{+0.07}$ & $\cdots$ & $4.77_{-0.05}^{+0.05}$ & $\cdots$ & $\cdots$ &
$0.62_{-0.03}^{+0.03}$ & $\cdots$ & $1.03_{-0.15}^{+0.15}$ & $\cdots$ & $0.00_{-0.00}^{+0.07}$ & $0.00_{-0.00}^{+0.21}$\\ \\[-5 pt] 
\multicolumn{10}{l}{1T+PL: (244.2/166/7.5e-5)}\\
$3.21_{-0.17}^{+0.18}$ & $\cdots$ & $2.59_{-0.27}^{+0.27}$ & $\cdots$ & $5.38_{-0.55}^{+0.51}$ & 
$1.36_{-0.20}^{+0.25}$ & $\cdots$ & $1.44_{-0.21}^{+0.23}$ & $\cdots$ & $0.35_{-0.26}^{+0.28}$ & $0.59_{-0.37}^{+0.39}$\\ \\[-5 pt] 
\multicolumn{10}{l}{2T: (176.4/165/0.26)}\\
$1.27_{-0.12}^{+0.13}$ & $6.12_{-0.26}^{+0.29}$ & $1.32_{-0.17}^{+0.15}$ & $3.74_{-0.15}^{+0.15}$ & $\cdots$ &
$0.70_{-0.04}^{+0.04}$ & $\cdots$ & $0.64_{-0.10}^{+0.11}$ & $\cdots$ & $0.51_{-0.25}^{+0.26}$ & $0.70_{-0.37}^{+0.38}$\\ \\[-5 pt]
\multicolumn{10}{l}{2T+PL: (168.8/164/0.38)}\\
$1.30_{-0.13}^{+0.15}$ & $5.59_{-0.47}^{+0.43}$ & $1.13_{-0.21}^{+0.19}$ & $3.05_{-0.42}^{+0.42}$ & $2.00_{-1.13}^{+1.09}$ &
$0.83_{-0.09}^{+0.12}$ & $\cdots$ & $0.77_{-0.14}^{+0.15}$ & $\cdots$ & $0.68_{-0.29}^{+0.31}$ & $0.89_{-0.43}^{+0.46}$\\ \\[-5 pt]
\multicolumn{10}{l}{Constant $\xi$: (183.8/166/0.16)}\\
4.94 & $9.03_{-0.40}^{+0.36}$ & $4.91_{-0.08}^{+0.08}$ & $\cdots$ & $\cdots$ &
$0.76_{-0.04}^{+0.04}$ & $\cdots$ & $0.79_{-0.10}^{+0.10}$ & $\cdots$ & $0.37_{-0.21}^{+0.21}$ & $0.55_{-0.32}^{+0.32}$\\ \\[-5 pt]
\end{tabular}

\medskip

\raggedright

Best-fitting and 90 per cent confidence limits on one interesting
parameter ($\Delta\chi^2=2.71$) for models fitted to the summed SIS
data between 1.6-9 keV. The models are defined similarly to those in
Table \ref{tab.models} with the following exceptions: (1) the values
of $\chi^2$, dof, and the null hypothesis probability are listed in
parentheses next to the model, (2) the emission measures $(\rm EM)$
are quoted in units of $10^{-15}n_en_pV/4\pi D^2$ similar to what is
done in {\sc XSPEC}, (3) PL indicates the normalization of the
power-law model (photon index 1.7) at 1 keV in units of $10^{-2}$
photons cm$^{-2}$ s$^{-1}$ keV$^{-1}$, and (4) the abundances are
quoted in terms of the meteoritic solar values of Feldman
\shortcite{feld} and where ``$\cdots$'' indicates that the abundance
of that element is tied to Fe in its meteoritic solar ratio. For
models with multiple temperature components the abundances of each
component are the same. Finally, the ``Constant $\xi$'' model
represents a plasma emitting over a range of temperatures having equal
emission measure over equal temperature intervals; for this model
$T_{\rm c}$ represents the temperature of the midpoint of the range
while $T_{\rm h}$ indicates the length of the temperature interval.
\end{table*}

The line ratios provide strong evidence for multiple temperature
components within the {\sl ASCA} apertures in each cluster. To
quantify more precisely the evidence for non-isothermal gas we must
fit plasma models convolved with the detector response directly to the
{\sl ASCA} spectra. As we have shown in section \ref{2tcf} it is vital
to consider as many line ratios as possible to distinguish between
simple two-temperature and cooling flow models. Hence, in this section
we examine models fitted over 1.6-9 keV which includes the strong Si,
S, Ar, Ca, and Fe K shell transitions.

We do not analyze energies below 1.6 keV since the emission near 1 keV
is dominated by Fe L emission, and intrinsic absorbing material needs
to be considered for lower energies (e.g. Fabian et al. 1994a). The
latter point is especially relevant for our analysis because this
allows us to assume (to a good approximation) that the same Galactic
column density (Table \ref{tab.obs}) modifies each spectral component;
i.e. this simplification limits the number of free parameters for
models that emit over a continuous range of temperatures. We account
for absorption by our Galaxy using the photo-electric absorption cross
sections according to Baluci\'{n}ska-Church \& McCammon
\shortcite{phabs}.

Since it is our purpose to distinguish between rival spectral models
we desire to optimize S/N by rebinning the {\sl ASCA} spectra above the
minimum 20 counts (see beginning of section \ref{id}) so that the
energy resolution of the detectors are not greatly over-sampled. For
the SIS data we settled on minimums of 50 counts for M87 and 100
counts for Centaurus and Perseus. Since the GIS resolution is poorer
than for the SIS we chose larger minimums: 100 counts for M87 and 150
counts for Centaurus and Perseus. The effects of using these larger PI
bins are actually small, and the results we present below are
consistent with the results obtained from using the standard 20-count
binning employed in the previous section.

We again focus on the SIS data to provide a consistent comparison with
the results of the previous section, and because they provide the best
constraints on the line properties. However, we discuss the additional
constraints provided by joint fitting of the SIS and GIS data below in
section \ref{pl}.

Unlike most previous studies we do not use the photospheric solar
abundances of Anders \& Grevesse \shortcite{ag}, but instead we use
the meteoritic solar abundances of Feldman \shortcite{feld} which are
more appropriate \cite{im}. The principal difference is that the
meteoritic Fe abundance (Fe/H of $3.24\times 10^{-5}$ by number) is
1/1.44 times the photospheric value. Below in section \ref{abun} we
discuss how this choice affects our results in comparison to some
previous studies.

\subsubsection{Constraints on $\xi(T)$}
\label{xi}

In Table \ref{tab.fits} we list the results of fitting isothermal
models (1T), two-temperature models (2T), and models with a constant
differential emission measure, $\xi(T)=$~constant, to the summed SIS
data. The best-fitting 1T and 2T models are displayed at the bottom of
Figures \ref{fig.m87} - \ref{fig.per}. The spike in the residuals of
both the 1T and 2T models of Perseus (Figure \ref{fig.per}) coincides
with the 2.2 keV feature discussed in section \ref{lineprop}. We find
that this feature cannot be removed by making simple alterations to
$\xi(T)$ which is consistent with the feature being dominated by the
systematic error at 2.2 keV arising from the optical constants in the
XRT \cite{gendreau}. Hence, the results listed in Table \ref{tab.fits}
for Perseus exclude the energies 2.-2.8 keV where this feature
dominates. 

From examination of the residuals in Figures \ref{fig.m87} -
\ref{fig.per} and the $\chi^2$ null hypothesis probabilities $(P)$ in
Table \ref{tab.fits} it is clear that 1T models give poor fits $(P\ll
1)$ to the SIS spectra of each cluster. The most prominent residuals
occur near the Si and Fe lines. We find that $P$ decreases from M87 to
Centaurus and then to Perseus which agrees with our conclusion reached
in section \ref{ratio} using line ratios that an isothermal gas is
most strongly excluded for Perseus and then Centaurus. The 1T models
for each cluster require $\alpha$/Fe ratios different from solar to
obtain a best fit.

The 2T models provide superior fits that are formally acceptable
($P>0.1$) for each cluster. Examination of Figures \ref{fig.m87} -
\ref{fig.per} reveals no large residuals for these models though some
excess emission at higher energies may be indicated which we discuss
below in section \ref{pl}. Unlike M87 the 2T models of Centaurus and
Perseus are not improved significantly when allowing elements
different from Fe to be varied separately: the 2T model having the
abundances of the elements tied to Fe in their (meteoritic) solar
ratios gives $\chi^2$/dof/$P$ values of 298.2/304/0.58 and
179.0/168/0.27 respectively for Centaurus and Perseus. (Note S is tied
to Fe for Perseus since we exclude the energies 2-2.8 keV from the
fits.)

For comparison to the 2T model we examine the simplest model of a gas
emitting over a range of temperatures with constant differential
emission measure, $\xi(T)=$~constant. This model provides fits that
are virtually identical to the 2T models for M87 and Centaurus, and
for Perseus the constant-$\xi$ model fits almost as well as the 2T
model. Again the fits are not improved significantly for Centaurus and
Perseus when allowing multiple abundances to be varied separately: we
obtain 310.1/305/0.41 and 195.4/169/0.08 respectively for Centaurus
and Perseus for the constant-$\xi$ model where only the total
metallicity is varied. We emphasize that these constant-$\xi$ models
have qualitatively different emission measure distributions from the
2T models. That is, for lower and upper temperatures denoted by
$T_{\rm low}$ and $T_{\rm hi}$, the constant-$\xi$ models indicate
that the hot gas emits at constant emission measure over the following
temperature ranges: $T_{\rm low}\approx 0.92$ keV and $T_{\rm
hi}\approx 3.46$ keV for M87; $T_{\rm low}\approx 0.92$ keV and
$T_{\rm hi}\approx 5.9$ keV for Centaurus; $T_{\rm low}\approx 0.43$
keV and $T_{\rm hi}\approx 9.4$ keV for Perseus.

The strong similarity between the fits of the 2T and constant-$\xi$
models highlights the difficulty in constraining the shape of a
general $\xi(T)$ with {\sl ASCA} data. Although we have not presented
results for specific multiphase cooling flow models (sections
\ref{cfs} and \ref{2tcf}) because of computational expense, the
similarity between the fits of the 2T and constant-$\xi$ models is
sufficient to demonstrate that multiphase cooling flows also cannot be
distinguished from the 2T models; i.e. multiphase cooling flows have
$\sigma_\xi$ intermediate between the 2T and constant-$\xi$
models. Therefore, the fits to the SIS data over 1.6-9 keV for each
cluster clearly rule out isothermal gas but cannot determine whether
the gas emits at only two temperatures or over a continuous range of
temperatures as expected for a multiphase cooling flow.

\subsubsection{Abundances}
\label{abun}

The Fe abundances obtained from isothermal models are very similar to
those obtained from the multitemperature models, whereas several of
the $\alpha$/Fe ratios differ from solar for each 1T model but not for
the multitemperature models of Centaurus and Perseus. (A possible
exception being the sub-solar Ar/Fe ratio for the constant-$\xi$ model
of Perseus.) The fact that the Si/Fe ratios exceed solar for 1T models
in contrast to the 2T and constant-$\xi$ models is another
demonstration of biased measurements of abundances which occur from
fitting a multitemperature spectrum with an isothermal model (e.g. see
Appendix of Buote 1999b).

If instead the photospheric solar abundances are used as done in most
previous studies, then $\alpha$/Fe ratios in excess of solar are
required to obtain fits of similar quality to the (meteoritic)
multitemperature models of Centaurus and Perseus; e.g. using
photospheric solar abundances for the 2T model of Centaurus we obtain
$Z_{\rm Fe}=0.83Z_{\sun}$ $Z_{\rm Si}=1.2Z_{\sun}$ $Z_{\rm
S}=1.1Z_{\sun}$ $Z_{\rm Ar}=1.4Z_{\sun}$ $Z_{\rm Ca}=1.4Z_{\sun}$.
This Si/Fe ratio of $\sim 1.5Z_{\sun}$ (photospheric) is similar to
the value of 2 solar (photospheric) obtained by Fukazawa et
al. \shortcite{fuk98} using isothermal models within an annulus of
$0.1h^{-1}_{50}$ - $0.4h^{-1}_{50}$ Mpc which excludes the region we
have analyzed (Table \ref{tab.obs}). When using the photospheric solar
abundances for the 1T model of Centaurus we obtain a Si/Fe ratio of 2
solar in good agreement with Fukazawa et al.. Hence, the Si/Fe
enhancements obtained by Fukazawa et al. are due in part to their
using photospheric solar abundances, and if the hot gas within their
apertures deviates significantly from isothermal then the assumption
of isothermal gas contributes as well.

When accounting for the differences between meteoritic and
photospheric solar abundances we find that the Fe abundances we have
obtained agree well with previous {\sl ASCA} studies; e.g. for M87
(Matsumoto et al. 1996; Hwang et al. 1997), for Centaurus (Fukazawa et
al. 1994; Fabian et al. 1994a), and for Perseus (Fabian et
al. 1994a). We mention that for the multitemperature models we do not
require the abundances to be different for different temperature
components. That is, our models do not require abundance gradients
within our apertures but they do not exclude them either. The
existence of abundance gradients within our apertures does not
compromise the abundances inferred by our simple models because, as we
have shown previously for ellipticals and groups (see section 4.1 in
Buote 1999b and section 5.3 of Buote 1999a), these models provide
accurate measurements of the average Fe abundances obtained from the
integrated spectrum within a large spatial region.

\subsubsection{Evidence for high energy excesses?}
\label{pl}

Examination of the residuals from the 2T models (Figures \ref{fig.m87}
- \ref{fig.per}) suggests excess emission for energies above $\sim 7$
keV for Centaurus and perhaps the other clusters as well. Using simple
models we have investigated the contribution of a non-thermal
component which manifests itself preferentially at higher energies in
the {\sl ASCA} bandpass. We find that the results for a canonical AGN
component (i.e. power law with photon index 1.7 -- Mushotzky et
al. 1993) are nearly identical to those of a bremsstrahlung component
with temperature greater than $\sim 10$ keV. We shall focus on the
results using the AGN model.

In Table \ref{tab.fits} we give the results of adding the simple AGN
power-law model (PL) to the 1T and 2T models.  The 1T models are
improved significantly when the PL component is added, but the
improvement in each case is not nearly as great as that observed for
the 2T models over the 1T models. The PL component only improves the
2T models slightly. A negligible improvement in the multitemperature
fits is expected because of the good agreement between the Fe
K$\alpha$ ratios and the local continuum temperatures discussed in
section \ref{ratio}.

It might be expected that when the GIS data are included in the fits
then, because of its better sensitivity to energies above $\sim 8$
keV, the need for a power-law component will be made clearer. In fact,
we find just the opposite. When the GIS data are included the relative
improvement of the 2T+PL model over the 2T models is {\sl less} for
Centaurus and Perseus although for M87 the results are similar to
those of the SIS data alone. We find that for the simultaneous
SIS+GIS fits for Centaurus and Perseus the SIS residuals exceed the 2T
model while the GIS residuals lie below the best-fitting 2T
model. Differences between the SIS and GIS at these energies are not
unexpected because of the calibration issues mentioned in section
\ref{lineprop} (see Gendreau \& Yaqoob 1997). At any rate, the weak
evidence for the power-law components obtained from the 2T models
(which also applies to constant-$\xi$ models) is unclear because
Gendreau \& Yaqoob \shortcite{gendreau} note that existing calibration
errors can give rise to a hard tail in the data. Hence, our spectral
analysis does not provide convincing evidence in support of excess
hard emission over that produced by the hot plasma components. The
lack of excess hard emission in M87 is supported by recent RXTE
observations \cite{csr}.

We mention as a caveat that if lower energies are included in the fits
(e.g. down to $\sim 0.6$ keV) then the temperatures of the 2T models
are lowered (due to the influence of cooler temperature components)
and there is more of a need for an additional higher energy component
which can be modeled with a (now intrinsically absorbed) power law.
However, disentangling the uncertain shape of $\xi(T)$ and the
relative contribution of a power-law component from issues of excess
absorption and possible uncertainties in the Fe L lines is beyond the
scope of this paper; i.e. we conclude that the multitemperature models
fitted over 1.6-9 keV, which do not require us to accurately model
excess absorption or the Fe L lines, also do not require high energy
excesses.

\section{Prospects for future missions}
\label{sims}

{\sl ASCA} has provided a tantalizing glimpse of the temperature
structure from emission line ratios in a few ellipticals and clusters.
We are fortunate that within the next few years satellites will
provide superior X-ray data capable of determining $\xi(T)$ with
unprecedented precision. Our intention in this section is to give some
indication of how well the multiphase structure of ellipticals and
clusters can be probed by these new missions.

We continue with our prototype examples of NGC 4472 and Centaurus to
examine how well these new missions can distinguish a cooling flow
from a two-temperature plasma (see Section \ref{2tcf}). That is, we
simulate observations of cooling flow models of these clusters (Table
\ref{tab.models}) and then fit two-temperature models to the simulated
data. We then compare ratios of the best-fitting two-temperature
models to the simulated data.

\subsection{ {\sl Chandra} and {\sl XMM} }
\label{axaf}

\begin{figure*}
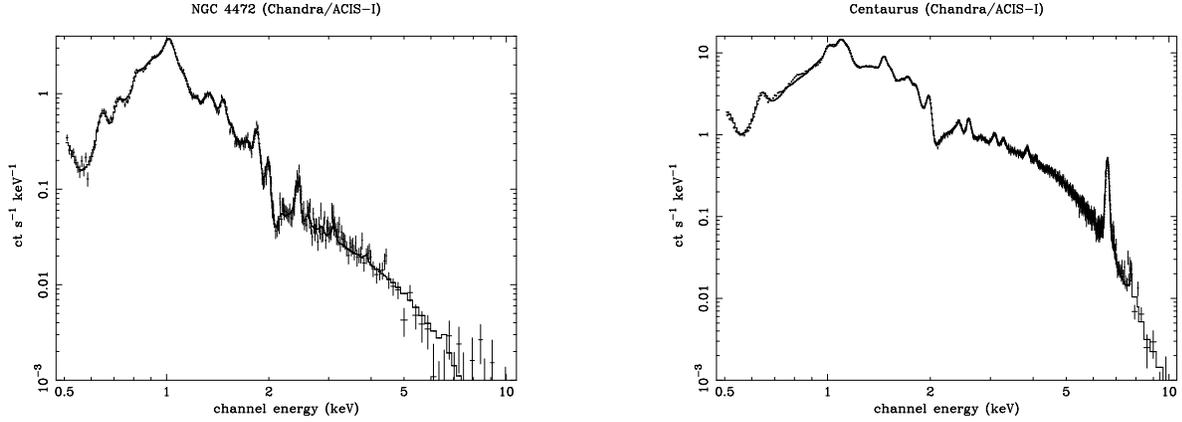

\parbox{0.49\textwidth}{
\centerline{\psfig{figure=fig15a.ps,angle=-90,height=0.23\textheight}}
}
\parbox{0.49\textwidth}{
\centerline{\psfig{figure=fig15b.ps,angle=-90,height=0.23\textheight}}
}
\caption{\label{fig.axaf} Simulated 100 ks observations of NGC 4472
and Centaurus with the {\sl Chandra} ACIS-I following the cooling flow
models in Table \ref{tab.models}. The best-fitting two-temperature
models are also shown.}
\end{figure*}

\begin{figure*}
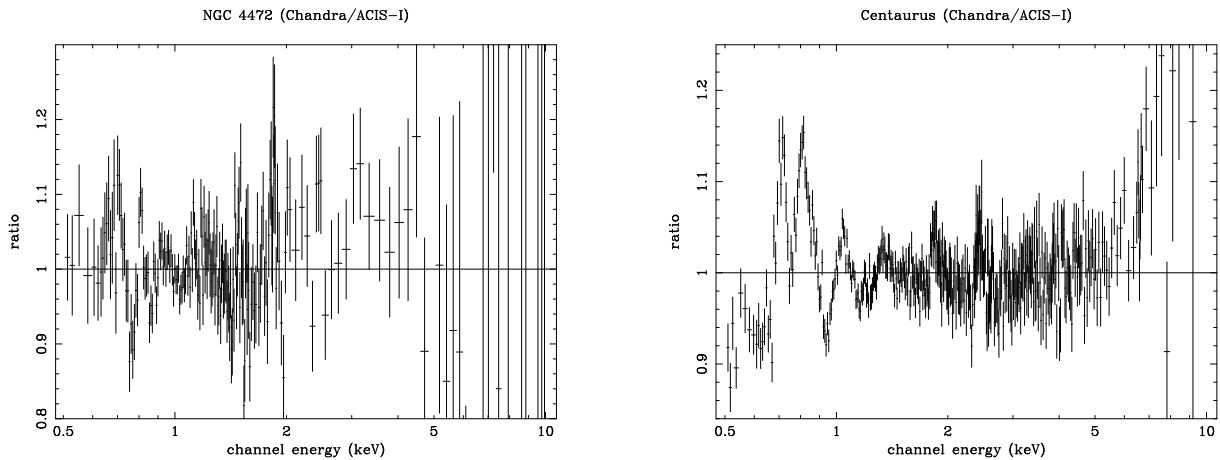

\parbox{0.49\textwidth}{
\centerline{\psfig{figure=fig16a.ps,angle=-90,height=0.25\textheight}}
}
\parbox{0.49\textwidth}{
\centerline{\psfig{figure=fig16b.ps,angle=-90,height=0.25\textheight}}
}
\caption{\label{fig.axaf.ratio} Same as Figure \ref{fig.axaf}
  except we plot the ratio of the best-fitting two-temperature model to
  the simulated data arising from the cooling flow model. Note that we
  increased the energy bin sizes with respect to Figure
  \ref{fig.axaf} for aesthetic reasons.}
\end{figure*}

Our probe into the future begins with the {\sl Chandra X-Ray
Observatory}\footnote{http://chandra.harvard.edu} which is scheduled
to launch in 1999 and the {\sl X-Ray Multi-Mirror Mission
(XMM)}\footnote{http://astro.estec.esa.nl/XMM/} which is scheduled to
launch in January 2000. The ccds on board these satellites have
similar energy resolution, and thus we shall focus on simulations of
{\sl Chandra} observations and note any important differences expected
from {\sl XMM}.

There are several instruments on board {\sl Chandra}, but we focus our
attention on the AXAF CCD Imaging Spectrometer (ACIS) since it is most
appropriate for high resolution spectral analysis of extended
sources. The energy resolution (and bandwidth) of the four front
illuminated CCDs employed in the ACIS imaging array (ACIS-I) is
similar to the pre-flight {\sl ASCA} SIS: e.g.  $\Delta E\sim 60$ eV
(FWHM) at $E\sim 1$ keV and $\Delta E\sim 140$ eV (FWHM) at $E\sim
6.5$ keV. However, the spatial resolution of the ACIS-I is vastly
superior: 90 per cent encircled energy fraction within $1\arcsec$ of a
1.5 keV monochromatic point source.

\begin{figure*}
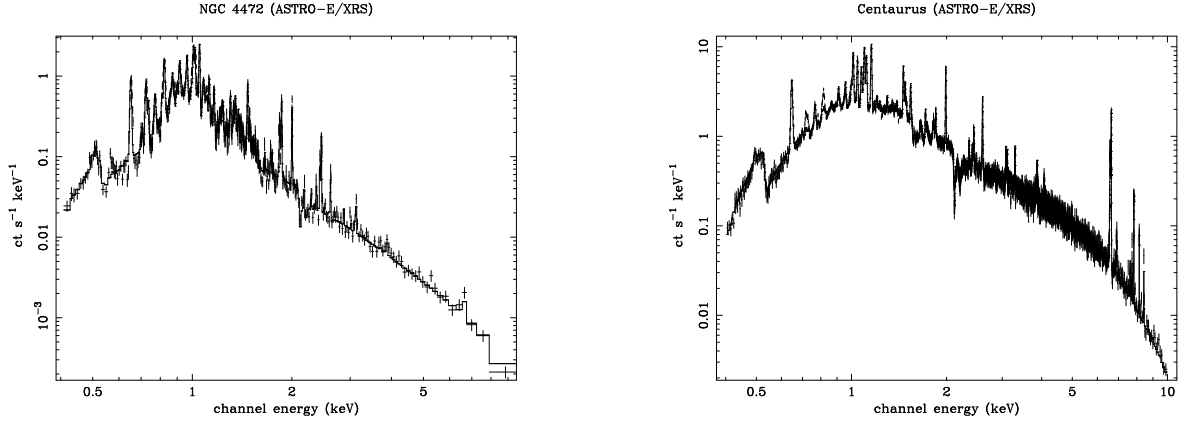

\parbox{0.49\textwidth}{
\centerline{\psfig{figure=fig17a.ps,angle=-90,height=0.23\textheight}}
}
\parbox{0.49\textwidth}{
\centerline{\psfig{figure=fig17b.ps,angle=-90,height=0.23\textheight}}
}
\caption{\label{fig.astroe} Same as Figure \ref{fig.axaf} except the
  observations are simulated for the {\sl ASTRO-E} XRS.}
\end{figure*}

\begin{figure*}
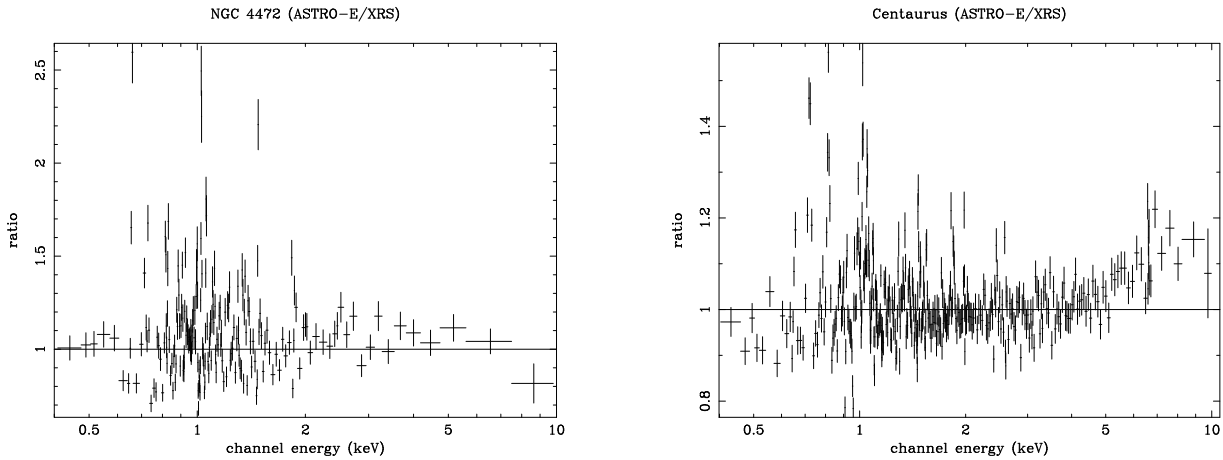

\parbox{0.49\textwidth}{
\centerline{\psfig{figure=fig18a.ps,angle=-90,height=0.25\textheight}}
}
\parbox{0.49\textwidth}{
\centerline{\psfig{figure=fig18b.ps,angle=-90,height=0.25\textheight}}
}
\caption{\label{fig.astroe.ratio} Same as Figure \ref{fig.astroe}
  except we plot the ratio of the best-fit two-temperature model to
  the simulated data arising from the cooling flow model. Note that we
  increased the energy bin sizes with respect to Figure
  \ref{fig.astroe} for aesthetic reasons.}
\end{figure*}

With XSPEC we simulated 100 ks ACIS-I observations of NGC 4472 and
Centaurus using the cooling flow models listed in Table
\ref{tab.models}; we used the default RMF and ARF files supplied for
simulating AO1 observations. We included background in these
simulations using models for the diffuse X-ray background and
instrumental background following Figure 5.14 of the AXAF Proposers'
Guide
Rev. 1.0\footnote{http://asc.harvard.edu/USG/docs/docs.html}. The
background-subtracted spectrum of NGC 4472 is shown in Figure
\ref{fig.axaf}. Also shown is the best-fitting 2T+BREM model (variable
$N_{\rm H}$ for each plasma component, abundances are tied
together). In Figure \ref{fig.axaf.ratio} we plot the ratio of the
best-fitting 2T+BREM model to the simulated data.

The simulated ACIS-I data for NGC 4472 are of substantially higher
quality than the real {\sl ASCA} data (see Buote 1999a). Line blends
of O VIII (H), Mg XI (He) and XII (H), Si XIII (He) and XIV (H), S XV
(He) and XVI (He), and Ar XVII (He) are readily apparent. Deviations
of the model and simulated data are apparent for the K$\alpha$ lines
of Si and S, but the discrepancies are only significant at $\sim 90$
per cent level. However, the 2T+BREM model is clearly rejected by
lines at lower energies. As expected (section \ref{2tcf}) the most
important residuals occur for the Fe L lines from $\sim 0.7 - 0.9$ keV
and the O VIII (H1) line.

We note that {\sl XMM} is able to reject the 2T+BREM model with higher
confidence than {\sl Chandra}. Firstly, since {\sl XMM} has
considerably larger effective area the constraints on the K$\alpha$
lines of Si and S are more precise; i.e. it is not so necessary to
appeal to the Fe L lines to reject the two-temperature
model. Secondly, the larger bandwidth (0.1-12 keV) of {\sl XMM} makes
it more difficult for the two-temperature model to mimic the cooling
flow spectrum.

The simulated 100 ks observation of Centaurus and the best-fitting 2T
model are also shown in Figure \ref{fig.axaf}; similarly the ratio of
the best-fitting 2T model to the simulated data is plotted in Figure
\ref{fig.axaf.ratio}.  All of the line blends detected with the {\sl
ASCA} data (Figure \ref{fig.cen}) are also clearly detected by {\sl
Chandra}. Similar to NGC 4472, the line fluxes are better constrained
than with {\sl ASCA}. The H/He K$\alpha$ ratios of Si and S are both
marginally inconsistent with the two temperature models at the $\sim
2\sigma$ level as is apparent from the excesses in Figure
\ref{fig.axaf.ratio}. Again the most substantial deviations of the
two-temperature model occur for energies $\la 1$ keV where the Fe L
and oxygen lines dominate the spectrum and easily reject the
two-temperature fit. Nevertheless, it is significant that the H/He
K$\alpha$ ratios of Si, S, and Fe can independently differentiate the
two-temperature and cooling flow models for Centaurus so that issues
like the accuracy of the Fe L transitions in the plasma codes and the
amount of absorption are not an issue.

\subsection{ {\sl ASTRO-E} }
\label{astroe}

After {\sl Chandra} and {\sl XMM} a significant leap forward in energy
resolution will be provided by the X-Ray Spectrometer (XRS) on board
the {\sl ASTRO-E}
satellite\footnote{http://lheawww.gsfc.nasa.gov/docs/xray/astroe/}
scheduled for launch in March 2000.  Unlike previous X-ray detectors,
the XRS uses an array of X-ray micro-calorimeters which will provide
an unprecedented combination of high energy resolution ($\Delta E\sim
10$ eV FWHM) and large collecting area. Although its spatial
resolution is limited and its expected lifetime is only $\sim 2$
years, the XRS will have the ability to measure individual emission
lines of extended sources such as ellipticals, groups, and clusters
like no previous X-ray mission.

In analogy to the {\sl Chandra} cases, we simulated 100 ks
observations with the XRS for the cooling flow models of NGC 4472 and
Centaurus.  We used the default RMF, ARF, and background files
provided by the {\sl ASTRO-E} GOF web
site\footnote{http://heasarc.gsfc.nasa.gov/docs/astroe/astroegof.html}.
We show in Figure \ref{fig.astroe} the simulated XRS spectra and
best-fitting two-temperature models of NGC 4472 (2T+BREM) and
Centaurus (2T). The ratios of the two-temperature models to the
simulated data are displayed in Figure \ref{fig.astroe.ratio}

With the vastly improved resolution many blends of lines in the {\sl
Chandra} and {\sl XMM} (and {\sl ASCA}) data separate. The He-like
K$\alpha$ transitions of Si and S become separate He4 lines and blends
of He4-5 and satellites; similar transitions of Mg are also
resolved. Many new transitions are resolved including the K$\beta$
lines of Si XIII (He3) and Si XIV (H2). The Fe XXV complex for
Centaurus, though still unresolved, shows significant structure
indicating separate contributions from the He4, He5-6, and satellites.
Large residuals in the two-temperature fits at the energies of these K
shell lines shows that finally at the resolution of the XRS the K
shell lines can clearly differentiate between two-temperature and
cooling flow models for NGC 4472 in addition to Centaurus. The novel
constraints on ratios of K$\beta$/K$\alpha$ lines of the same element
can be used to verify fundamental assumptions of the coronal model;
e.g. the assumptions of ionization equilibrium and an optically thin
plasma.

Moreover, the oxygen lines and many of the key Fe L blends are easily
measured with the XRS. The O VII (He) and OVIII (H1) lines for NGC
4472 clearly distinguish the simulated cooling flow from the
two-temperature fit. (Note that allowing the relative abundances of,
e.g., O, Ne, Mg, Si, S, Ar, and Ca to be free does not improve the fit
substantially.) Of more importance is the Fe XVII blend at $\sim 0.7$
keV (see Table \ref{tab.fel}) and other Fe L lines near 1 keV. It is
clear that data of this quality will provide exciting new insights
into the temperature structure of the hot plasma in these systems.

\section{Discussion}
\label{disc}

\subsection{Resonance Scattering}
\label{res}

It is usually assumed that the hot gas in ellipticals and clusters is
optically thin to X-rays. However, Gil'fanov, Syunyaev, \& Churazov
\shortcite{gilfanov} have argued that resonance scattering of photons
produced by strong transitions to the ground state of an abundant ion
may be important in these systems.  Gil'fanov et al. estimated
significant optical depths $\tau\sim 4.0$ and 2.7 to the Fe XXV (He4)
line in the centers of Perseus and M87 respectively.  They predicted
that steeply rising abundance profiles with increasing radius should
be observed if the intrinsic abundances are constant with radius.

In fact, observations of clusters almost always show flat abundance
profiles or, particularly for cooling flow clusters, abundance
profiles that increase towards the center (e.g. Fukazawa et al.
1994). It would seem unlikely that in most cases the nearly flat
abundance profiles of observed clusters are actually due to large
intrinsic negative abundance gradients which almost exactly cancel the
predicted effect due to resonance scattering. As noted by Gil'fanov et
al., turbulent motions in the gas may significantly reduce the
importance of resonance scattering.  (Turbulent velocities of $\sim
300$ km s$^{-1}$ should be sufficient to reduce the optical depth of
the Fe XXV He4 line below unity in a massive cluster.)  Fortunately,
as shown by Gil'fanov et al., resonance scattering modifies the
observed line shapes and thus its importance may be directly measured
by future missions with sufficient energy resolution (e.g. {\sl
Constellation-X}).

As shown in section \ref{ratio} the Fe K$\beta$/K$\alpha$ ratio
measured from the {\sl ASCA} data of Perseus is consistent with
optically thin plasma and does not suggest resonance scattering in
contrast to the result obtained by Molendi et al. \shortcite{molendi}
using {\sl SAX} data. Although resolving the discrepancy between the
{\sl ASCA} and {\sl SAX} results is beyond the scope of our paper, we
believe the consistent results we have obtained for M87, Centaurus,
and Perseus using {\sl ASCA} data argue that resonance scattering has
a negligible impact on the Fe abundances inferred from current X-ray
data.

\subsection{Multiphase Strengths of Clusters and $\Omega_0$}
\label{omega}

Since structure formation in the Universe is believed to proceed by
hierarchical clustering, galaxy clusters should show evidence of merging
depending on their dynamical ``age''. The relative fraction of
dynamically young clusters in a complete cluster sample should be
sensitive to the cosmological density parameter $\Omega_0$ as first
suggested by Richstone, Loeb, \& Turner \shortcite{rlt}. Recent
analysis of X-ray cluster morphologies favors a universe with low
matter density $\Omega_0\sim 0.35$ \cite{bx}.

Cluster morphologies, however, are only one indicator of the dynamical
states of clusters. The dynamical state indicated by cluster
morphology is the same as that indicated by the quantity, $\Delta M
/\, \overline{M}$, the fractional amount of mass a cluster accreted
during the previous crossing time, but the connection between cluster
morphology and $\Delta M /\, \overline{M}$ (and thus $\Omega_0$) is
rigorously valid only for clusters that are not very close to a
virialized state (see section 2.1 of Buote 1998).

In contrast, the temperature structure of clusters can provide a
sensitive measure of the dynamical states of nearly relaxed clusters.
For illustrative purposes, let us consider a massive cluster ($T\sim
8$ keV) which has recently formed. At this time there will be little
or no contribution to the emission measure $(\xi(T)dT)$ from a cooling
flow since arguments \cite{mcglynn} and {\sl ROSAT} data (Buote \&
Tsai 1996; see also Jones \& Forman 1999) show that the presence of
cooling flows anti-correlates with substantial substructure.  Within
the central few hundred kpc there will be significant temperature
fluctuations arising from incomplete relaxation (e.g.  as in A754 and
A2256 -- Henry \& Briel 1995 and Briel \& Henry 1995).  When the
cluster has relaxed to a state that is close to the onset of a cooling
flow, the emission measure distribution will be peaked near $T\sim 8$
keV along with residual temperature fluctuations. In terms of the
multiphase strength (Section \ref{mps}), such a cluster about to start
a cooling flow will have typically $\sigma_\xi\la 0.1$.

As the cluster relaxes a cooling flow will gradually develop. The
emission measure distribution will widen until it resembles those in
Figure \ref{fig.cf} with $\sigma_\xi\sim 0.7$ - 0.8. (Within the
central few hundred kpc the cooling flow can contribute up to $\sim
80$ - 90 per cent of the X-ray emission; e.g. Fabian 1994.)  Notice
that the dynamic range in $\sigma_\xi$ between a cooling flow cluster
$(\sim 0.7)$ and an incipient cooling flow $(\la 0.1)$ is considerably
larger than the difference between an incipient cooling flow and an
isothermal gas ($\sigma_\xi=0$); i.e. the temperature structure,
expressed in terms of $\sigma_\xi$, is a sensitive discriminant of the
``age'' of nearly relaxed clusters unlike cluster morphologies.

However, clusters which have recently experienced major mergers can
have a wide distribution of temperature components contributing to the
emission measure, which in some cases may give $\sigma_\xi$ comparable
to cooling-flow clusters.  It will be necessary, as a result, to
consider the joint $\sigma_\xi$ and morphology (e.g.
$\log_{10}P_m/P_0$ -- Buote \& Tsai 1995) distributions to remove this
potential degeneracy of cluster ages with large $\sigma_\xi$.  Hence,
with future X-ray satellites capable of detailed mapping of the
emission measure distributions of clusters (e.g. {\sl ASTRO-E,
  Constellation-X}), the multiphase strength $\sigma_\xi$ can be
measured for many clusters and be compared to those produced in
N-body/ Hydrodynamical simulations (and combined with information from
cluster morphologies) in order to constrain $\Omega_0$.

It is possible with currently available data from {\sl ASCA} to
estimate $\sigma_\xi$ of many clusters without detailed mapping of
$\xi(T)$ for each cluster.  This is possible if one assumes that the
true $\xi(T)$ of clusters is accurately represented by a two component
model consisting of a constant pressure cooling flow and an isothermal
ambient gas component (e.g. cooling flow models in Table
\ref{tab.models}). As described in section \ref{mps}, the multiphase
strength of such models is just $f\sigma_\xi$, where $\sigma_\xi$ is
the multiphase strength of the cooling flow component (Figure
\ref{fig.sigxi_cf}) and $f$ is the fractional contribution of the
cooling flow to the total emission measure. Since these two-component
models describe the available data of clusters quite well, this
procedure for estimating $\sigma_\xi$ should be a good approximation
in most cases.

\section{Conclusions}
\label{conc}

We have studied the ability of X-ray emission line ratios to probe the
temperature structure of elliptical galaxies and galaxy clusters using
theoretical models, the best available {\sl ASCA} data, and simulated
observations with {\sl Chandra} and {\sl ASTRO-E}.  We have emphasized
multiphase spectra consisting of either two temperatures or cooling
flow models and examined the best means of distinguishing between
them. Ratios of lines of K shell transitions have been emphasized over
the Fe L lines because of possible remaining inaccuracies in the
available plasma codes.

From analysis of isothermal MEKAL models computed with energy
resolution 2 eV (FWHM) appropriate for the proposed {\sl
Constellation-X} mission, we found that the most useful ratios of K
shell transitions for probing the emission measure distributions of
ellipticals and clusters are the H1/He4 (K$\alpha$) and satellite/He4
ratios of the same element (notation described in Table
\ref{tab.notation}). The H2/H1 and He3/He4 ratios (K$\beta$/K$\alpha$)
are much less temperature sensitive for temperatures where the
transitions are strong. Ratios of similar transitions of different
elements (e.g., He4 of S/Fe) are very temperature sensitive but must
be used with caution for interpreting temperatures structure since
they depend on the relative abundances of the elements. Finally, we
also discuss the very strong temperature dependence of a few ratios of
Fe L lines between $\sim 0.7$ - 1.4 keV which are typically the
strongest lines and most temperature sensitive ratios of all.

We considered a few simple multiphase models to evaluate how well the
various line ratios can constrain the general differential emission
measure $\xi(T)$. To facilitate quantitative comparison of ellipticals
and clusters having different $\xi(T)$ we introduced the ``multiphase
strength'' parameter, $\sigma_\xi$, which measures the fractional
width of $\xi(T)$: $\sigma_\xi=0$ for an isothermal gas and
$\sigma_\xi\cong 1$ for a constant emission measure distributed over a
large temperature range.  Of the simple multiphase models examined
(two temperature, constant emission measure, gaussian emission
measure, cooling flows) it is found that the two-temperature models
have the most flexibility; i.e.  for a given value of $\langle
T\rangle$ (emission measure weighted temperature) the two-temperature
models can generally provide the best approximation to a given line
ratio.

The flexibility of the two-temperature model is illustrated by its
ability to match very closely a model consisting of a cooling flow
component and a single temperature component corresponding to ambient
gas (Figure \ref{fig.2tvscf}). Cooling flows have nearly maximal
multiphase strength ($\sigma_\xi\sim 0.7$), but their K$\alpha$ line
ratios (aside from O) can be approximated by two temperature models
within $\sim 20$ per cent for elliptical galaxies like NGC 4472 and
within $\sim 50$ per cent for clusters like Centaurus. The best means
to distinguish between two-temperature models and cooling flows is to
use oxygen lines or Fe L lines whose ratios can differ by a factor of
$\sim 10$ between the models.

We have re-analyzed the {\sl ASCA} spectra of three of the brightest
galaxy clusters to assess the evidence for multiphase gas in their
cores: M87 (Virgo), the Centaurus cluster, and the Perseus
cluster. K$\alpha$ emission line blends of Si, S, Ar, Ca, and Fe are
detected in each system as is significant Fe K$\beta$ emission.  The
Fe K$\beta$/K$\alpha$ ratios are consistent with optically thin plasma
models and do not suggest resonance scattering in these systems.
Consideration of both the ratios of H-like to He-like K$\alpha$ lines
and the continuum temperatures clearly rules out isothermal gas in
each case.

To obtain more detailed constraints we fitted plasma models over 1.6-9
keV where the emission is dominated by these K shell lines and by
continuum. In each case the {\sl ASCA} spectra cannot determine
whether the gas emits at only two temperatures or over a continuous
range of temperatures as expected for a cooling flow. The metal
abundances are near solar for all of the multiphase models. For
Centaurus and Perseus the Si/Fe ratios are consistent with the
meteoritic solar values. We also find that our multitemperature models
fitted over 1.6-9 keV, which do not require us to accurately model
excess absorption or the Fe L lines, also do not require any excess
emission at higher energies from, e.g, an AGN component.

We examined the ability of future X-ray missions to constrain $\xi(T)$
in ellipticals and clusters. By simulating observations with the {\sl
Chandra} ACIS-I and the {\sl ASTRO-E} XRS of cooling flow models of
NGC 4472 and Centaurus we found that these future instruments will
have the ability to distinguish between two-temperature models and
cooling flows. However, the lower resolution {\sl
Chandra} ACIS-I (and also the {\sl XMM} ccds) must rely on constraints
from O and Fe L lines for definitive results whereas the {\sl
ASTRO-E} XRS only requires the K$\alpha$ lines above $\sim 2$ keV
(e.g. Si and S).

We have discussed the evidence for resonance scattering in light of
our results for the K$\beta$/K$\alpha$ ratios. We also discuss the
multiphase strength, $\sigma_\xi$, as an indicator of the dynamical
states of clusters and its potential use as a probe of $\Omega_0$ in
conjunction with galaxy cluster morphologies.

The difficulty in distinguishing between different multiphase models,
specifically a two-temperature plasma from a cooling flow, has been
emphasized in this paper.  Presently, there are only a few cases where
line ratios have been measured with sufficient precision to merely
indicate that the gas is not isothermal. It is imperative that the
temperature structures of ellipticals and clusters be considered when
interpreting properties (e.g. abundances) derived from spectral fits
in these systems.

\section*{Acknowledgments}

We thank J. Kaastra and the referee, M. Loewenstein, for helpful
comments.  This research has made use of ASCA data obtained from the
High Energy Astrophysics Science Archive Research Center (HEASARC),
provided by NASA's Goddard Space Flight Center. DAB acknowledges
support for this work provided by NASA through Chandra Fellowship grant
PF8-10001 awarded by the Chandra Science Center, which is operated by the
Smithsonian Astrophysical Observatory for NASA under contract
NAS8-39073.

\end{document}